\documentclass[twocolumn,aps,prb,superscriptaddress,showpacs,floatfix]{revtex4-2}
\usepackage{graphicx}
\usepackage{hyperref} 
\usepackage{subcaption}
\usepackage{tabularx}
\usepackage{array}
\usepackage{makecell}
\usepackage{footmisc}
\usepackage{enumitem}
\usepackage[utf8]{inputenc}
\renewcommand{\selectlanguage}[1]{}
\usepackage{amsmath, amssymb, amsfonts}
\usepackage{cellspace}
\DeclareSymbolFont{extraup}{U}{zavm}{m}{n}
\DeclareMathSymbol{\varheart}{\mathalpha}{extraup}{86}
\DeclareMathSymbol{\vardiamond}{\mathalpha}{extraup}{87}
\usepackage{physics}
\usepackage{xcolor}
\usepackage[normalem]{ulem}
\newcolumntype{Y}{>{\raggedright\arraybackslash}X}

\usepackage{siunitx}
\usepackage{mathrsfs}
\usepackage[version=4]{mhchem}
\usepackage{tikz}
\usetikzlibrary{shapes.geometric, arrows, positioning, fit, calc, backgrounds}
\usepackage{booktabs}
\usepackage{multirow}

\RenewDocumentCommand{\paragraph}{s m}{%
  \par\addvspace{0.4\baselineskip}%
  \noindent\textit{#2}\textemdash\ignorespaces
}
\newcommand{\done}[1]{}
\begin{document}
\newcommand{\magImg}[1]{\raisebox{-0.5\height}{\includegraphics[width=1.2cm]{#1}}}
\newcommand{\tabImg}[1]{\raisebox{-0.5\height}{\includegraphics[width=2.8cm]{#1}}}
\title{Symmetry Classification of Multipolar Orders in Crystals: Theory, Property Tensors and Automated Analysis with \textit{MagSymMultipoles}}

\author{Maxime Braun}
\email{maxime.braun@neel.cnrs.fr}
\affiliation{Univ. Grenoble Alpes, CNRS, Institut Néel, 25 Rue des Martyrs, 38042, Grenoble, France}

\author{Quintin N. Meier}
\email{quintin.meier@neel.cnrs.fr}
\affiliation{Univ. Grenoble Alpes, CNRS, Institut Néel, 25 Rue des Martyrs, 38042, Grenoble, France}

\date{\today}

\begin{abstract}
Many structural and magnetic phases and properties of crystalline materials can be related to the ordering of electric and magnetic multipoles. Examples include ferroelectricity, linear magnetoelectricity, and altermagnetism. Determining the symmetry-allowed multipoles in a crystal is therefore useful for characterizing its ordered phases and physical properties. Here, we present a unified Cartesian framework for decomposing moment tensors of arbitrary rank into ordinary, toroidal, and poloidal multipoles, and for determining, from crystallographic and magnetic symmetry, their ferroic, antiferroic, and noncollinear arrangements within the crystal. We further establish the direct symmetry relationship between multipolar order and the allowed components of associated physical-property tensors in both relativistic and non-relativistic settings.

We implement this methodology in \textit{MagSymMultipoles} (\url{https://mag-sym-multipoles.com}), an interactive web application for calculating and visualizing symmetry-adapted electric and magnetic multipoles. The application uses magnetic symmetries both with and without spin-orbit coupling to derive the allowed multipoles as well as the underlying Cartesian moment tensors, making it easy to explore the connection between a given multipolar order and the corresponding physical response tensors. 
We demonstrate the approach for ferroelectric \ce{BaTiO3},
antiferroelectric \ce{PbZrO3}, magnetoelectric \ce{Cr2O3},
and the altermagnets \ce{MnF2}, \ce{MnTe}, and \ce{Mn3IrSi}.
These examples demonstrate how our method allows us to obtain an intuitive picture linking the multipolar order of a material directly to its physical properties, thereby facilitating the interpretation of theoretical and experimental results.

\end{abstract}

\maketitle
\section{Introduction}
The multipole expansion is a standard tool of electromagnetism for describing the fields of localized charge
and current distributions in terms of contributions of successively higher multipolar order~\cite{jacksonClassicalElectrodynamics1998}.
Multipole moments characterize the spatial structure of charge, current, polarization, or magnetization distributions and
determine their contributions to the resulting fields, as well as their coupling to external fields and field gradients. In a crystal, the crystallographic symmetry operations fix the allowed multipole components. Experimental probes that resolve individual
multipolar contributions therefore provide valuable information about the symmetry and ordering of materials~\cite{
matteoResonantXrayDiffraction2012,
WilliamLovesey2013XrayDiffractionMagnetic,
WilliamLovesey2010ParityOddAtomicMultipoles,
VanDerLaan2021ElectronicMultipolesSecond,
Yanagisawa2024UnlockingHiddenMultipole,
Suzuki2018FirstprinciplesTheoryMagnetic,
Urru2023NeutronScatteringLocal}.

At the same time, the irreducible multipoles themselves can serve as order parameters for a variety of ordered phases: ferroelectrics and ferromagnets are
traditionally characterized by ordered electric and magnetic dipoles, and higher-rank moments capture more complex order. For example, electric quadrupoles serve as order
parameters for ferroelastic, electronic-nematic, and Jahn--Teller
phases~\cite{rosenbergDivergenceQuadrupolestrainSusceptibility2019,massatFieldtunedFerroquadrupolarQuantum2022,sartbaevaQuadrupolarOrderingLaMnO2007}, while magnetic quadrupoles and magnetoelectric monopoles characterize magnetoelectric
order~\cite{hehlRelativisticNatureMagnetoelectric2008,spaldinMonopolebasedFormalismDiagonal2013}. Higher-order magnetic multipoles have
recently been invoked to characterize altermagnets, which are collinear magnets with non-relativistic spin
splitting~\cite{smejkalEmergingResearchLandscape2022,krempaskyAltermagneticLiftingKramers2024,bhowalFerroicallyOrderedMagnetic2024,Mizumaki2025DetectionFerroicOctupole,verbeekNonrelativisticFerromagnetotriakontadipolarOrder2024}. Toroidal
multipoles, which describe vortex-like arrangements of magnetic dipoles, characterize ferrotoroidic
order~\cite{spaldinToroidalMomentCondensedmatter2008,
VanAken2007ObservationFerrotoroidicDomains}, while electric toroidal multipoles describe
ferroaxial phases~\cite{Hlinka2016SymmetryGuideFerroaxial,
Jin2020ObservationFerrorotationalOrder}, and electric toroidal monopoles serve as order parameters for structural chirality~\cite{Inda2024QuantificationChiralityToroidalMonopole}.

Electric and magnetic multipoles have historically been introduced in different physical contexts and with different conventions.
Consequently, the relations between Cartesian moment tensors, their irreducible multipole decomposition,
and macroscopic property tensors are not always transparent.
For magnetic systems, the relevant symmetry constraints also depend on whether spin--orbit coupling is included or neglected, giving rise to distinct relativistic and non-relativistic
settings~\cite{
Litvin1974SpinGroups,chenEnumerationRepresentationTheory2024,Orgawa2026}.

Several tools exist to automate and facilitate parts of such analyses.
The Bilbao Crystallographic Server provides symmetry-adapted
tensors with and without spin--orbit coupling through
\textsc{TENSOR},
\textsc{MTENSOR}~\cite{Gallego2019AutomaticCalculationSymmetryadapted},
and \textsc{STENSOR}~\cite{
Elcoro2026AutomaticCalculationSymmetryadapted,
Etxebarria2025CrystalTensorProperties}, while \textsc{MSITESYM}~\cite{Aroyo2006BilbaoCrystallographicServer,Aroyo2006BilbaoCrystallographicServerb}
provides site-symmetry groups and magnetic corepresentations for individual Wyckoff positions.
In parallel, symmetry-adapted multipole bases have been developed for magnetic structures
and model Hamiltonians~\cite{
Suzuki2019MultipoleExpansionMagnetic,
Kusunose2023SymmetryadaptedModelingMolecules}.
These approaches provide complementary descriptions of tensor symmetries and multipolar order.
However, a systematic approach that directly connects local multipoles, their symmetry-enforced arrangements across a crystal, and the resulting macroscopic response tensors is still lacking.

Here, we develop a general Cartesian formulation of moment tensors and their multipole decomposition in crystals.
The formulation provides a multipolar decomposition at arbitrary rank and treats ordinary, toroidal, and poloidal electric and magnetic multipoles on the same footing.
We determine their symmetry-allowed components and ordering under
crystallographic, magnetic, and spin-group symmetries.
The resulting correspondence allows symmetry-allowed components, their real-space representations, and their relation to
macroscopic responses to be examined together, thereby connecting symmetry analysis with first-principles
calculations and experimental observables.

We implement the approach in \textit{MagSymMultipoles} (\url{https://mag-sym-multipoles.com}), a browser-based
program that runs the analysis directly from the crystallographic and magnetic
symmetries of a structure, requires no installation, and accepts either a magnetic
space group from built-in tables or a structure in CIF or mCIF format. The program
\begin{itemize}[itemsep=0pt]
    \item determines symmetry-allowed electric and magnetic parent moment tensors
          up to rank~5 for all 1{,}651 magnetic space groups and decomposes them
          into ordinary and dual-derived toroidal and poloidal multipoles,
          visualized in the spherical-harmonic basis;
    \item resolves the analysis globally or over a chosen Wyckoff orbit,
          identifying ferroic, antiferroic, and non-collinear arrangements;
    \item separates relativistic (SOC) and non-relativistic (SOC-free)
          contributions;
    \item maps the allowed moment tensors onto physical property tensors in
          Cartesian form or a compact tensor shorthand.
\end{itemize}

We demonstrate these capabilities on a sequence of materials of increasing
complexity, first in multipole rank and then in the spatial arrangement of the
ordered moments, comparing the symmetry-allowed multipoles with first-principles
calculations in each case:
\begin{itemize}[itemsep=0pt]
    \item \ce{BaTiO3} --- ferroic ordering of electric dipoles (ferroelectricity);
    \item \ce{PbZrO3} --- antiferroic ordering of electric dipoles
          (antiferroelectricity);
    \item \ce{Cr2O3} --- ferroic magnetoelectric-monopole and magnetic-quadrupole order;
    \item \ce{MnF2} --- magnetic octupolar order and $d$-wave altermagnetism;
    \item \ce{MnTe} --- magnetic triakontadipolar order and $g$-wave
          altermagnetism;
    \item \ce{Mn3IrSi} --- magnetic octupolar order in a non-collinear
          altermagnet.
\end{itemize}
Sec.~\ref{sec:theory} introduces the moment tensors, their multipole
decomposition, and their symmetry constraints; Sec.~\ref{sec:order_response} maps
the order parameters onto response functions; Sec.~\ref{sec:architecture}
describes the implementation, including visualization
(Sec.~\ref{sec:visualization}); and Sec.~\ref{sec:examples} presents the examples
above.

\section{Theoretical Background}
\label{sec:theory}

 This section introduces electric and magnetic moment tensors and their decomposition into ordinary, toroidal, and poloidal multipoles. It then derives the symmetry constraints on their local components and ordering in crystals.

\subsection{The multipole expansion}
\label{subsec:multipole_expansion}
The basis of our analysis is the multipole expansion of the charge and magnetization densities~\cite{jacksonClassicalElectrodynamics1998,KaraKurkiSuonio1981,spaldinToroidalMomentCondensedmatter2008,Kuramoto2008ElectronicHigherMultipoles,Kuramoto2009MultipoleOrdersFluctuations,Hayami2024SymmetryClassificationAntiferromagnets,hayamiUnifiedDescriptionElectronic2024}.

\begin{figure*}[htb]
    \centering
\makebox[\linewidth][c]{%
        \includegraphics[width=0.7\linewidth]{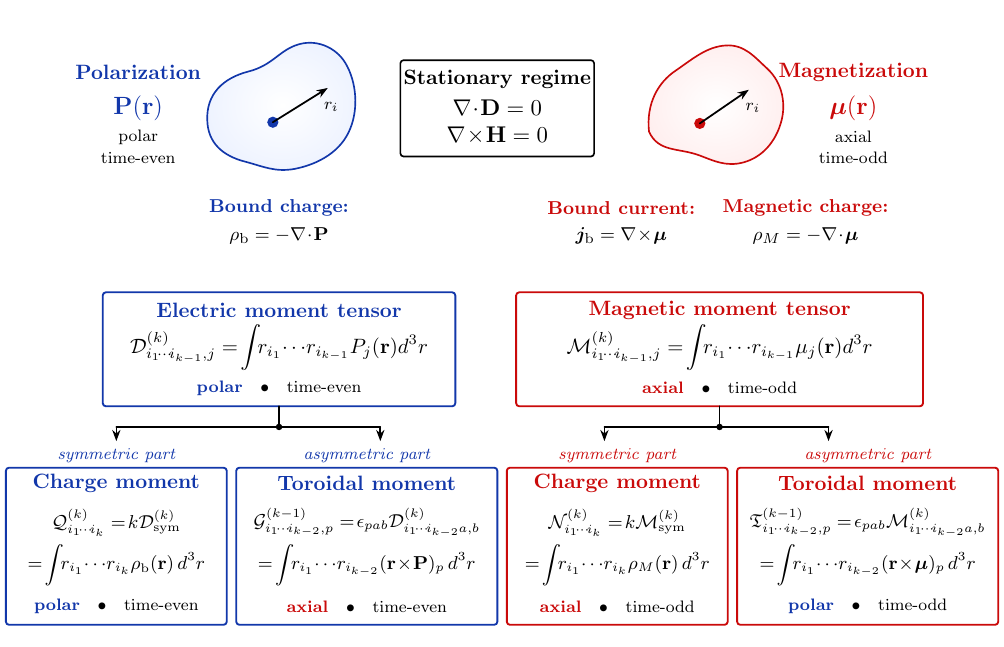}%
    }
    \caption{Relationship between the different macroscopic moment tensors in the stationary (time-independent) regime, in the absence of free currents or free charges.}
    \label{fig:moment_tensors}
\end{figure*}

Figure~\ref{fig:moment_tensors} provides an overview of the relationships
between the electric and magnetic moment tensors introduced below, in the
stationary regime without free charges or free currents.

\subsubsection{Electric moment tensors}
For a localized charge distribution $\rho(\mathbf r)$, the Cartesian electric charge moment tensor of rank $k$ can be defined as
\begin{equation}\label{eq:elemp}
    \mathcal Q^{(k)}_{i_1\dots i_k}
    =
    \int d^3r\;
    r_{i_1}\cdots r_{i_k}\rho(\mathbf r),
\end{equation}
where $\mathbf r$ denotes the position vector relative to a chosen expansion
center. Since the position vectors can be freely permuted,
$\mathcal Q^{(k)}$ is a fully symmetric polar tensor of rank $k$. The cases
$k=0$ and $k=1$ correspond to the isotropic charge and the electric dipole moment, respectively. These tensors arise naturally from the Taylor expansion of the electrostatic interaction energy,
\begin{equation}
    U_{\rm int}^{\rm el}
    =
    \int d^3r\,\rho(\mathbf r)\phi(\mathbf r)
    =
    \sum_{k=0}^{\infty}
    \frac{1}{k!}\,
    \mathcal Q^{(k)}_{i_1\dots i_k}
     \partial_{i_1}\cdots\partial_{i_k}\phi\big|_{\mathbf r=\mathbf 0}.
    \label{eq:electrostatic_energy}
\end{equation}
Thus, the rank-$k$ electric charge moment tensor couples to the $k$-th spatial derivative of the electrostatic potential $\phi$. In particular,
$\mathcal Q^{(0)}\phi(\mathbf 0)$ gives the point-charge energy associated
with the total charge of the distribution, while $\mathcal Q^{(1)}_i \partial_i \phi\big|_{\mathbf r=\mathbf 0}=-p_i E_i$ gives the usual electric dipole energy.

In a macroscopic description, the total charge density separates into free and bound contributions, $\rho=\rho_{\rm f}+\rho_{\rm b}$, where the bound charge is related to the polarization through $\rho_{\rm b}(\mathbf r)=-\nabla\cdot\mathbf P(\mathbf r)$. The bound charge encodes the internal charge distribution in a material, and thus provides a natural starting point for a moment expansion. We first consider an insulator with no free charge, $\rho_{\rm f}=0$, and return to the general case at the end of this subsection. Integration by parts of the bound-charge contribution in
Eq.~\eqref{eq:electrostatic_energy} gives the macroscopic electric interaction
energy

\begin{equation}
    U_{\rm int}^{\rm el}
    =
    -\int d^3r\;
    \mathbf P(\mathbf r)\cdot\mathbf E(\mathbf r).
\end{equation}
The expansion can now be done in terms of the polarization $\mathbf P(\mathbf r)$ rather than the electrostatic potential. The resulting \emph{electric moment tensor} for a macroscopic system is given by
\begin{equation}\label{eq:polmom}
    \mathcal D^{(k)}_{i_1\dots i_{k-1},\,j}
    =
    \int d^3r\;
    r_{i_1}\cdots r_{i_{k-1}}P_j(\mathbf r),
    \qquad k\ge1 .
\end{equation}
We use the convention that $\mathcal D^{(k)}$ carries $k-1$ spatial indices and one polarization index, in order to preserve the same total rank as for $\mathcal{Q}^{(k)}$. The corresponding expansion of the interaction energy is given by
\begin{equation}
    U_{\rm int}^{\rm el}
    =
    -\sum_{k=1}^{\infty}
    \frac{1}{(k-1)!}\,
    \mathcal D^{(k)}_{i_1\dots i_{k-1},\,j}
    \partial_{i_1}\cdots\partial_{i_{k-1}}E_j\big|_{\mathbf r=\mathbf 0}.
\end{equation}
The electric moment tensor is a generalized version of the charge moment tensor for macroscopic systems. Its symmetric part corresponds exactly, up to a renormalization factor $k$, to the charge moment tensor $\mathcal{Q}^{(k)}$,
\begin{equation}
    \mathcal Q^{(k)}_{i_1\dots i_k}
    =k\mathcal D^{(k)}_{(i_1\dots i_k)}=
    k\,\mathcal D^{(k)}_\text{sym} ,
\end{equation}
where parentheses denote symmetrization over all $k$ indices. (Explicit tensor operations are given in Appendix~\ref{app:tensor_formulas}.)

In addition to the symmetric part, $\mathcal D^{(k)}$ also contains an asymmetric component, obtained by removing the fully symmetric part from the total tensor. This can be done by contracting the polarization index and one spatial index with the Levi-Civita tensor $\epsilon_{ijk}$. We obtain an axial rank-$(k-1)$ tensor that is symmetric in its first $k-2$ indices.

\begin{align}
\label{eq:G_from_D}
    \mathcal G^{(k-1)}_{i_1\dots i_{k-2},p}
    &=
    \epsilon_{pab}\,
    \mathcal D^{(k)}_{i_1\dots i_{k-2}a,b}
    =
    \int d^3r\;
    r_{i_1}\cdots r_{i_{k-2}}\,
    (\mathbf r\times\mathbf P)_p,
\end{align}
The tensor $\mathcal G^{(k-1)}$ is called the \emph{electric toroidal moment} and characterizes rotational patterns of the polarization density. The relationship between the different tensors is visualized in Fig.~\ref{fig:moment_tensors}.

If the free charge ($\rho_f$) contributes to the multipoles of the system, it adds an additional contribution to the charge moment tensor,
\begin{equation}\label{eq:Q_free}
    \mathcal Q^{(k)}_{i_1\dots i_k}
    =
    k\,\mathcal D^{(k)}_\text{sym}
    +
    \int d^3r\;
    r_{i_1}\cdots r_{i_k}\,\rho_{\rm f}(\mathbf r).
\end{equation}
Moments of a scalar density are fully symmetric, so the free charge contributes to $\mathcal Q^{(k)}$ alone and cannot generate an electric toroidal moment.

\subsubsection{Magnetic moment tensors}

The magnetic counterpart of Eq.~\eqref{eq:electrostatic_energy} is the interaction energy of a stationary current distribution $\mathbf j(\mathbf r)$, with a vector potential $\mathbf A(\mathbf r)$,
\begin{equation}
    U_{\rm int}^{\rm mag}
    = -\int d^3r\;\mathbf j(\mathbf r)\cdot\mathbf A(\mathbf r),
    \label{eq:magnetostatic_energy}
\end{equation}
In the absence of free currents, $\mathbf j$ contains the bound currents stemming from spin and orbital magnetizations, which we can write as
$\mathbf j_{\rm b}=\nabla\times\boldsymbol\mu$~\cite{jacksonClassicalElectrodynamics1998},
where $\boldsymbol\mu(\mathbf r)$ is the magnetization density. Integration by parts
of Eq.~\eqref{eq:magnetostatic_energy} then leads to
\begin{equation}
    U_{\rm int}^{\rm mag}
    = -\int d^3r\;\boldsymbol\mu(\mathbf r)\cdot\mathbf B(\mathbf r),
    \label{eq:muB_energy}
\end{equation}
where the magnetic flux density $\mathbf B=\nabla\times\mathbf A$ appears as the
field conjugate to $\boldsymbol\mu$. Taylor expansion of this term yields the \emph{magnetic moment tensor} $\mathcal M^{(k)}$,
\begin{equation}\label{eq:magmp}
    \mathcal M^{(k)}_{i_1\dots i_{k-1},\,j}
    = \int d^3r\; r_{i_1}\cdots r_{i_{k-1}}\,\mu_j(\mathbf r),
    \qquad k\ge1,
\end{equation}
which couples to the magnetic flux via the interaction energy
\begin{equation}
    U_{\rm int}^{\rm mag}
    = -\sum_{k=1}^{\infty}\frac{1}{(k-1)!}\,
      \mathcal M^{(k)}_{i_1\dots i_{k-1},\,j}\,
      \partial_{i_1}\cdots\partial_{i_{k-1}}B_j\big|_{\mathbf r=\mathbf 0}.
    \label{eq:magmp_energy_B}
\end{equation}
The magnetic moment tensor carries $k-1$ spatial indices and one magnetization
index.  $\mathcal M^{(k)}$  transforms as an axial, time-odd tensor, and it is symmetric in its spatial indices. It is the magnetic analogue of $\mathcal{D}^{(k)}$. The divergence of the magnetization density defines an effective magnetic charge $\rho_M=-\nabla\cdot\boldsymbol\mu$, from which we construct a
\emph{magnetic charge moment tensor}
\begin{equation}\label{eq:NfromM}
    \mathcal N^{(k)}_{i_1\dots i_k}
    =\int d^3r\;r_{i_1}\cdots r_{i_k}\,\rho_M(\mathbf r)
    =k\,\mathcal M^{(k)}_{(i_1\dots i_k)},
\end{equation}
which captures the fully symmetric part of $\mathcal M^{(k)}$ and plays the role of $\mathcal Q^{(k)}$ in the electric case. (See Fig.~\ref{fig:moment_tensors}.)
The tensor $\mathcal M^{(k)}$ also contains an asymmetric component,
obtained by the antisymmetrization of one spatial index with the magnetic index.
It can be represented by the rank $k-1$ \emph{magnetic toroidal moment} tensor
\begin{align}
\label{eq:T_from_M}
    \mathfrak T^{(k-1)}_{i_1\dots i_{k-2},p}
    &=
    \epsilon_{pab}\,
    \mathcal M^{(k)}_{i_1\dots i_{k-2}a,b}
    =
    \int d^3r\;
    r_{i_1}\cdots r_{i_{k-2}}\,
    (\mathbf r\times\boldsymbol\mu)_p ,
\end{align}
which characterizes rotational patterns of the magnetization density. The relationships between the different tensors and the analogies to the electric tensors are visualized in Fig.~\ref{fig:moment_tensors}.

In the literature, magnetic moment tensors are often introduced through a
thermodynamic free-energy functional expressed in terms of the magnetic field $\mathbf H$~\cite{hehlRelativisticNatureMagnetoelectric2008,spaldinMonopolebasedFormalismDiagonal2013,urruMagneticOctupoleTensor2022}, rather than the flux density $\mathbf B$, since $\mathbf H$ is the variable controlled by the current in a coil. The interaction term reads
\begin{equation}
    F_{\rm int}^{\rm mag}
    = -\mu_0\sum_{k=1}^{\infty}\frac{1}{(k-1)!}\,
      \mathcal M^{(k)}_{i_1\dots i_{k-1},\,j}\,
      \partial_{i_1}\cdots\partial_{i_{k-1}}H_j\big|_{\mathbf r=\mathbf 0}.
    \label{eq:magmp_energy_H}
\end{equation}
Which components of $\mathcal M^{(k)}$ enter the interaction energy depend on the conjugate field. Since $\nabla\times\mathbf H=\mathbf j_{\rm f}=\mathbf 0$, the magnetic field derives from a scalar potential $\mathbf H=-\nabla\phi_M$ and as a result its gradient tensors are fully symmetric. Equation~\eqref{eq:magmp_energy_H} therefore couples to the fully symmetric part of $\mathcal M^{(k)}$, and thus to $\mathcal N^{(k)}$, alone. In other words, $\mathbf{H}$ does not couple to the magnetic toroidal moments (Sec.~\ref{subsec:reducibility}). 
In contrast, if $(\nabla\times\mathbf B\neq 0)$, the magnetic flux density directly couples to the magnetic toroidal moments through Eq.~\eqref{eq:magmp_energy_B}~\cite{gaoMicroscopicTheorySpin2018}.  

\subsection{Decomposition of moment tensors into multipoles}
\label{subsec:reducibility}

\begin{figure}[htb!]
    \centering
    \includegraphics[width=1.05\linewidth]{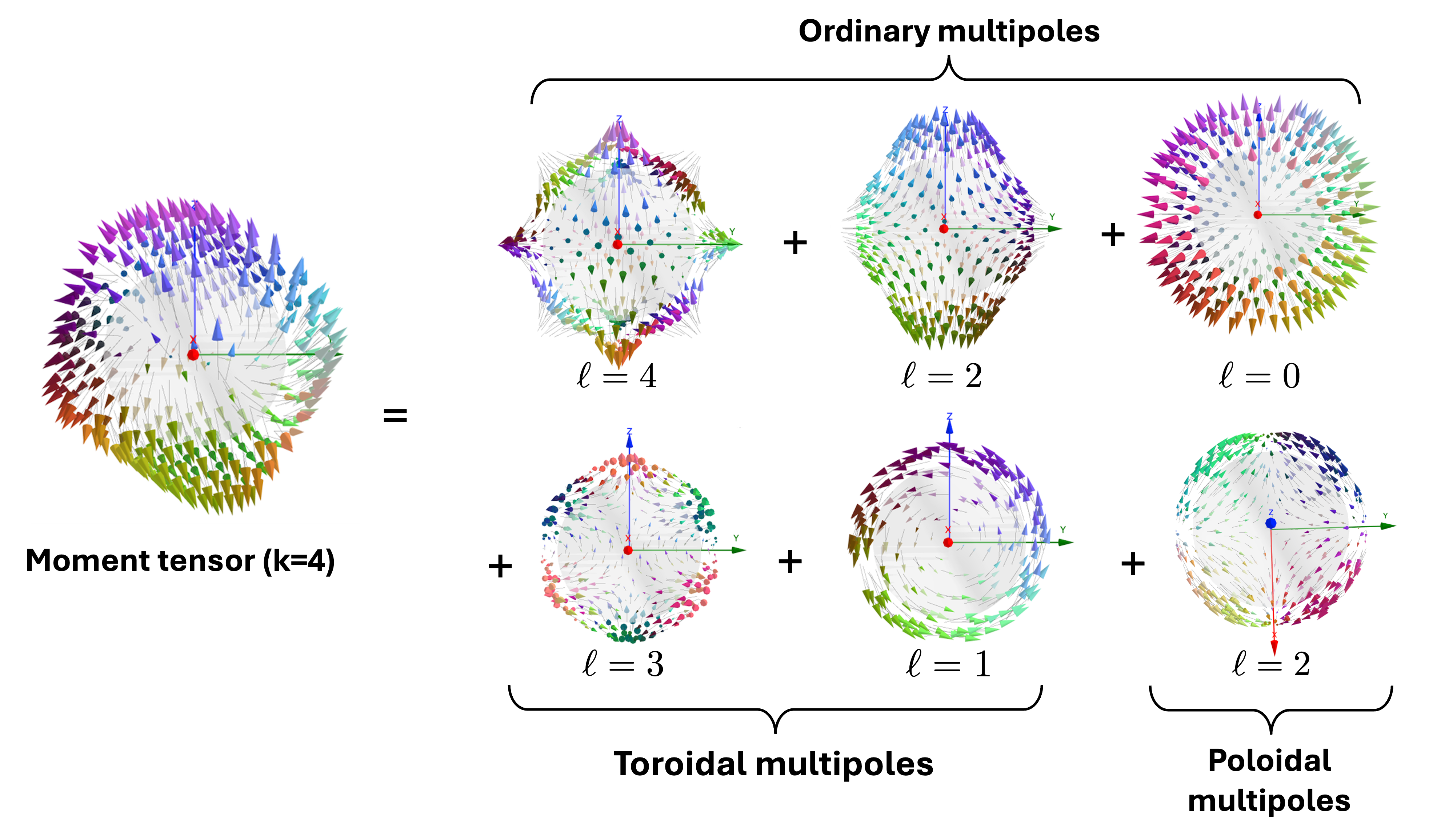}
    \caption{Decomposition of a rank-4 magnetic moment tensor into ordinary and toroidal multipoles, as well as poloidal multipoles.}
    \label{fig:decomposition}
\end{figure}

The moment tensors introduced in the previous section are the general
rank-$k$ moments that arise from the expansion of localized charge and
current densities.  Their components can be decomposed into a combination
of \emph{multipoles}, which are the symmetry-adapted components of the moment tensor that transform as irreducible representations of the three-dimensional
rotation group SO(3) [see
Fig.~\ref{fig:decomposition}]~\cite{thorneMultipoleExpansionsGravitational1980,
dubovikToroidMomentsElectrodynamics1990,urruMagneticOctupoleTensor2022}.
To decompose the tensors, let us first consider the index symmetries of the
different moment tensors.
The charge moments $\mathcal Q^{(k)}$ and $\mathcal N^{(k)}$ are the tensor product of $k$ position indices (vectors) and thus are symmetric under all index permutations.  The electric and magnetic moment tensors $\mathcal D^{(k)}$ and $\mathcal M^{(k)}$ contain one moment index (the polarization component $P_j$ or the magnetization component $\mu_j$) and $k-1$ symmetric position indices. The tensor is generally not symmetric under the exchange of the moment index with a spatial index.  The corresponding vector spaces of the moment tensors are thus given by:
\begin{equation}
\label{eq:moment_index_structure}
\begin{aligned}
    \mathcal Q^{(k)},\mathcal N^{(k)}
    &\in
    \biggl(\underbrace{\mathbf 3\otimes\cdots\otimes\mathbf 3}_{k}\biggr)_{\!\mathrm{sym}},
    \\[2pt]
    \mathcal D^{(k)},\,\mathcal M^{(k)}
    &\in
    \biggl(\underbrace{\mathbf 3\otimes\cdots\otimes\mathbf 3}_{k-1}\biggr)_{\!\mathrm{sym}}\!\!\!\!\!\!\!
    \otimes\mathbf 3 ,
\end{aligned}
\end{equation}
where a boldface $\mathbf d$ denotes a $d$-dimensional irreducible
representation of $\mathrm{SO}(3)$ (where $\mathbf 3$ is the vector representation) and the subscript ``sym'' denotes the space of fully symmetric tensors. 

Let us now decompose a generic moment tensor $\mathcal T^{(k)}_{i_1\cdots i_{k-1},j}$, where the comma separates the $k-1$ symmetric position indices from the moment index (this definition includes the charge moments as special symmetric cases where the last index is symmetric as well).

We first separate the fully symmetric part of the moment tensor from its asymmetric part. The remainder has mixed symmetry: it is symmetric in the first $k-1$ spatial indices, but generally not under exchange of a spatial index with the moment index. The symmetric part contains the ordinary multipoles, while the remainder contains the toroidal and, for $k\geq3$, poloidal sectors.
\begin{equation}
\label{eq:sym_antisym_split}
    \mathcal T^{(k)}
    =
    \underbrace{\mathcal T^{(k)}_{\mathrm{sym}}}_{\text{ordinary multipoles}}
    +
    \underbrace{\mathcal T^{(k)}_{\mathrm{asym}}}_{\text{toroidal and poloidal sectors}} ,
\end{equation}
where $\mathcal T^{(k)}_{\mathrm{sym}}=\mathcal T^{(k)}_{(i_1\cdots i_k)}$ denotes the symmetrization over all $k$ indices [Eq.~\eqref{eq:Tsym_def}]. The asymmetric part is absent for the charge moments, since they are fully symmetric by construction.

\paragraph*{Ordinary multipoles}
Starting with the fully symmetric part $\mathcal T^{(k)}_{\mathrm{sym}}$, we make use of the fact that a symmetric trace-free (STF) tensor of rank $k$ corresponds to the irreducible representation of SO(3) with angular momentum $\ell=k$.  These irreducible representations (the multipoles) can thus straightforwardly be obtained by the subtraction of all possible traces from $\mathcal T^{(k)}_{\mathrm{sym}}$
[Eq.~\eqref{eq:STF_projection}].  Contributions from lower-order multipoles are contained in the traces of the rank-$k$ tensor, and can be obtained using the same methodology: the multipoles of angular momentum $\ell=k-2p$ are obtained
by first taking $p$ traces of the original tensor, and then taking the STF part of that trace.  Since every trace lowers the rank by two, this yields STF tensors of rank $\ell=k,k-2,\ldots$ down to $\ell=0$ for even $k$ and $\ell=1$ for odd $k$.  Thus, we can write the symmetric tensor as the sum
of irreducible multipole tensors,
\begin{equation}
\label{eq:sym_decomposition}
    \mathcal T^{(k)}_{\mathrm{sym}}
    =
    \sum_{\ell=k,\,k-2,\ldots}
    \mathcal T^{(k)}_{\mathrm{STF},(\ell)} ,
\end{equation}
where $\mathcal T^{(k)}_{\mathrm{STF},(\ell)}$ is the rank-$\ell$ multipole
tensor, re-embedded in the rank-$k$ tensor according to
Eq.~\eqref{eq:embedded_component}.  The decomposition of a rank-4 moment tensor, whose leading ordinary multipole is a hexadecapole, is shown in Fig.~\ref{fig:decomposition}, where the ordinary multipoles yield
three multipolar components with $\ell=4,2,0$.

\paragraph*{Toroidal moments and toroidal multipoles}
Next we decompose the asymmetric part $\mathcal T^{(k)}_{\mathrm{asym}}$.
Its content is captured by a lower-rank tensor, given by the Levi--Civita dual of
the original tensor. This tensor is commonly referred to as the \emph{toroidal
moment}~\cite{dubovikToroidMomentsElectrodynamics1990}.  For $k\geq 2$,
\begin{equation}
\label{eq:toroidal_dual}
    \tau^{(k-1)}_{i_1\cdots i_{k-2},c}
    =
    \epsilon_{cab}\,
    \mathcal T^{(k)}_{i_1\cdots i_{k-2}a,\,b} .
\end{equation}
The contraction with $\epsilon_{cab}$ annihilates the fully symmetric part,
so $\tau^{(k-1)}$ contains all components of $\mathcal T^{(k)}_{\mathrm{asym}}$.  The toroidal moment is itself reducible: the
traces between the toroidal moment index $c$ and a position index vanish identically, whereas traces among the position indices do not have to be zero.

As a result, the toroidal moment can be decomposed into toroidal multipoles using the same procedure: subtracting the traces of its symmetric part gives the toroidal
multipole tensors $\tau^{(k-1)}_{\mathrm{STF},(\ell)}$ [Eqs.~\eqref{eq:Tsym_def}, \eqref{eq:STF_projection}], with a leading multipole of $\ell=k-1$ and lower ones $\ell=k-3,k-5,\ldots$ down to
$\ell\geq 1$ ($\ell=0$ is not allowed because it contains the trace of a position and $c$). 
In our example, the rank-4 moment in Fig.~\ref{fig:decomposition} thus yields both a toroidal octupole and a toroidal dipole.  

\paragraph*{Poloidal moments and poloidal multipoles}
So far we have separated the symmetric part of the rank-$k$ moment tensor, which contains the ordinary multipoles, from the asymmetric part, whose Levi--Civita dual, the toroidal moment, yields the toroidal multipoles.
For rank $k\ge3$, the toroidal moment $\tau$ itself is generally not fully symmetric, which means that we
can perform a second dualization,
\begin{equation}
\label{eq:hypertoroidal_double_dual}
   \pi^{(k-2)}_{i_1\cdots i_{k-3},d}
    =
    \epsilon_{dpc}\,
    \tau^{(k-1)}_{i_1\cdots i_{k-3}p,c}.
\end{equation}
 We refer to $\pi^{(k-2)}$ as the
\emph{poloidal moment} in reference to the toroidal--poloidal decomposition of solenoidal fields~\cite{dubovikToroidMomentsElectrodynamics1990,backusPoloidalToroidalFields1986,elsasserInductionEffectsTerrestrial1946}.
To obtain the poloidal multipoles, we can again construct the STF tensors $\pi^{(k-2)}_{\mathrm{STF},i_1\cdots i_{k-3},d}$ by symmetrization and subtraction of the traces, and obtain multipoles of rank $\ell=k-2,k-4,\ldots$ with $\ell\geq1$. 
While $\pi^{(k-2)}$ is not generally a symmetric tensor, its antisymmetric part generates no new multipoles.  Its dual corresponds to the trace of the
toroidal moment, $\epsilon_{qsd}\,\pi^{(k-2)}_{i_1\cdots i_{k-4}s,d}
=-\tau^{(k-1)}_{i_1\cdots i_{k-4}pp,q}$, a multipole component we have already obtained (Appendix~\ref{app:poloidal_antisymmetry}).

\paragraph*{Full decomposition} As we have shown, the decomposition of a rank $k$ moment yields ordinary, toroidal, and poloidal multipoles. This set of multipoles corresponds to the full Clebsch--Gordan decomposition of the rank-$k$ moment tensor:
\begin{equation}
\label{eq:full_dimension_decomposition}
\begin{split}
    \biggl(\underbrace{\mathbf 3\otimes\cdots\otimes\mathbf 3}_{k-1}\biggr)_{\!\mathrm{sym}}
    \otimes\mathbf 3
    =\;
    &\underbrace{\textstyle\bigoplus_{\ell=k,\,k-2,\ldots}(\mathbf{2\ell+1})}_{\text{ordinary multipoles}} \\[2pt]
    \oplus\;
    &\underbrace{\textstyle\bigoplus_{\ell=k-1,\,k-3,\ldots;\,\ell\ge 1}(\mathbf{2\ell+1})}_{\text{toroidal multipoles}} \\[2pt]
    \oplus\;    
    &\underbrace{\textstyle\bigoplus_{\ell=k-2,\,k-4,\ldots;\,\ell\ge 1}(\mathbf{2\ell+1})}_{\text{poloidal multipoles}} .
\end{split}
\end{equation}

For the rank-$4$ tensor shown in Fig.~\ref{fig:decomposition}, this becomes
\begin{equation}
\label{eq:rank4_dimension_decomposition}
    \underbrace{(\mathbf9\oplus\mathbf5\oplus\mathbf1)}_{
        \text{ordinary: }\ell=4,2,0}
    \oplus
    \underbrace{(\mathbf7\oplus\mathbf3)}_{
        \text{toroidal: }\ell=3,1}
    \oplus
    \underbrace{\mathbf5}_{
        \text{poloidal: }\ell=2}
\end{equation}
Thus, the 30 independent components of the tensor can be decomposed into a nine-component hexadecapole, a five-component quadrupole, and a scalar; the toroidal sector contains a seven-component octupole and a three-component dipole; and the poloidal sector contains a second, independent five-component quadrupole.

\subsection{Ordinary, toroidal, and poloidal multipoles}
\label{sec:multipoles}
The tensor components derived in the previous section can be directly related to the real-space density distributions using a projection onto real spherical harmonics $Y_{\ell m}$. The pure multipoles are usually referred to as the top-rank sector of each moment tensor, with angular momentum $\ell$ equal to the tensor rank. Multipoles in the traces of higher-rank tensors pick up an additional radial factor of $r^2$ for each trace.

\paragraph*{Charge multipoles}
The decomposition of the fully symmetric rank-$k$ charge moment tensor $\mathcal Q^{(k)}$ into spherical multipole coefficients is obtained by contracting its rank-$\ell$ STF components with unit vectors projection onto real spherical harmonics $Y_{\ell m}$, with $\ell=k$. Equivalently, these coefficients can be related directly
directly to charge density integrals over real spherical harmonics
\begin{align}
    Q_{\ell m}
    &=
    \int_{S^2}\!d\Omega\;
    \mathcal Q^{(k)}_{\mathrm{STF},(\ell),\,i_1\cdots i_\ell}
    \hat r_{i_1}\cdots\hat r_{i_\ell}Y_{\ell m}
    \notag\\
    &\propto
    \int d^3r\;r^\ell Y_{\ell m}\rho(\mathbf r).
    \label{eq:charge_spherical_projection}
\end{align}
Here $Y_{\ell m}$ are real spherical harmonics.
For example, for $\ell=2$, $r^2Y_{2,-2}\propto xy$, and hence
\begin{equation}
    Q_{2,-2}
    \propto Q_{xy}
    =
    \int d^3r\;xy\,\rho(\mathbf r).
\end{equation}
This is the standard multipole expansion of the electronic
charge~\cite{jacksonClassicalElectrodynamics1998}.
The magnetic analog follows by replacing $\mathcal Q$ with $\mathcal N$ and $\rho$ with $\rho_M$.

\paragraph*{Channel-resolved multipoles}
For the electric and magnetic moment tensors $\mathcal D^{(k)}$ or $\mathcal M^{(k)}$ we can perform the same  expansion for each polarization $P_c(\mathbf r)$ or magnetization component $\mu_c(\mathbf r)$. This gives the spatial distribution within each vector component, and the contraction of these indices with unit vectors and projection onto real spherical harmonics defines the multipolar distribution within each channel.
\begin{align}
    Y_{\ell m}\hat P_c
    &\propto
    \int_{S^2}\!d\Omega\;
    \mathcal D^{(k)}_{i_1\cdots i_{k-1},c}
    \hat r_{i_1}\cdots\hat r_{i_{k-1}}Y_{\ell m}
    \notag\\
    &\propto
    \int d^3r\;r^\ell Y_{\ell m}P_c(\mathbf r),
    \label{eq:channel_polarization_projection}
    \\
    Y_{\ell m}\hat M_c
    &\propto
    \int_{S^2}\!d\Omega\;
    \mathcal M^{(k)}_{i_1\cdots i_{k-1},c}
    \hat r_{i_1}\cdots\hat r_{i_{k-1}}Y_{\ell m}
    \notag\\
    &\propto
    \int d^3r\;r^\ell Y_{\ell m}\mu_c(\mathbf r).
    \label{eq:channel_magnetization_projection}
\end{align}

These \emph{channel multipoles} describe the angular structure within each Cartesian component of the moment distribution.
Unlike the charge multipoles, they are not irreducible under three-dimensional rotations, since rotations can mix the different  moment channels. Coupling the spatial and moment indices with the corresponding Clebsch--Gordan coefficients yields the ordinary, toroidal, and poloidal multipoles.

\begin{figure}[htb!]
    \centering
    \includegraphics[width=\linewidth]{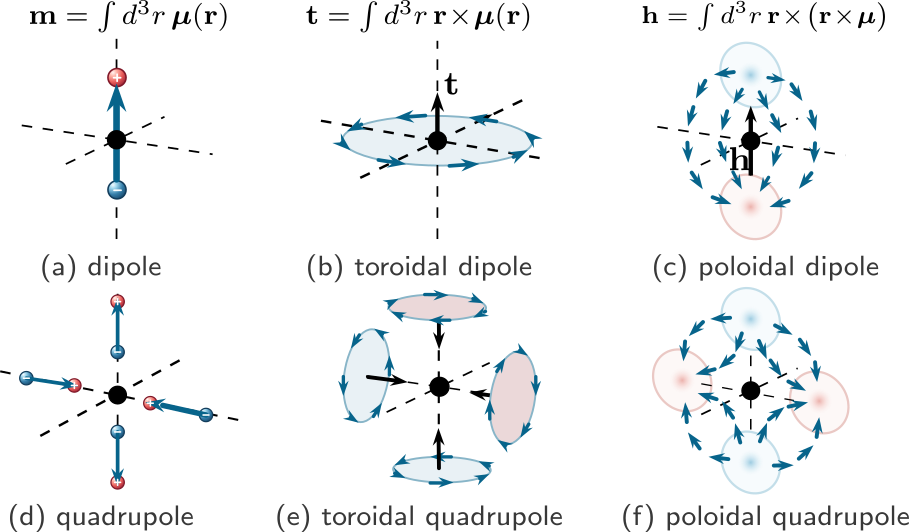}
    \caption{
        Real-space textures of ordinary, toroidal, and poloidal magnetic
        multipoles. Blue arrows denote the local magnetic-moment density
        $\boldsymbol{\mu}(\mathbf r)$ and dashed lines the Cartesian axes.
        Panels (a)--(c) show the ordinary, toroidal, and poloidal dipoles,
        and panels (d)--(f) the corresponding quadrupoles. Black arrows in
        the dipolar panels give the resulting moment direction; those
        centered on the loops in panel (e) indicate the local circulation
        axes.
    }
    \label{fig:multipole_textures}
\end{figure}
\paragraph*{Ordinary multipoles}
The ordinary multipoles of the electric and magnetic  moment tensors are the leading STF sectors of the symmetrized $\mathcal D^{(k)}$ and $\mathcal M^{(k)}$, at $\ell=k$. Their spherical coefficients follow from these STF components or, equivalently, directly from the polarization
and magnetization distributions~\cite{Kuramoto2008ElectronicHigherMultipoles,
Kuramoto2009MultipoleOrdersFluctuations,
Yatsushiro2021MultipoleClassification122,
Suzuki2018FirstprinciplesTheoryMagnetic}:
\begin{align}
    Q_{\ell m}
    &=
    \int_{S^2}\!d\Omega\;
    \mathcal D^{(k)}_{\mathrm{STF},(\ell),\,i_1\cdots i_\ell}
    \hat r_{i_1}\cdots\hat r_{i_\ell}Y_{\ell m}
    \notag\\
    &\propto
    \int d^3r\;\mathbf P(\mathbf r)\cdot
    \boldsymbol\nabla[r^\ell Y_{\ell m}],
    \label{eq:electric_ordinary_projection}
    \\
    M_{\ell m}
    &=
    \int_{S^2}\!d\Omega\;
    \mathcal M^{(k)}_{\mathrm{STF},(\ell),\,i_1\cdots i_\ell}
    \hat r_{i_1}\cdots\hat r_{i_\ell}Y_{\ell m}
    \notag\\
    &\propto
    \int d^3r\;\boldsymbol\mu(\mathbf r)\cdot
    \boldsymbol\nabla[r^\ell Y_{\ell m}].
    \label{eq:magnetic_ordinary_projection}
\end{align}
The gradient $\boldsymbol\nabla[r^\ell Y_{\ell m}]$ encodes the
Clebsch--Gordan coefficients linking the channel-resolved multipoles
to irreducible multipoles of total angular momentum $(\ell,m)$.
It therefore provides an alternative way to obtain the
irreducible multipoles directly from the polarization and
magnetization densities.
Equations~\eqref{eq:electric_ordinary_projection}
and~\eqref{eq:magnetic_ordinary_projection} recover the
charge-density form of Eq.~\eqref{eq:charge_spherical_projection}
by partial integration, using
$\rho_{\rm b}=-\boldsymbol\nabla\cdot\mathbf P$ and
$\rho_M=-\boldsymbol\nabla\cdot\boldsymbol\mu$, respectively.
The ordinary multipoles are thus equivalent up to a factor $\ell$ to the charge multipoles and correspond to the source--sink
structure of the dipole distribution [Fig.~\ref{fig:multipole_textures}(a,d)].
Higher orders introduce additional nodal sectors, as illustrated
by the quadrupolar texture in
Fig.~\ref{fig:multipole_textures}(d). 

\paragraph*{Toroidal multipoles}
Taking the Levi--Civita dual of $\mathcal D^{(k)}$ and
$\mathcal M^{(k)}$, as defined in the previous section, gives
the Cartesian toroidal moment tensors
\begin{align}
    \mathcal G^{(k-1)}_{i_1\cdots i_{k-2},c}
    &=
    \int d^3r\;
    r_{i_1}\cdots r_{i_{k-2}}\,
    [\mathbf r\times\mathbf P(\mathbf r)]_c ,
    \\
    \mathfrak T^{(k-1)}_{i_1\cdots i_{k-2},c}
    &=
    \int d^3r\;
    r_{i_1}\cdots r_{i_{k-2}}\,
    [\mathbf r\times\boldsymbol\mu(\mathbf r)]_c .
\end{align}
The Levi--Civita contraction removes the fully symmetric part
and therefore the ordinary multipoles. The free index $c$
specifies the toroidal moment components $g_c$ and $t_c$.
The toroidal multipole occurs at $\ell=k-1$ and is obtained
from the rank-$\ell$ STF component or directly from the moment
distributions:
\begin{align}
    G_{\ell m}
    &=
    \int_{S^2}\!d\Omega\;
    \mathcal G^{(k-1)}_{\mathrm{STF},(\ell),\,i_1\cdots i_\ell}
    \hat r_{i_1}\cdots\hat r_{i_\ell}Y_{\ell m}
    \notag\\
    &\propto
    \int d^3r\;[\mathbf r\times\mathbf P(\mathbf r)]
    \cdot\boldsymbol\nabla[r^\ell Y_{\ell m}],
    \label{eq:electric_toroidal_projection}
    \\
    T_{\ell m}
    &=
    \int_{S^2}\!d\Omega\;
    \mathfrak T^{(k-1)}_{\mathrm{STF},(\ell),\,i_1\cdots i_\ell}
    \hat r_{i_1}\cdots\hat r_{i_\ell}Y_{\ell m}
    \notag\\
    &\propto
    \int d^3r\;[\mathbf r\times\boldsymbol\mu(\mathbf r)]
    \cdot\boldsymbol\nabla[r^\ell Y_{\ell m}].
    \label{eq:magnetic_toroidal_projection}
\end{align}

In real space, the first dual selects the component of the local
moment perpendicular to $\hat{\mathbf r}$ that circulates. Since $[\mathbf r\times\mathbf X]\cdot\hat{\mathbf r}=0$,
a moment tensor cannot contain a toroidal monopole. The lowest members are the toroidal dipoles shown in Fig.~\ref{fig:multipole_textures}(b), directed along the corresponding circulation axis~\cite{
dubovikToroidMomentsElectrodynamics1990,
spaldinToroidalMomentCondensedmatter2008,
Matsumoto2021NonreciprocalMagnonExcitations}.
At higher order, the coefficients $G_{\ell m}$ and $T_{\ell m}$ describe the different nodal structures of the moment toroidal multipoles
[Fig.~\ref{fig:multipole_textures}(e)].

\paragraph*{Poloidal multipoles}
For $k\geq3$, taking a second Levi--Civita dual gives the poloidal moment tensors
\begin{align}
    \mathcal E^{(k-2)}_{i_1\cdots i_{k-3},d}
    &= \int d^3r\;
    r_{i_1}\cdots r_{i_{k-3}}\,
    [\mathbf r\times(\mathbf r\times\mathbf P(\mathbf r))]_d ,
    \\
    \mathcal H^{(k-2)}_{i_1\cdots i_{k-3},d}
    &= \int d^3r\;
    r_{i_1}\cdots r_{i_{k-3}}\,
    [\mathbf r\times(\mathbf r\times\boldsymbol\mu(\mathbf r))]_d .
\end{align}
The free index $d$ is the poloidal moment index. For $k=3$,
these tensors reduce directly to the poloidal dipoles $e_d$
and $h_d$. Their leading component occurs at $\ell=k-2$. The
spherical coefficients follow from the rank-$\ell$ STF component
or directly from the original polarization and magnetization distributions:
\begin{align}
    E_{\ell m}
    &=
    \int_{S^2}\!d\Omega\;
    \mathcal E^{(k-2)}_{\mathrm{STF},(\ell),\,i_1\cdots i_\ell}
    \hat r_{i_1}\cdots\hat r_{i_\ell}Y_{\ell m}
    \notag\\
    &\propto
    \int d^3r\;
    [\mathbf r\times(\mathbf r\times\mathbf P(\mathbf r))]
    \cdot\boldsymbol\nabla[r^\ell Y_{\ell m}],
    \label{eq:electric_poloidal_projection}
    \\
    H_{\ell m}
    &=
    \int_{S^2}\!d\Omega\;
    \mathcal H^{(k-2)}_{\mathrm{STF},(\ell),\,i_1\cdots i_\ell}
    \hat r_{i_1}\cdots\hat r_{i_\ell}Y_{\ell m}
    \notag\\
    &\propto
    \int d^3r\;
    [\mathbf r\times(\mathbf r\times\boldsymbol\mu(\mathbf r))] \cdot\boldsymbol\nabla[r^\ell Y_{\ell m}].
    \label{eq:magnetic_poloidal_projection}
\end{align}
\begin{table*}[htb!]
\centering
\caption{The multipole families, their Cartesian moment tensors, rank-$\ell$ STF
multipole tensors, degree-$\ell$ spherical multipole coefficients and transformation signatures
under spatial inversion ($\mathcal P$) and time reversal ($\mathcal T$).
Calligraphic symbols denote Cartesian tensors and italic symbols spherical
components.}
\label{tab:four_multipole_families}
\begin{tabular}{l l l l l l}
\hline\hline
Family & Moment tensor & Multipole tensor & Multipole & Channel
& $(\mathcal P,\mathcal T)$ \\
\hline
Electric
& $\mathcal Q^{(\ell)}$, $\mathcal D^{(\ell)}$
& $\mathcal Q^{(\ell)}_{\mathrm{STF}}$, $\mathcal D^{(\ell)}_{\mathrm{STF}}$
& $Q_{\ell m}$
& ---
& $((-1)^{\ell},\,+)$ \\
Magnetic
& $\mathcal N^{(\ell)}$, $\mathcal M^{(\ell)}$
& $\mathcal N^{(\ell)}_{\mathrm{STF}}$, $\mathcal M^{(\ell)}_{\mathrm{STF}}$
& $M_{\ell m}$
& $M_s$
& $((-1)^{\ell+1},\,-)$ \\
Electric toroidal
& $\mathcal G^{(\ell)}\propto\epsilon\,\mathcal D^{(\ell+1)}$
& $\mathcal G^{(\ell)}_{\mathrm{STF}}$
& $G_{\ell m}$
& $g_p$
& $((-1)^{\ell+1},\,+)$ \\
Magnetic toroidal
& $\mathfrak T^{(\ell)}\propto\epsilon\,\mathcal M^{(\ell+1)}$
& $\mathfrak T^{(\ell)}_{\mathrm{STF}}$
& $T_{\ell m}$
& $t_p$
& $((-1)^{\ell},\,-)$ \\
Electric poloidal
& $\mathcal E^{(\ell)}\propto\epsilon\,\mathcal G^{(\ell+1)}$
& $\mathcal E^{(\ell)}_{\mathrm{STF}}$
& $E_{\ell m}$
& $e_d$
& $((-1)^{\ell},\,+)$ \\
Magnetic poloidal
& $\mathcal H^{(\ell)}\propto\epsilon\,\mathfrak T^{(\ell+1)}$
& $\mathcal H^{(\ell)}_{\mathrm{STF}}$
& $H_{\ell m}$
& $h_d$
& $((-1)^{\ell+1},\,-)$ \\
\hline\hline
\end{tabular}
\smallskip
\begin{flushleft}
\small
\textit{Note:} Although we distinguish moment tensors from the multipole tensors in
their decomposition, ``moment'' and ``multipole'' are often used interchangeably.
The poloidal multipoles share the transformation signatures of the electric and
magnetic rows and therefore do not define additional symmetry families.
\end{flushleft}
\end{table*}

The second dual retains the remaining component perpendicular
to $\hat{\mathbf r}$. Unlike the circulating toroidal texture,
the local moments spread away from one point of the sphere and
converge toward the opposite point
[Fig.~\ref{fig:multipole_textures}(c)], forming the two poles
of the poloidal texture while carrying no radial source at the
expansion center.

\begin{figure}[htb]
    \centering
    \includegraphics[width=\linewidth]{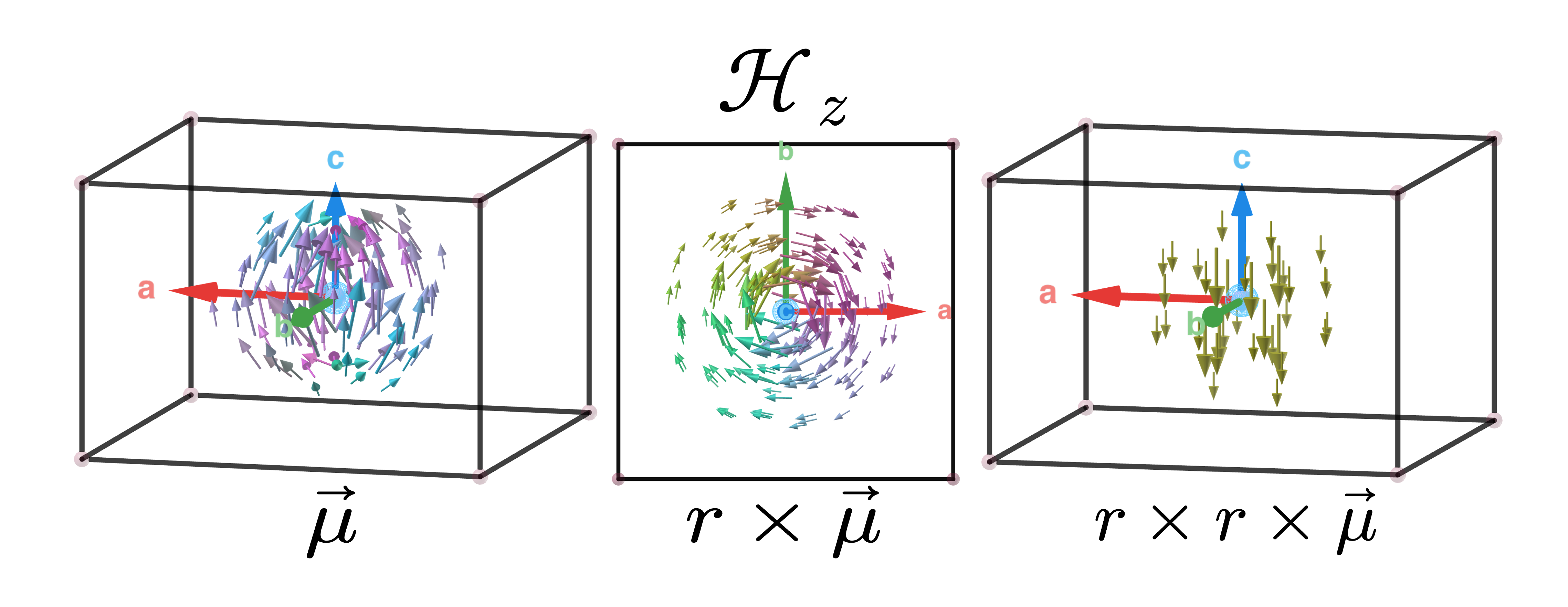}
    \caption{\textbf{Poloidal dipole texture in the different dual representations.}
    Plot of a magnetic poloidal dipole $\mathcal H_z$.
    The poloidal magnetization texture $\boldsymbol\mu(\mathbf r)$ (left) corresponds to a pure circulation in the first dual $\mathbf r\times\boldsymbol\mu(\mathbf r)$ (center), which corresponds to a pure dipolar pattern in the second dual
    $\mathbf r\times[\mathbf r\times\boldsymbol\mu(\mathbf r)]$ (right).}
    \label{fig:MnF2Poloidal}
\end{figure}

The poloidal dipoles $e_d$ and $h_d$ are sometimes referred to in the literature as hypertoroidal
moments~\cite{
prosandeevHypertoroidalMomentComplex2009,
planesRecentProgressThermodynamics2015,
thornerAxialHypertoroidalMoment2014}.
For higher $\ell$, the poles of the pattern depend on the nodal structure of the spherical harmonic, as illustrated
by the quadrupole in Fig.~\ref{fig:multipole_textures}(f).
The coefficients $E_{\ell m}$ and $H_{\ell m}$ describe
these higher-order poloidal textures. 
Figure~\ref{fig:MnF2Poloidal} shows the magnetization distribution for a magnetic poloidal dipole. The magnetization distribution ($\boldsymbol\mu(\mathbf r)$) shows a poloidal texture with no radial part. The first dual $\mathbf r\times\boldsymbol\mu(\mathbf r)$ shows a pure circulation around the $z$ axis,  while the second dual $\mathbf r\times[\mathbf r\times\boldsymbol\mu(\mathbf r)]$ reveals the dipolar character. This shows the field distribution of a poloidal dipole in the different steps of the dualization.

The second dual restores the tensor character of the parent,
so despite their different real-space textures, the poloidal
multipoles carry the same symmetry as the ordinary multipoles of equal rank.
The six sectors of Table~\ref{tab:four_multipole_families}
therefore realize four distinct symmetry families~\cite{Hayami2018ClassificationAtomicscaleMultipoles}.

\subsection{Multipoles, moments, and their symmetry constraints in periodic crystals}
\label{subsec:local_system}
In the preceding subsections, we defined
moment tensors and multipoles for a charge or magnetization distribution
about a general expansion point. In a periodic crystal, the natural expansion centers are the atomic sites.
Because the atomic sites are related by the operations of the crystallographic and magnetic
space group, both the allowed components of the multipoles and their relative ordering are
constrained by symmetry. Summing the contributions of each site in the unit cell allows us to define
macroscopic \emph{system moments and multipoles}, which characterize the net
multipole of the crystal.

\paragraph*{Site multipoles}
For each site $a$ at position $\mathbf R_a$, the rank-$k$ moment tensors of that site are
obtained by the expansion of the density about the site in local coordinates
$\mathbf u=\mathbf r-\mathbf R_a$ and integrating over an atom-centered region
$\Omega(a)$,
\begin{align}
    \!\!\!\!\!\mathcal D^{(k)}_{i_1\dots i_{k-1},\,j}(a)
    &= \int_{\Omega(a)} d^3u\; u_{i_1}\cdots u_{i_{k-1}}\,
       P_j(\mathbf R_a+\mathbf u), \\
   \!\!\!\!\! \mathcal M^{(k)}_{i_1\dots i_{k-1},\,j}(a)
    &= \int_{\Omega(a)} d^3u\; u_{i_1}\cdots u_{i_{k-1}}\,
       \mu_j(\mathbf R_a+\mathbf u),
\end{align}
for $k\ge1$, with $\mathcal Q^{(k)}(a)$ and $\mathcal N^{(k)}(a)$ defined analogously
from $\rho(\mathbf R_a+\mathbf u)$ and $\rho_M(\mathbf R_a+\mathbf u)$, and
$\mathcal G^{(k-1)}(a)$, $\mathfrak T^{(k-1)}(a)$ following from
Eqs.~\eqref{eq:G_from_D} and~\eqref{eq:T_from_M}. Within each family, the site
multipoles correspond to the symmetric trace-free (STF) parts
$\mathcal T^{(\ell)}_{\mathrm{STF}}(a)$ of the corresponding site moment tensor. Their
values can be obtained directly from first-principles calculations by decomposing the
local density matrix in the spherical-harmonic basis $Y_{\ell m}$, as implemented, for
example, in \texttt{multipyles} and
\texttt{ProDenCeR}~\cite{Merkel2023MultipylesV110,bultmarkMultipoleDecompositionLDA2009,Buiarelli2025ProDenCeR}.

\paragraph*{Transformation under crystal symmetries}
Which components of a site moment tensor may be nonzero, and how the multipoles on
symmetry-equivalent sites are related, follow from the transformation of the local moment tensor
$\mathcal T^{(k)}(a)$ under the space-group symmetries of the crystal. A crystallographic or magnetic space-group operation may be written as
$g=\{\mathcal R\,|\,\mathbf t\}$, where $\mathcal R$ is a point-group operation and
$\mathbf t$ a translation. For a magnetic space group, $g$ may in addition contain
time reversal. Its action on a rank-$k$ moment tensor is represented by
$\hat O(g)$, which acts componentwise as
\begin{equation}
\label{eq:tensor_transformation}
    \!\!\!\bigl[\hat O(g)\,\mathcal T^{(k)}\bigr]_{i_1\dots i_k}
    \!\!\!= \sigma(g)^{\tau}\,\det(\mathcal R)^{\delta}
     \!\! \sum_{j_1\dots j_k}\!\!\!
      \mathcal R_{i_1 j_1}\cdots \mathcal R_{i_k j_k}\,
      \mathcal T^{(k)}_{j_1\dots j_k},
\end{equation}
where $\sigma(g)=-1$ if $g$ contains time reversal and $+1$ otherwise. The binary
indices $(\delta,\tau)$ specify the transformation character of the tensor family. A
tensor is called \emph{axial} if it acquires the factor $\det(\mathcal R)$ under improper
spatial operations ($\delta=1$) and \emph{polar} otherwise ($\delta=0$). It is
\emph{time-odd} if it changes sign under time reversal ($\tau=1$) and
\emph{time-even} otherwise ($\tau=0$). The values of $(\delta,\tau)$ for the four
moment tensor families are listed in Table~\ref{tab:tensor_indices}.

\begin{table}[b]
\centering
\caption{Binary indices entering the tensor transformation
law~\eqref{eq:tensor_transformation}. The index $\delta$ governs the response to
improper spatial operations, while $\tau$ governs the response to time reversal. The
superscript denotes the rank of the moment tensor.}
\label{tab:tensor_indices}
\begin{ruledtabular}
\begin{tabular}{llcc}
Family & Multipoles and moment tensors & $\delta$ & $\tau$ \\
\colrule
Electric          & $Q_{\ell m}$, $\mathcal Q^{(k)}$, $\mathcal D^{(k)}$ & 0 & 0 \\
Magnetic          & $M_{\ell m}$, $\mathcal N^{(k)}$, $\mathcal M^{(k)}$ & 1 & 1 \\
Electric toroidal & $G_{\ell m}$, $\mathcal G^{(k)}$                     & 1 & 0 \\
Magnetic toroidal & $T_{\ell m}$, $\mathfrak T^{(k)}$                    & 0 & 1 \\
\end{tabular}
\end{ruledtabular}
\end{table}

For magnetic multipoles, the choice $(\delta,\tau)=(1,1)$ reflects the role of
spin--orbit coupling. In the presence of SOC the spin and spatial sectors are locked
together, so a magnetic multipole transforms as a single time-odd axial tensor, with
all indices acted on by the same matrix $\mathcal R$.

\paragraph*{Magnetic tensors without SOC} In the absence of SOC, the spin and orbital degrees of freedom are fully independent. Consequently, the rank $k$ magnetic moment $\mathcal M^{(k)}\sim\mathcal Q^{(k-1)}\otimes\boldsymbol\mu$ combines a polar tensor of rank $k-1$, acted on by the lattice rotation, with a spin channel $\boldsymbol\mu$ that transforms under an independent spin-space rotation:
\begin{align}
 &\bigl[\hat O(g)\mathcal T^{(k)}\bigr]_{i_1\dots i_{k-1};\,s}
 \notag\\
 &\quad=
 U_{ss'}(g)\,
 \mathcal R_{i_1j_1}\cdots
 \mathcal R_{i_{k-1}j_{k-1}}\,
 \mathcal T^{(k)}_{j_1\dots j_{k-1};\,s'} .
 \label{eq:neumann_nosoc}
\end{align}
Here $g=\mathcal R$ is an element of the crystallographic (paramagnetic) point group. Its spatial action permutes the
atomic sites, $a\to g(a)$, while the associated spin-space rotation $U_{ss'}(g)$ is not affected by $\mathcal R$.

Since spin and lattice are independent, $U_{ss'}(g)$ is determined separately:
it is the spin-space rotation that recovers the magnetic configuration after
application of a point group element $\mathcal R$, thus
$U(g)\,\mathbf m_a = \mathbf m_{g(a)}$. The pairs $(U,\mathcal R)$ for which
this condition holds constitute the spin point
group~\cite{Litvin1974SpinGroups,Jiang2024EnumerationSpinSpaceGroups}. Time
reversal needs no separate treatment, as it acts on $\boldsymbol\mu$ as the
spin-space inversion and is therefore absorbed into $U$.

In the collinear case $U_{ss'}(g)$ reduces to a sign $\pm1$ (time reversal), depending on
whether $g$ maps onto the same or the opposite sublattice. This decoupling of spin and lattice rotations
distinguishes non-relativistic magnetic tensors from their relativistic
counterparts, where SOC locks $U_{ss'}(g)$ to $\det(\mathcal R)\,\mathcal R$.
Although the spatial and spin indices transform independently without SOC, 
expressing magnetic multipole components in the crystal frame still requires
specifying the spin orientation relative to the lattice. This choice is
consistent with the notion of an oriented spin space
group~\cite{Liu2026SymmetryClassificationMagnetic}.

\paragraph*{Determination of symmetry-allowed multipoles and tensor elements}
In order to determine the allowed tensor elements and multipoles, we make use of the generalized Neumann principle, which states that any property tensor must be invariant under all point-group operations of the system. In the case of an atomic site, the subgroup of space-group elements that map a site $a_0$ onto itself is called the \emph{site-symmetry group}, or \emph{site stabilizer} $S_{a_0}$. For every
operation $g\in S_{a_0}$ the transformed tensor must coincide with the original at the same site, so the local moment tensor obeys
\begin{equation}
\label{eq:neumann}
    \mathcal T^{(k)}_{(i_1\dots i_{k-1}),\,j}(a_0)
    =
    \left[\hat O(g)\,\mathcal T^{(k)}(a_0)\right]_{(i_1\dots i_{k-1}),\,j},
\end{equation}
with  $g\in S_{a_0}$ and $\hat O(g)$ evaluated componentwise using Eq.~\eqref{eq:tensor_transformation} and the appropriate $(\delta,\tau)$ from Table~\ref{tab:tensor_indices}, or \eqref{eq:neumann_nosoc} in the case of non-relativistic magnetism. Equation~\eqref{eq:neumann} is the generalized Neumann principle~\cite{Nye1985PhysicalPropertiesCrystals,Post1978MagneticSymmetryImproper}: a set of linear constraints on the tensor components, one for each operation of the site-symmetry group, whose solution space is the set of symmetry-allowed local tensors. Symmetrization over the spatial indices, denoted by parentheses (Appendix~\ref{app:tensor_formulas}), adds another set of constraints to Eq.~\eqref{eq:neumann}. For the charge moment tensors the symmetrization runs over all indices. Components outside the solution space vanish identically, while those inside it may be locked into fixed linear combinations with one another.

\paragraph*{Symmetry-adapted multipoles with and without SOC}
The linear constraint system discussed in the previous section enables a complete decomposition of a moment tensor into its symmetry-allowed independent degrees of freedom and their associated multipoles. The appropriate multipolar basis depends primarily on the treatment of spin--orbit coupling (SOC), which determines how the spatial and moment indices transform under symmetry.

With SOC, as in the electric case, all indices of $\mathcal T^{(k)}$ transform under the same rotation matrix $\mathcal R$. The resulting symmetry constraints therefore couple different Cartesian components of the tensor, so that the independent solutions are generally not confined to a single Cartesian channel. Instead, the tensor components are reorganized according to their transformation under the full rotation group. The corresponding basis thus is equivalent to the Clebsch--Gordan decomposition into the irreducible ordinary, toroidal, and poloidal multipole families defined in Sec.~\ref{sec:multipoles}. In the relativistic limit, the number of independent degrees of freedom is therefore equal to the number of symmetry-allowed irreducible multipoles.

In the absence of SOC, in contrast, the transformation in Eq.~\eqref{eq:neumann_nosoc} factorizes into a spatial rotation $\mathcal R$ and an independent spin-space operation $U(g)$. Since $U(g)$ acts only on the spin index and does not mix the spatial indices, Eq.~\eqref{eq:neumann} decomposes into three independent linear systems, one for each Cartesian channel $c\in{x,y,z}$. Each channel therefore admits an independent multipolar decomposition of its spatial indices, for which the real spherical harmonics $Y_{\ell m}$ [Eq.~\eqref{eq:channel_magnetization_projection}] provide the natural irreducible angular basis. Consequently, the number of independent degrees of freedom is exactly equal to the number of symmetry-allowed \emph{channel multipoles} $Y_{\ell m}\hat M_c$ defined in Sec.~\ref{sec:multipoles}.

\paragraph*{Ordering across an orbit}
While the site stabilizer $S_{a_0}$ governs the allowed form of the local tensor $\mathcal T^{(k)}(a_0)$, space-group operations $g$ mapping outside $S_{a_0}$ generate the other sites $a_i = g(a_0)$ in the Wyckoff orbit $\mathcal O = \{a_i\}$. The local tensor at each equivalent site is obtained by orbit propagation,
\begin{equation}
\label{eq:orbit_propagation}
    \mathcal T^{(k)}(a_i) = \hat O(g)\,\mathcal T^{(k)}(a_0).
\end{equation}
Symmetry-equivalent sites have the same number of independent tensor components, and their allowed tensors are related by $\hat O(g)$. Furthermore, because index symmetrization and trace removal commute with the transformation operator $\hat O(g)$, the multipolar decomposition is preserved along the entire orbit: each rank-$\ell$ STF multipole $\mathcal T^{(\ell)}_{\mathrm{STF}}$ propagates independently under Eq.~\eqref{eq:orbit_propagation} without mixing ranks or channels.

\begin{figure}[tb]
\centering
\includegraphics[width=\linewidth]{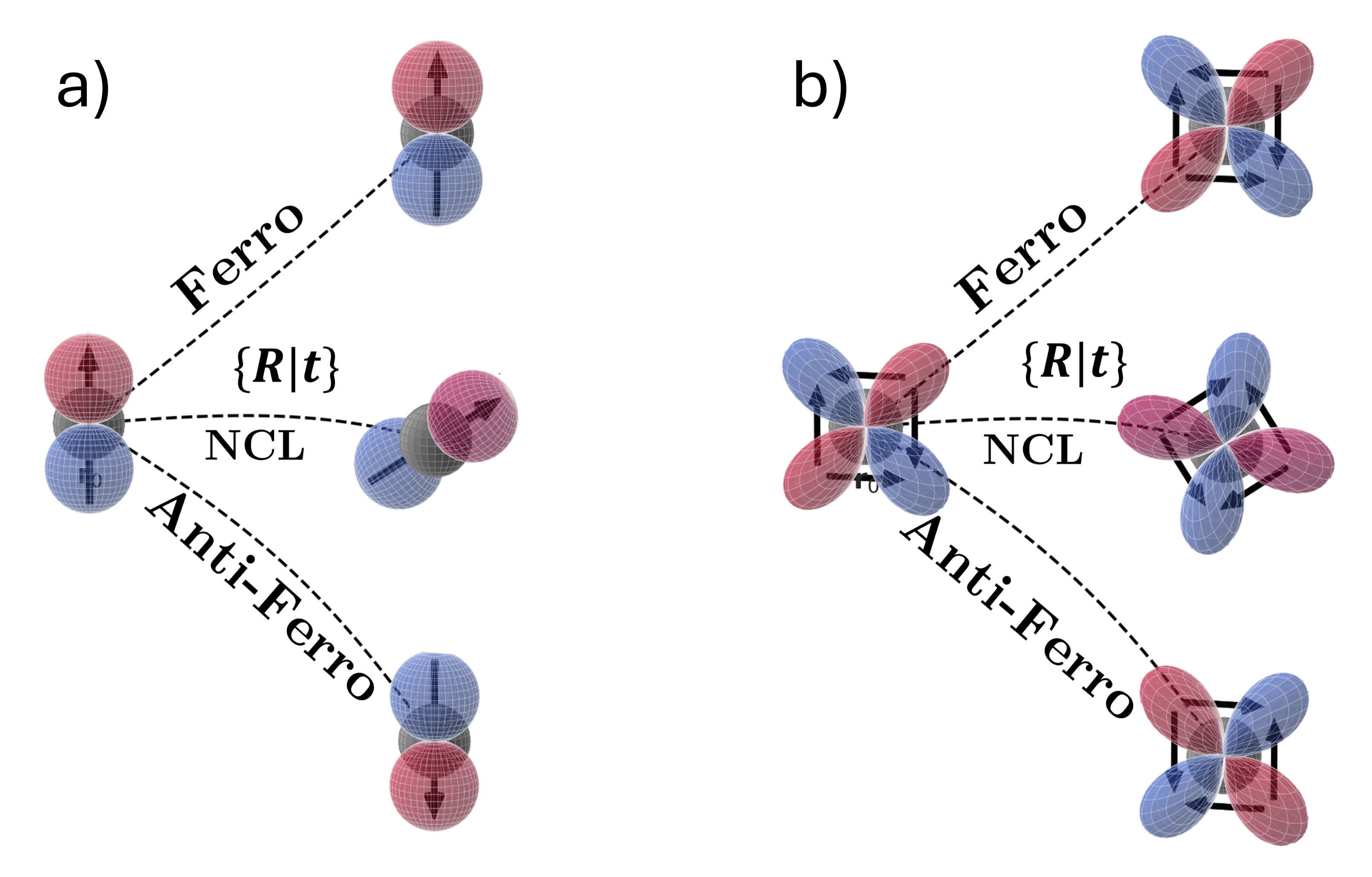}
\caption{Schematic orbit propagation for (a) dipolar and (b) quadrupolar local
tensors. The representative STF tensor
$\mathcal T^{(\ell)}_{\mathrm{STF}}(a_0)$ is propagated to every site of the
Wyckoff orbit. Aligned, sign-reversed, and otherwise rotated tensors correspond
to ferroic, antiferroic, and non-collinear arrangements, respectively.}
\label{fig:multipole_alignment}
\end{figure}

The collective ordering of each multipole over the Wyckoff orbit is given by its \emph{orbit sum},
\begin{equation}
\label{eq:orbit_sum_def}
    \mathcal T^{(k)}_{\mathcal O}
    = \sum_{a_i\in\mathcal O}\mathcal T^{(k)}(a_i)
    = \sum_{g}\hat O(g)\,\mathcal T^{(k)}(a_0),
\end{equation}
Here, the sum includes one symmetry operation $g$ mapping the reference site $a_0$ to each distinct site in the orbit. Together with the local tensor orientations, the orbit sum classifies the order into three canonical patterns (Fig.~\ref{fig:multipole_alignment}): \emph{ferroic}, where tensors align and add constructively ($\mathcal T_{\mathcal O} \neq 0$); \emph{antiferroic}, where sign-reversed local tensors compensate across the cell ($\mathcal T_{\mathcal O} = 0$); and \emph{non-collinear}, where tensors are related by rotations other than simple sign reversals. In particular, a vanishing orbit sum identifies compensated or hidden multipolar order, which remains locally active despite having zero net macroscopic moment.

\paragraph*{System moments and system multipoles}
The \emph{system} tensor is the volume-normalized sum of the local tensors over all occupied atomic sites $a$ of the unit cell,
\begin{equation}
\label{eq:system_tensor}
    \mathcal T^{(k)}_{\mathrm{sys}}
    =
    \frac{1}{\Omega_{\mathrm{cell}}}
    \int_{\mathrm{cell}} d^3r\,\mathcal T^{(k)}(\mathbf r)
    =
    \frac{1}{\Omega_{\mathrm{cell}}}
    \sum_{a=1}^{N_{\mathrm{at}}}\mathcal T^{(k)}(a) .
\end{equation}
For $k=1$, the electric and magnetic system tensors correspond to the macroscopic polarization and magnetization, respectively. For $k\ge2$, Eq.~\eqref{eq:system_tensor} differs from the moment tensor of the total density about a single origin by cross terms involving lower-rank local moments and the site positions $\mathbf R_a$.

The symmetry-allowed components of the system tensor follow from Neumann's principle, Eq.~\eqref{eq:neumann}, applied to the full crystallographic or magnetic point group. Equivalently, the sum over a complete Wyckoff orbit has the same symmetry-allowed structure,
\begin{equation}
\label{eq:orbit_sum}
    \mathcal T^{(k)}_{\mathrm{sys}}
    \sim
    \mathcal T^{(k)}_{\mathcal O},
\end{equation}
where $\sim$ denotes equality of the symmetry-allowed components for a generic site tensor. Particular configurations may instead cancel across the orbit, $\mathcal T^{(k)}_{\mathcal O}=0$, corresponding to antiferroic order. A proof is given in Supp.~Appendix~SI.

The STF components of the system tensor define the corresponding system multipoles, or \emph{multipolizations}, i.e., multipole moments per unit volume. These can also be formulated directly as bulk properties, and first-principles methods have been developed to evaluate them within density-functional theory. For the polarization, this has been formulated in terms of the modern theory of polarization~\cite{restaModernTheoryPolarization1994,king-smithTheoryPolarizationCrystalline1993}.  The approach can be  generalized to other electric and magnetic multipoles~\cite{tholeFirstprinciplesCalculationBulk2016,satoQuantumTheoryMagnetic2026}. In such bulk formulations, the multipolization is multivalued and defined only modulo a \emph{multipolization quantum}.

\section{Multipolar order parameters and corresponding response functions}
\label{sec:order_response}

\begin{table*}[t]
\centering
\caption{Symmetry classification of the moment tensors and their relation to
macroscopic property tensors. Each moment tensor is symmetric in its first $k-1$
polar indices, giving the Jahn symbol $[V^{k-1}]V$, where square brackets denote
index symmetrization (Jahn
convention~\cite{Jahn1949NoteBhagavantamSuranarayana}). The last index carries the
prefactors $e$ for axial and $a$ for time-odd character. Property and transport
tensors with the same Jahn symbol share the symmetry of the corresponding moment
tensor, and we list examples drawn from the Bilbao MTENSOR
tabulation~\cite{Gallego2019AutomaticCalculationSymmetryadapted} and related
work~\cite{Hayami2018ClassificationAtomicscaleMultipoles,Etxebarria2025CrystalTensorProperties}.
For $k\geq2$ the moment tensor decomposes into a leading STF irrep $\langle
V^k\rangle$, the order parameter of Table~\ref{tab:order_parameters}, together with
lower-rank STF terms, and an asymmetric part $[V^{k-2}]\{V^2\}$ carrying the
lower-rank moment tensor of the next family along the dualization ladder: electric
$\to$ electric toroidal $\to$ electric poloidal, and magnetic $\to$ magnetic
toroidal $\to$ magnetic poloidal, with $\{V^2\}\cong eV$ in three dimensions. The
poloidal tensors $\mathcal E^{(k-1)}$ and $\mathcal H^{(k-1)}$ appear in the 
asymmetry column of the toroidal blocks and recover the Jahn symbols of the electric
and magnetic tensors of equal rank. The general-$k$ row gives the pattern at all higher ranks. Index order conventions may differ between a moment tensor and its property tensor, and where the index
symmetries differ the moment tensor captures only part of the property tensor.}
\label{tab:moment_tensors}
\renewcommand{\arraystretch}{1.35}
\begin{ruledtabular}
\begin{tabular}{llcll}
\textbf{Moment tensor} &
$(\mathcal{P},\mathcal{T})$ &
\textbf{Jahn symbol} &
\textbf{Related property tensor} &
\textbf{Asymmetric part} \\
\hline
\multicolumn{5}{c}{\textit{Electric}\quad $(\delta,\tau)=(0,0)$} \\
\hline
$\mathcal{D}^{(1)}_{i}$      & $(-,+)$     & $V$          & Polarization $P_i$ & — \\
$\mathcal{D}^{(2)}_{i,j}$    & $(+,+)$     & $[V^1]V$     & Dielectric, strain, longitudinal conductivity & $\mathcal{G}^{(1)}$\; $\{V^2\}\cong eV$ \\
$\mathcal{D}^{(3)}_{ij,k}$   & $(-,+)$     & $[V^2]V$     & Piezoelectricity, second-harmonic generation, Pockels effect & $\mathcal{G}^{(2)}$\; $[V^1]\{V^2\}$ \\
$\mathcal{D}^{(4)}_{ijk,l}$  & $(+,+)$     & $[V^3]V$     & Flexoelectricity, Kerr effect & $\mathcal{G}^{(3)}$\; $[V^2]\{V^2\}$ \\
$\mathcal{D}^{(k)}$          & $((-)^k,+)$ & $[V^{k-1}]V$ & higher-rank responses & $\mathcal{G}^{(k-1)}$\; $[V^{k-2}]\{V^2\}$ \\
\hline
\multicolumn{5}{c}{\textit{Electric toroidal}\quad $(\delta,\tau)=(1,0)$} \\
\hline
$\mathcal{G}^{(1)}_{i}$      & $(+,+)$         & $eV$          & Axial toroidal moment & — \\
$\mathcal{G}^{(2)}_{i,j}$    & $(-,+)$         & $e[V^1]V$     & Gyration / optical activity, magnetotoroidic effect & $\mathcal{E}^{(1)}$\; $e\{V^2\}\cong V$ \\
$\mathcal{G}^{(3)}_{ij,k}$   & $(+,+)$         & $e[V^2]V$     & Electrogyration, piezoaxiality & $\mathcal{E}^{(2)}$\; $e[V^1]\{V^2\}$ \\
$\mathcal{G}^{(k)}$          & $((-)^{k+1},+)$ & $e[V^{k-1}]V$ &  higher-rank responses  & $\mathcal{E}^{(k-1)}$\; $e[V^{k-2}]\{V^2\}$ \\
\hline
\multicolumn{5}{c}{\textit{Magnetic}\quad $(\delta,\tau)=(1,1)$} \\
\hline
$\mathcal{M}^{(1)}_{i}$      & $(+,-)$         & $aeV$          & Magnetization; anomalous Hall & — \\
$\mathcal{M}^{(2)}_{i,j}$    & $(-,-)$         & $ae[V^1]V$     & Magnetoelectric & $\mathfrak{T}^{(1)}$\; $ae\{V^2\}\cong aV$ \\
$\mathcal{M}^{(3)}_{ij,k}$   & $(+,-)$         & $ae[V^2]V$     & Piezomagnetic, spin/orbital Hall, magneto-optic Kerr effect & $\mathfrak{T}^{(2)}$\; $ae[V^1]\{V^2\}$ \\
$\mathcal{M}^{(4)}_{ijk,l}$  & $(-,-)$         & $ae[V^3]V$     & Flexomagnetic, piezomagnetoelectric & $\mathfrak{T}^{(3)}$\; $ae[V^2]\{V^2\}$ \\
$\mathcal{M}^{(k)}$          & $((-)^{k+1},-)$ & $ae[V^{k-1}]V$ &  higher-rank responses  & $\mathfrak{T}^{(k-1)}$\; $ae[V^{k-2}]\{V^2\}$ \\
\hline
\multicolumn{5}{c}{\textit{Magnetic toroidal}\quad $(\delta,\tau)=(0,1)$} \\
\hline
$\mathfrak{T}^{(1)}_{i}$     & $(-,-)$     & $aV$          & Polar toroidal moment & — \\
$\mathfrak{T}^{(2)}_{i,j}$   & $(+,-)$     & $a[V^1]V$     & Electrotoroidic tensor, Nernst effect & $\mathcal{H}^{(1)}$\; $a\{V^2\}\cong aeV$ \\
$\mathfrak{T}^{(3)}_{ij,k}$  & $(-,-)$     & $a[V^2]V$     & Nernst effect, piezotoroidic & $\mathcal{H}^{(2)}$\; $a[V^1]\{V^2\}$ \\
$\mathfrak{T}^{(k)}$         & $((-)^k,-)$ & $a[V^{k-1}]V$ &  higher-rank responses  & $\mathcal{H}^{(k-1)}$\; $a[V^{k-2}]\{V^2\}$ \\
\end{tabular}
\end{ruledtabular}
\end{table*}

\begin{table*}[t]
\centering
\caption{The symmetric trace-free (STF) part $\langle V^\ell\rangle$ of each
moment tensor (Table~\ref{tab:moment_tensors}) is the irreducible
degree-$\ell$ multipole, serving as the ferroic order parameter for the
corresponding ordered phase, following the multipole classification
of Refs.~\cite{Watanabe2018GrouptheoreticalClassificationMultipole,Hayami2018ClassificationAtomicscaleMultipoles,Suzuki2019MultipoleExpansionMagnetic}.  Angle brackets
$\langle\cdots\rangle$ denote the STF irrep~\cite{thorneMultipoleExpansionsGravitational1980}; $a$ and $e$
are the time-odd and axial prefactors.  Toroidal multipoles are listed up
to the quadrupole.}
\label{tab:order_parameters}
\renewcommand{\arraystretch}{1.35}
\begin{ruledtabular}
\begin{tabular}{llccll}
\textbf{Name} &
\textbf{Multipole} &
$(\mathcal{P},\mathcal{T})$ &
$\ell$ &
\textbf{STF irrep} &
\textbf{Ferroic order parameter for} \\
\hline
\multicolumn{6}{c}{\textit{Electric}\quad $(\delta,\tau)=(0,0)$} \\
\hline
Monopole        & $Q_{00}$ & $(+,+)$ & 0 & $\langle V^0\rangle$ & Charge order \\
Dipole          & $Q_{1m}$ & $(-,+)$ & 1 & $\langle V^1\rangle$ & Ferroelectric order \\
Quadrupole      & $Q_{2m}$ & $(+,+)$ & 2 & $\langle V^2\rangle$ & Ferroelastic, nematic order, electric quadrupole order \\
Octupole        & $Q_{3m}$ & $(-,+)$ & 3 & $\langle V^3\rangle$ & Electric octupolar order \\
Hexadecapole    & $Q_{4m}$ & $(+,+)$ & 4 & $\langle V^4\rangle$ & Electric hexadecapolar order \\
Triakontadipole & $Q_{5m}$ & $(-,+)$ & 5 & $\langle V^5\rangle$ & Electric triakontadipolar order \\
\hline
\multicolumn{6}{c}{\textit{Electric toroidal}\quad $(\delta,\tau)=(1,0)$} \\
\hline
Monopole   & $G_{00}$ & $(-,+)$ & 0 & $e\langle V^0\rangle$ & Structural chirality; separately defined scalar \\
Dipole     & $G_{1m}$ & $(+,+)$ & 1 & $e\langle V^1\rangle$ & Ferroaxial / ferro-rotational order \\
Quadrupole & $G_{2m}$ & $(-,+)$ & 2 & $e\langle V^2\rangle$ & Uniaxial (anisotropic) chirality; ET quadrupolar order \\
\hline
\multicolumn{6}{c}{\textit{Magnetic}\quad $(\delta,\tau)=(1,1)$} \\
\hline
Monopole        & $M_{00}$ & $(-,-)$ & 0 & $ae\langle V^0\rangle$ & Isotropic linear magnetoelectricity, axion order \\
Dipole          & $M_{1m}$ & $(+,-)$ & 1 & $ae\langle V^1\rangle$ & Ferromagnetic/ferrimagnetic order\\
Quadrupole      & $M_{2m}$ & $(-,-)$ & 2 & $ae\langle V^2\rangle$ & Magnetic quadrupolar order, linear magnetoelectricity \\
Octupole        & $M_{3m}$ & $(+,-)$ & 3 & $ae\langle V^3\rangle$ & Magnetic octupole order / $d$-wave altermagnetic order (without SOC)\\
Hexadecapole    & $M_{4m}$ & $(-,-)$ & 4 & $ae\langle V^4\rangle$ & Magnetic hexadecapolar order \\
Triakontadipole & $M_{5m}$ & $(+,-)$ & 5 & $ae\langle V^5\rangle$ & Magnetic triakontadipolar order, $g$-wave altermagnetic order (without SOC)\\
\hline
\multicolumn{6}{c}{\textit{Magnetic toroidal}\quad $(\delta,\tau)=(0,1)$} \\
\hline
Monopole   & $T_{00}$ & $(+,-)$ & 0 & $a\langle V^0\rangle$ & Magnetic toroidal monopole; separately defined scalar \\
Dipole     & $T_{1m}$ & $(-,-)$ & 1 & $a\langle V^1\rangle$ & Ferrotoroidic / anapole order \\
Quadrupole & $T_{2m}$ & $(+,-)$ & 2 & $a\langle V^2\rangle$ & Magnetic toroidal quadrupolar order \\
\end{tabular}
\end{ruledtabular}
\end{table*}
Since the moment tensors and multipoles are linked directly to the crystal
symmetries, they stand in direct relationship to many material properties.
In particular, the moment tensors $\mathcal{D},\mathcal{M},\mathcal{G},\mathfrak{T}$
share their symmetry with several response and property tensors.  For example, the
symmetry of $\mathcal{M}^{(2)}$ is identical to that of the magnetoelectric
response, while $\mathcal{D}^{(3)}$ is symmetry-wise identical to the piezoelectric or SHG
response.  In Table~\ref{tab:moment_tensors} we list the moment tensors
together with symmetry-equivalent response tensors. We give the Jahn symbol for each tensor to facilitate the comparison with different responses, for example using
MTENSOR~\cite{Gallego2019AutomaticCalculationSymmetryadapted}.

Since the moment tensors share the symmetry of the response tensors, the
multipoles---their irreducible parts---are natural candidates to act as
order parameters for ordered phases.  In Table~\ref{tab:order_parameters} we
list representative structural and magnetic orders for which multipoles can act as order parameters. The examples below connect these symmetry classifications to observations in materials.

\paragraph*{Monopoles ($k=0$)}
The rank-0 electric multipole $Q_{00}$ represents the isotropic component of the charge distribution and is proportional to the total charge. For a charge-neutral system, the corresponding total monopole therefore vanishes. If the nuclear charges are not included, however, this component can represent the isotropic part of the electronic charge distribution. A uniform change of this scalar throughout the system is generally constrained by charge neutrality, whereas antiferroic arrangements of local scalar charge components can describe charge order, such as charge disproportionation~\cite{alonsoChargeDisproportionation$mathitRmathrmNiO_3$1999a}.

In the magnetic case, the fundamental absence of magnetic monopoles forbids a ferroic ordering of this scalar. But, if allowed by symmetry, the rank-0 magnetic multipole can occur in the trace of the rank-two or higher order magnetic moment tensors. This inversion-odd, time-reversal-odd quantity $M_{0}$ is often referred to as the magnetoelectric monopole and cannot be interpreted as a fundamental magnetic charge~\cite{spaldinMonopolebasedFormalismDiagonal2013}.
Scalar components also occur in the traces of higher even-rank moment tensors. The magnetoelectric monopole shares the symmetry of the isotropic part of the linear magnetoelectric response and the topological axion~\cite{wilczekTwoApplicationsAxion1987,essinMagnetoelectricPolarizabilityAxion2009}. In $\alpha$-\ce{Cr2O3}, the measured linear magnetoelectric response contains an isotropic component with magnetoelectric-monopole symmetry~\cite{hehlRelativisticNatureMagnetoelectric2008}. Microscopic monopolar order has also been characterized by first-principles calculations~\cite{Verbeek2023HiddenOrdersAntimagnetoelectric}.  Antiferroic arrangements of local magnetoelectric monopoles can instead give antimagnetoelectric order, with vanishing net linear magnetoelectric response. Such order has been characterized by first-principles calculations in $\alpha$-\ce{Fe2O3}, \ce{BiCoO3}, and \ce{KCoF3}~\cite{Verbeek2023HiddenOrdersAntimagnetoelectric,braunLargeDynamicalMagnetic2024,FirstprinciplesStudyJahnTeller}. A vanishing bulk response does not preclude surface magnetoelectricity: the reduced symmetry at a surface can allow for the existence of the magnetoelectric monopole, and thus a surface magnetoelectric effect~\cite{Bhowal2025EmergentSurfaceMultiferroicity}.

\paragraph*{Dipoles ($k=1$)}
Dipolar order is the most widely studied form of multipolar order in materials.
The electric dipoles $Q_{1m}$ are time-even polar vectors whose macroscopic density is the polarization
$P_i$. A spontaneous polarization is allowed in pyroelectric
materials, including ferroelectrics such as \ce{BaTiO3} and \ce{PbTiO3}.  Magnetic dipoles $M_{1m}$ are time-odd axial vectors whose macroscopic density is the magnetization $M_i$.
Ferroic alignment of magnetic dipoles breaks time-reversal symmetry ($\mathcal T$) and characterizes conventional ferromagnets such as \ce{Fe} and \ce{CrO2}.  Local dipoles can also form antiferroic arrangements, giving rise to antiferromagnetism~\cite{neelProprietesMagnetiquesLetat1936,
neelPreuvesExperimentalesFerromagnetisme1949}
and antiferroelectricity~\cite{kittelTheoryAntiferroelectricCrystals1951}.

\paragraph*{Quadrupoles ($k=2$)}
The rank-2 electric moment tensor $\mathcal{D}^{(2)}_{i,j}$ is time- and
parity-even, and its symmetric part transforms as the strain tensor
$\varepsilon_{ij}$. Its irreducible components, the electric quadrupoles $Q_{2m}$, therefore couple to strain and can act as order parameters for ferroelasticity, electronic nematicity~\cite{rosenbergDivergenceQuadrupolestrainSusceptibility2019,massatFieldtunedFerroquadrupolarQuantum2022,vinogradSecondOrderZeeman2022}, and Jahn--Teller distortions~\cite{sartbaevaQuadrupolarOrderingLaMnO2007}, or $f$ electron systems \ce{TmVO4}~\cite{massatFieldtunedFerroquadrupolarQuantum2022,vinogradSecondOrderZeeman2022}, \ce{DyB2C2}~\cite{Hirota2000DirectObservationAntiferroQuadrupolar}, and \ce{TbPO4}~\cite{forinoInducingFerroquadrupolarOrder2026}. Due to the weaker long-range elastic coupling, antiferroquadrupolar order is more common than ferro order.  The magnetic moment tensor $\mathcal{M}^{(2)}_{i,j}$ is time-odd and parity-odd and transforms like the linear magnetoelectric tensor, which describes the magnetization induced by an applied electric field, or equivalently the polarization induced by an applied magnetic field. We discussed the isotropic $\ell=0$ monopole previously, but the decomposition also yields magnetic quadrupoles $M_{2m}$, which correspond to the anisotropic part of the linear magnetoelectric response~\cite{spaldinMonopolebasedFormalismDiagonal2013}. In \ce{CuO}, spherical neutron polarimetry combined with first-principles calculations provides evidence consistent with local magnetoelectric multipoles~\cite{Urru2023NeutronScatteringLocal}, and antiferroic magnetoquadrupolar order has been detected through magnetodielectric measurements in \ce{Ba(TiO)Cu4(PO4)4}~\cite{Kimura2016MagnetodielectricDetectionMagnetic}.

\paragraph*{Octupoles ($k=3$)}
The electric moment tensor $\mathcal{D}^{(3)}_{ij,k}$ is time-even and parity-odd, with the symmetry of third-rank polar tensors such as the piezoelectric response $d_{ijk}=\partial P_i/\partial\sigma_{jk}$ and the second-harmonic-generation
tensor $\chi^{(2)}_{ijk}$. These tensors can carry both an $\ell=1$ and an $\ell=3$
part, so they can contain contributions from both the polar dipole and the electric octupole $Q_{3m}$, the irreducible part that characterizes
non-centrosymmetric order beyond a polar dipole~\cite{VanDerLaan2021ElectronicMultipolesSecond}.

The magnetic rank-3 moment tensor $\mathcal{M}^{(3)}_{ij,c}$ is time-odd and parity-even and transforms identically to the piezomagnetic response tensor
$\Lambda_{ij,c}=\partial M_c/\partial\sigma_{ij}$. The magnetic octupole has also been introduced as the order parameter of $d$-wave altermagnetism, which is characterized by a momentum-dependent spin splitting of the electronic bands despite a vanishing net magnetization~\cite{bhowalFerroicallyOrderedMagnetic2024,mcclartyLandauTheoryAltermagnetism2024}. The spin splitting is controlled by the octupole amplitude~\cite{Martinelli2025MultipolesQuantitativeOrder}. Because the same octupole contributes to the piezomagnetic tensor, $d$-wave altermagnetism is naturally linked to piezomagnetism at the level of symmetry~\cite{Khodas2026TuningAltermagnetismStrain,Radaelli2024TensorialApproachAltermagnetism}. This connection is particularly relevant in compensated magnets, where the dipolar magnetic moment vanishes and the octupole provides the lowest-order contribution to the symmetric piezomagnetic response.
The antiferroic alignment of magnetic octupoles is predicted to induce anti-piezomagnetic
responses, with \ce{CoF2} identified as a particularly promising case where the
effect is expected to be much stronger than in
\ce{MnF2}~\cite{bhowalFerroicallyOrderedMagnetic2024}. Related antiferroic
multipole alignments have also been proposed to produce anti-altermagnetic
phases~\cite{meierNetCompensatedAltermagnetism2026}.

Particularly interesting cases are materials such as \ce{PrTi2Al20}, \ce{PrV2Al20}, and the double perovskite \ce{Ba2CaOsO6}~\cite{Sakai2011PrTr2Al20KondoMultipolar,Freyer2018PrT2Al20MultipolarOrdering,Ye2024PrV2Al20OctupoleSusceptibility,jaeschke-ubiergoAtomicAltermagnetism2025}, in which the ground state hosts magnetic octupoles even though magnetic dipoles are absent. In \ce{Ba2CaOsO6}, the ferro-octupolar ordering on the Os $5d^2$ sites realizes a pure form of ``atomic altermagnetism'', demonstrating that characteristic $d$-wave spin splittings can emerge entirely from local multipolar form factors without requiring a staggered Néel vector of magnetic dipoles~\cite{jaeschke-ubiergoAtomicAltermagnetism2025}.

Additional methods to observe octupolar order include magnetoacoustic resonance~\cite{Koga2026NovelMagnetoacousticResonance}, elastocaloric measurements~\cite{Ye2024PrV2Al20OctupoleSusceptibility}, as well as magnetic Compton scattering and x-ray emission magnetic circular dichroism~\cite{bhowalFerroicallyOrderedMagnetic2024,Mizumaki2025DetectionFerroicOctupole}.

\paragraph*{Hexadecapoles ($k=4$)}
The electric moment tensor $\mathcal{D}^{(4)}_{ijk,l}$ is time- and
parity-even; cooperative ordering of its irreducible part, the electric
hexadecapole $Q_{4m}$, can define a primary order parameter without dipolar
or quadrupolar order.  A prominent context is hidden order in \ce{URu2Si2},
for which antiferroic electric-hexadecapolar order of $xy(x^2-y^2)$ symmetry
is among the proposed scenarios~\cite{Kusunose2011HiddenOrderURu2Si2,Toth2011HexadecapolarKondoEffect}. The magnetic moment tensor $\mathcal{M}^{(4)}_{ijk,l}$ shares its symmetry with the flexomagnetic and piezomagnetic tensors.  Ferroic order of the magnetic hexadecapole $M_{4m}$ has been identified in \ce{BaMn2As2}, where a current-induced magnetopiezoelectric response in its doped metallic state is predicted~\cite{Watanabe2017MagneticHexadecapoleOrder}.

\paragraph*{Triakontadipoles ($k=5$)} The rank-five electric moment tensor $\mathcal D^{(5)}_{ijkl,m}$ is time-even and parity-odd. Its leading irreducible component, the electric triakontadipole $Q_{5m}$, describes a rank-five electric anisotropy. The magnetic rank-five moment tensor $\mathcal M^{(5)}_{ijkl,m}$ is instead time-odd and parity-even. 
Magnetic triakontadipolar order has been proposed as a hidden-order scenario for \ce{URu2Si2}~\cite{Cricchio2009URu2Si2Triakontadipole,Ikeda2014URu2Si2Triakontadipole}, and candidate triakontadipolar order was identified in \ce{NpO2} using resonant x-ray Bragg diffraction~\cite{Lovesey2012NeptuniumMultipolesResonant}.
More prominently, the magnetic triakontadipole has been identified as the order parameter for $g$-wave altermagnetism~\cite{verbeekNonrelativisticFerromagnetotriakontadipolarOrder2024}, characterized by non-relativistic momentum-dependent spin splitting with $g$-wave symmetry. 
Representative materials include $\alpha$-\ce{Fe2O3}~\cite{verbeekNonrelativisticFerromagnetotriakontadipolarOrder2024}, \ce{MnTe}~\cite{krempaskyAltermagneticLiftingKramers2024,Amin2024MnTeAltermagnetism,Sunko2025LinearMagnetobirefringenceProbe}, and \ce{CrSb}~\cite{Ding2024CrSbLargeBandSplitting}.
In \ce{CrSb}, recent work combines x-ray reconstruction of antiferroically arranged electric hexadecapoles with polarized-neutron evidence for the associated ferroic magnetic multipoles~\cite{misawaObservationGwaveAltermagnetic2026}. A separate polarized-neutron study of \ce{MnF2} reports a reconstructed magnetization density consistent with a rank-five magnetic multipole~\cite{kibalinRealspaceManifestationFerroic2026}. In \ce{MnTe}, photoemission and x-ray dichroic imaging probe spin splitting and altermagnetic domains~\cite{krempaskyAltermagneticLiftingKramers2024,Amin2024MnTeAltermagnetism}, while linear magnetobirefringence has been proposed as a symmetry-sensitive probe of the altermagnetic multipoles~\cite{Sunko2025LinearMagnetobirefringenceProbe}.

\paragraph*{Toroidal moments}
Toroidal multipoles describe ordered states whose symmetry is not captured by ordinary electric or magnetic multipoles, but by the antisymmetric part of the moment tensors.  The canonical example is the magnetic toroidal dipole $T_{1m}$, the order parameter of
\emph{ferrotoroidicity}~\cite{spaldinToroidalMomentCondensedmatter2008,
VanAken2007ObservationFerrotoroidicDomains,
Zimmermann2014FerroicNatureMagnetic}.  Microscopically, these correspond to
vortex-like arrangements of localized magnetic moments,
$\mathbf t\propto\sum_n\mathbf{r}_n\times\mathbf{m}_n$, and are odd under both
$\mathcal{P}$ and $\mathcal{T}$.  Ferrotoroidic domains have been observed in
magnetoelectric materials by optical second-harmonic generation, notably in
\ce{LiCoPO4}, while microscopic toroidal moments have been evaluated from
spin arrangements and first-principles calculations in systems such as
\ce{GaFeO3}~\cite{VanAken2007ObservationFerrotoroidicDomains,
EdererSpaldin2007MicroscopicToroidalCrystals}.  Atomic-scale magnetic toroidal
dipoles can also be accessed through resonant x-ray diffraction: in canted
antiferromagnetic hematite, parity-odd magnetoelectric multipoles including the
Fe anapole were inferred from resonant Bragg intensities~\cite{Lovesey2011ParityoddMultipolesMagnetic}.
In hexagonal \ce{YMnO3}, resonant x-ray diffraction reveals spin canting, while calculations suggest that parity-odd multipoles contribute to the observed spectral changes~\cite{ramakrishnanAntiferromagneticSpinCanting2023}.

Higher-rank toroidal multipoles provide order parameters for hidden structural
and magnetic patterns.  Magnetic toroidal quadrupoles describe antiferroic
arrangements of local toroidal dipoles and have been identified theoretically as
the relevant order parameter for spin-current generation without uniform
magnetization or SOC~\cite{Hayami2022SpinConductivityBaseda}.  Electric
toroidal quadrupoles play an analogous role in structural phases: they have
been proposed as the order parameters of the inversion-breaking transitions in
\ce{Cd2Re2O7}~\cite{Hayami2019ElectricToroidalQuadrupoles}.  In \ce{TiSe2}, circular dichroism observed in resonant x-ray diffraction was interpreted as evidence for electronic chirality~\cite{Xiao2024ObservationCircularDichroism}. Subsequent first-principles calculations reproduced the observations using a centrosymmetric, nonchiral structure and interference between atomic multipole scattering amplitudes~\cite{Ueda2025Nonchiral1TiS}. Thus, circular dichroism in diffraction alone need not establish chiral order.

The electric toroidal dipole $G_{1m}$ is the order parameter of ferroaxial
(ferrorotational) order~\cite{Hlinka2016SymmetryGuideFerroaxial,
Jin2020ObservationFerrorotationalOrder}.  It is even under both $\mathcal{P}$
and $\mathcal{T}$ but odd under mirror operations containing the rotation axis,
and therefore describes a macroscopic structural rotation rather than a polar or
magnetic moment.  Ferroaxial order has been probed by rotational-anisotropy second-harmonic generation in \ce{RbFe(MoO4)2}~\cite{Jin2020ObservationFerrorotationalOrder}, while domains in \ce{NiTiO3} have been imaged by electrogyration and electron diffraction~\cite{Hayashida2020VisualizationFerroaxialNiTiO3}. Ferroaxial coupling also explains magnetically induced polarization in \ce{CaMn7O12}~\cite{Johnson2012CaMn7O12Ferroaxial}; structural and magnetoelectric measurements characterize coexisting ferroic orders in \ce{BaCoSiO4}~\cite{Xu2022BaCoSiO4ChiralMagnet}.  First-principles calculations have quantified microscopic electric toroidal dipoles in ferroaxial \ce{NiTiO3} and \ce{K2Zr(PO4)2}~\cite{BhowalSpaldin2024ElectricToroidalDipole}.  Related
electronic toroidal multipoles have been quantified in the elemental chiral
crystals Te and Se using symmetry-adapted Wannier
models~\cite{Hayami2018MicroscopicToroidalHybridOrbitals,
Oiwa2025PredominantChiralityToroidalMultipoles}.

The rank-0 electric toroidal monopole $G_{00}$, with
$(\mathcal{P},\mathcal{T})=(-,+)$, is the pseudoscalar order parameter for
molecular and crystal chirality.  At the continuum level, it can be represented
by a polarization helicity,
$\int\mathbf{P}\cdot(\nabla\times\mathbf{P})\,d^3r$~\cite{Hlinka2014EightTypesVectorlike,
bousquetStructuralChiralityRelated2025,GomezOrtiz2024ChiralityHandedness},
while microscopically it may be written in terms of the atomic-scale operator
$\hat{G}_0\propto\hat{\mathbf{t}}\cdot\hat{\boldsymbol{\sigma}}$.  This monopole has been evaluated as a chirality measure in a model of twisted \ce{CH4}~\cite{Inda2024QuantificationChiralityToroidalMonopole} and provides a theoretical language for chirality-induced spin selectivity~\cite{Kusunose2024EmergenceChiralityMultipole,Kishine2022DefinitionChiralityEnantioselective}.  It has also been proposed as
a hidden-order parameter in \ce{URu2Si2}~\cite{Sato2023URu2Si2ChiralCharge}.
Alternative proposals for the hidden order include triakontadipolar
order~\cite{Cricchio2009URu2Si2Triakontadipole,Takagi2012SymmetryHiddenOrder} and hastatic
order~\cite{Chandra2013HastaticOrderHeavyfermion}.

The complementary magnetic toroidal monopole $T_{00}$, with
$(\mathcal{P},\mathcal{T})=(+,-)$, is a time-reversal-odd scalar order
parameter.  In antiferromagnetic \ce{Co2SiO4}, electric-field-induced nonreciprocal directional dichroism provides experimental evidence for a time-reversal-odd scalar response compatible with magnetic-toroidal-monopole symmetry~\cite{Hayashida2025ElectricFieldInducedNonreciprocal}.  Recent work has proposed
additional antiferromagnetic candidates and a Berry-curvature-based
first-principles route for evaluating this monopole in crystals~\cite{Hayami2023TimeReversalSwitching,
Yamanaka2026MagneticToroidalMonopolePolarization}.

\paragraph*{Dynamical response tensors}
The site-resolved moment tensors (see Sec.~\ref{subsec:local_system}) can be used to resolve the symmetry-allowed components of site-dependent response tensors.  For example, the Born
effective charge $Z^*_{ij}=\partial P_i/\partial u_j$ has the same symmetry as the electric moment tensor $\mathcal D^{(2)}$. The symmetry-allowed components of the Born effective charge tensor for an atom at site $a$ therefore coincide with those of the local moment tensor on that site $\mathcal D^{(2)}(a)$. The magnetic analog, the dynamical magnetic charge $Z^{*m}_{ij}=\partial M_i/\partial u_j$~\cite{yeDynamicalMagneticCharges2014}, relates a time-odd axial vector to a time-even polar one and is a parity-odd, time-odd rank-2 pseudotensor, the symmetry class of the magnetic moment tensor $\mathcal M^{(2)}$. Its allowed components at each site coincide with those of $\mathcal M^{(2)}(a)$.

\section{Computational Implementation}
\label{sec:architecture}
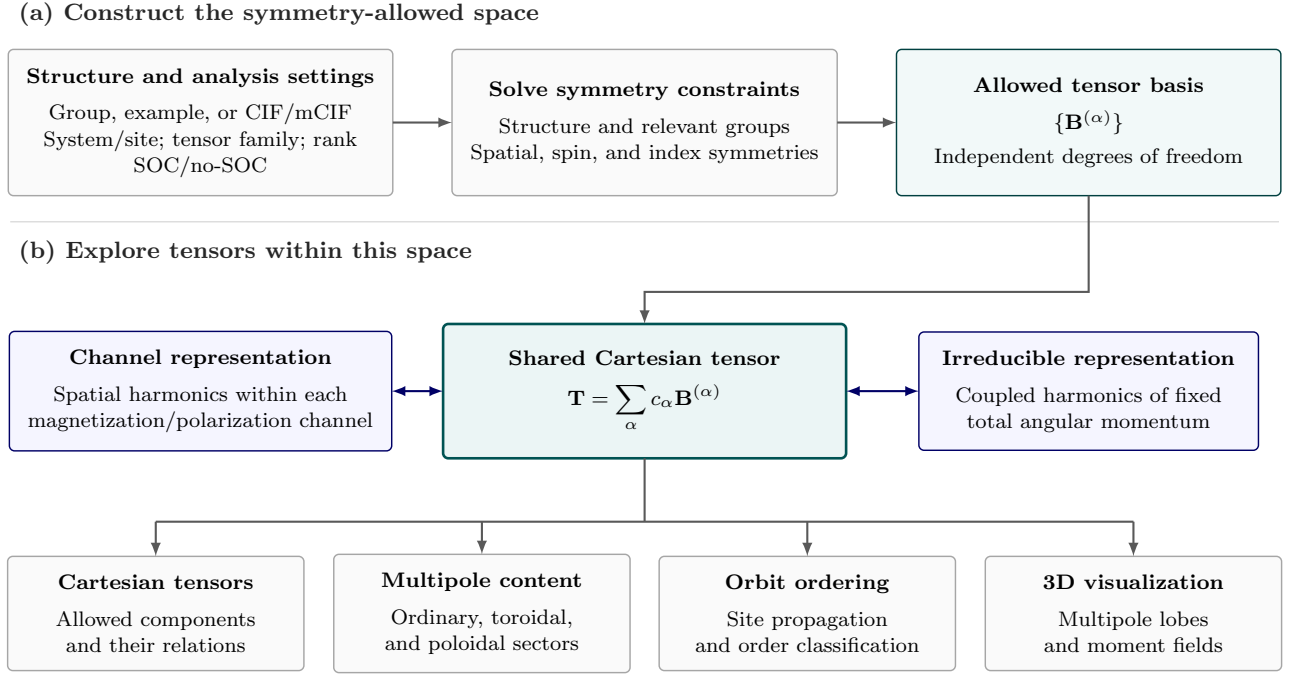
\begin{figure*}[t]
\centering
\resizebox{0.96\textwidth}{!}{%
\begin{tikzpicture}[
  x=1cm,y=1cm,
  font=\small,
  box/.style={rectangle,rounded corners=2pt,draw=black!35,
    line width=.55pt,align=center,inner sep=7pt,outer sep=0pt,font=\fontsize{8}{10}\selectfont},
  stagebox/.style={box,minimum width=5.15cm,minimum height=1.96cm,fill=black!2},
  control/.style={box,minimum width=4.55cm,minimum height=1.45cm,fill=blue!4,draw=blue!40!black},
  tensor/.style={box,minimum width=5.4cm,minimum height=1.8cm,fill=teal!8,draw=teal!65!black,line width=1pt},
  output/.style={box,minimum width=3.97cm,minimum height=1.35cm,fill=black!2},
  flow/.style={-latex,line width=.8pt,draw=black!65},
  sync/.style={latex-latex,line width=.8pt,draw=blue!40!black},
  heading/.style={anchor=west,font=\small\bfseries,text=black!85}
]
\path[use as bounding box] (-.2,.38) rectangle (17.25,-8.92);

\node[heading] at (0,0) {(a) Construct the symmetry-allowed space};

\node[stagebox] (input) at (2.55,-1.45) {
  \textbf{Structure and analysis settings}\\[4pt]
  Group, example, or CIF/mCIF\\
  System/site; tensor family; rank\\
  SOC/no-SOC
};
\node[stagebox] (solve) at (8.50,-1.45) {
  \textbf{Solve symmetry constraints}\\[4pt]
  Structure and relevant groups\\
  Spatial, spin, and index symmetries
};
\node[stagebox,fill=teal!5,draw=teal!50!black] (basis) at (14.45,-1.45) {
  \textbf{Allowed tensor basis}\\[4pt]
  $\{\mathbf B^{(\alpha)}\}$\\[3pt]
  Independent degrees of freedom
};
\draw[flow] (input) -- (solve);
\draw[flow] (solve) -- (basis);

\draw[black!15,line width=.5pt] (0,-2.78) -- (17,-2.78);
\node[heading] at (0,-3.18) {(b) Explore tensors within this space};

\node[tensor] (tensor) at (8.50,-5.05) {
  \textbf{Shared Cartesian tensor}\\[5pt]
  $\displaystyle\mathbf T=\sum_{\alpha}c_\alpha\mathbf B^{(\alpha)}$
};
\node[control] (channel) at (2.55,-5.05) {
  \textbf{Channel representation}\\[4pt]
  Spatial harmonics within each \\magnetization/polarization channel
};
\node[control] (irreducible) at (14.45,-5.05) {
  \textbf{Irreducible representation}\\[4pt]
  Coupled harmonics of fixed\\
  total angular momentum
};
\draw[flow] (basis.south) -- (14.45,-3.72) -- (8.50,-3.72) -- (tensor.north);
\draw[sync] (channel.east) -- (tensor.west);
\draw[sync] (tensor.east) -- (irreducible.west);
\node[font=\footnotesize,text=blue!40!black] at (2.55,-6.23) {};
\node[font=\footnotesize,text=blue!40!black] at (14.45,-6.23) {};

\node[output] (tables) at (1.95,-8.03) {
  \textbf{Cartesian tensors}\\[4pt]
  Allowed components\\
  and their relations
};
\node[output] (multipoles) at (6.32,-8.03) {
  \textbf{Multipole content}\\[4pt]
  Ordinary, toroidal,\\
  and poloidal sectors
};
\node[output] (ordering) at (10.68,-8.03) {
  \textbf{Orbit ordering}\\[4pt]
  Site propagation\\
  and order classification
};
\node[output] (visual) at (15.05,-8.03) {
  \textbf{3D visualization}\\[4pt]
  Multipole lobes\\
  and moment fields
};
\draw[line width=.8pt,draw=black!65] (tensor.south) -- (8.50,-6.80);
\draw[line width=.8pt,draw=black!65] (1.95,-6.80) -- (15.05,-6.80);
\foreach \dest in {tables,multipoles,ordering,visual}
  \draw[flow] (\dest.north |- 0,-6.80) -- (\dest.north);
\end{tikzpicture}}

\caption{Symmetry-constrained computation and interactive exploration in
\textit{MagSymMultipoles}. (a) Structure and analysis settings determine the
symmetry-allowed tensor basis. (b) Channel and Irreducible controls modify
the coefficients of a shared Cartesian tensor and update one another while
preserving the symmetry constraints. For parent-derived calculations, the
same tensor supplies the component tables, multipole decomposition, orbit ordering, and three-dimensional
visualizations. Symmetry fixes the allowed space; the amplitudes within it
remain free parameters.}
\label{fig:flowchart}
\end{figure*}
We developed \textit{MagSymMultipoles}
(\emph{Magnetic Symmetry-Adapted Multipoles}), an online tool
that automates the symmetry analysis and multipole decomposition
described in Secs.~\ref{sec:theory}\textendash\ref{sec:order_response}.
For any crystallographic or magnetic space group, the tool
constructs the symmetry-allowed Cartesian tensor basis and
decomposes it into ordinary, toroidal, and poloidal multipole components, expressed in the real spherical-harmonic basis.
A Python backend handles the group-theoretical linear algebra, while a JavaScript frontend displays the output tables and
interactive 3D visualizations. The tool is available at \url{https://mag-sym-multipoles.com}.

Figure~\ref{fig:flowchart} summarizes the implemented workflow.
Panel~(a) outlines the construction of the symmetry-allowed tensor basis from the input structure and analysis settings, using the relevant spatial, spin, and index symmetries described
in Secs.~\ref{subsec:solver} and~\ref{subsec:nosoc_solver}.
Panel~(b) shows how to examine the resulting tensors through their channel-resolved and
irreducible representations. These include tools for inspecting Cartesian components and their relations to property tensors, resolving ordinary, toroidal, and poloidal multipole content,
and identifying ordering across crystallographic orbits, alongside interactive 3D visualizations of multipole lobes and moment fields.
\subsection{User interface}

\begin{figure*}[t]
    \centering
    \includegraphics[width=1\linewidth]{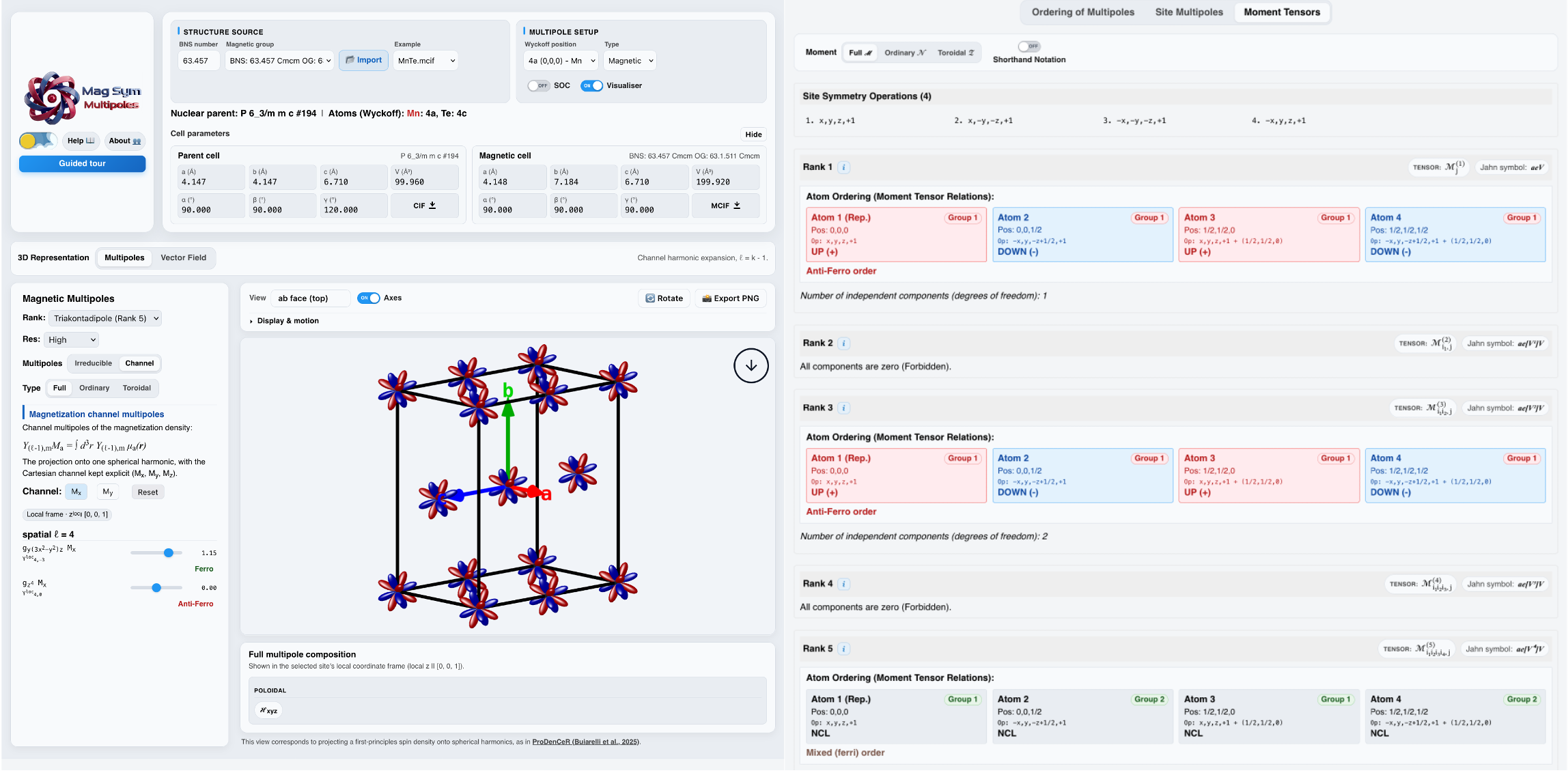}
    \caption{\textbf{Overview of the \textit{MagSymMultipoles} interface.}
    Users can import a crystal structure, select an atomic site, adjust
    multipole amplitudes, and inspect their real-space representation.
    The results panel provides complementary views of multipole ordering,
    site multipoles, and moment tensors. The interface is shown after
    importing an mCIF file for \ce{MnTe}, with results calculated without SOC.}
    \label{fig:interface}
\end{figure*}
Figure~\ref{fig:interface} illustrates the user interface for the \ce{MnTe} system:
the input controls at the top of the page define the symmetry problem, while the
multipole controls and three-dimensional interactive viewer allow users to explore tensors within the
resulting symmetry-allowed space.

The input stage in Fig.~\ref{fig:flowchart}(a) begins by selecting one
of the 1,651 magnetic space groups, loading a built-in example, or uploading
a CIF or mCIF file. When a file is uploaded,
the backend executes FINDSYM~\cite{Stokes2005FINDSYMProgramIdentifying} from the
ISOTROPY suite~\cite{ISOTROPYSoftwareSuite,Lian2021AlgorithmxIsosuite} to
identify the magnetic space group and, when required, its non-magnetic parent.

Uploading a non-magnetic CIF prompts an \emph{Initial Magnetic Moments} dialog,
where initial moments are entered in the crystallographic basis $(m_a, m_b, m_c)$ of the input cell for FINDSYM processing. While this input step uses the crystal frame, all resulting moment tensors---starting from rank-1 dipole vectors $(m_x, m_y, m_z)$---are converted and reported in an orthonormal Cartesian frame (Sec.~\ref{subsec:metric}). A real-time 3D preview updates the spin configuration as components are entered, allowing immediate visual verification before FINDSYM identifies the resulting magnetic space group and parent cell.

The user can then select a specific Wyckoff orbit or the macroscopic ``System'', choose the tensor family (electric, magnetic), and specify whether spin--orbit coupling (SOC) is included, defining the symmetry problem passed
to the backend.

The four output branches in Fig.~\ref{fig:flowchart}(b) correspond to
complementary postprocessing panels. \emph{Moment Tensors} displays the allowed Cartesian components and their
symmetry-enforced relations for each tensor rank, and allows users to explore
their relationship to common response tensors. For ranks $k\geq3$, the
component tables use the compact tensor-index shorthand defined in
Appendix~\ref{app:tensor_shorthand};
\emph{Site Multipoles} tabulates the Channel and Irreducible components,
resolving the ordinary, toroidal, and poloidal decomposition; and
\emph{Ordering of Multipoles} allows users to inspect the spatial arrangement across the
selected orbit. 
Finally, the \emph{3D visualizers} display spherical-harmonic lobes
(\emph{Multipoles}) and their ordering within the crystal for each symmetry-allowed multipole, while \emph{Vector Field} provides a real-space visualization of the decomposition of the moment tensor of a chosen rank.
\subsection{Null-space solver}
\label{subsec:solver}

The basis of our method is the determination of symmetry-allowed tensor elements. Mathematically, this amounts to finding the null space of a system of linear constraints. Each symmetry operation and index permutation restricts the allowed combinations of tensor components, and the solver gathers these constraints into a single constraint matrix. For the calculation of a site tensor, the relevant symmetries are given by the site stabilizer (point group of the chosen Wyckoff site). For macroscopic, or ``System'', calculations with spin--orbit coupling (SOC), the imposed symmetry constraints correspond to the operations of the full magnetic point group.

The solver starts with a general rank-$k$ Cartesian tensor $\mathbf T$ with $3^k$ components, which are flattened into a column vector $\mathbf v=\operatorname{vec}(\mathbf T)$. Next, we use the constraints from the generalized Neumann principle (Sec.~\ref{subsec:local_system}), which requires that a tensor remains invariant under all symmetry operations $g$ of the relevant point group. In other words, for any tensorial vector $\mathbf v$ the equality $(\hat{\mathbf D}(g)-\mathbf I)\mathbf v=0$ must hold, where
\begin{equation}
    \hat{\mathbf{D}}(g) = \sigma(g)^{\tau} \det(R)^{\delta} \underbrace{R \otimes R \otimes \dots \otimes R}_{k \text{ times}}.
\end{equation}
Here $R$ is the crystallographic point group element, while $\delta$ and $\tau$ account for the polar or axial character and the time-reversal parity, respectively (Table~\ref{tab:tensor_indices}).

Additionally, the intrinsic index-permutation symmetries of the targeted tensor must be taken into account. This provides an additional set of constraints: if components are interchangeable under permutation of indices $a$ and $b$, $\mathbf{v}$ must satisfy $(\hat P_{ab}-\mathbf I)\mathbf v=0$, where $\hat P_{ab}$ is the corresponding permutation operator. Charge moment tensors are fully symmetric under any permutation of their $k$ indices, whereas the electric and magnetic moment tensors are symmetric only in the first $k-1$ spatial indices. In practice, only nearest-neighbor permutations $\hat P_{a,a+1}$ are required to generate all necessary permutations. 

Having defined all the constraints on our tensors, we must now obtain all the vectors $\mathbf{v}$ that belong to the null space. This can in principle be done by stacking all the constraints $(\hat{\mathbf D}(g)-\mathbf I)$ and $(\hat P_{ab}-\mathbf I)$ in a large rectangular matrix $\mathbf A$ and solving the linear system $\mathbf A\mathbf v = 0$. In order to avoid large matrices, the solver uses the property $\ker(\mathbf A) = \ker(\mathbf A^\top \mathbf A)$, since $\mathbf v^\top \mathbf A^\top \mathbf A \mathbf v = \sum_i \|\mathbf A_i \mathbf v\|^2 = 0 \iff \mathbf A_i \mathbf v = 0 \;\forall i$. This allows us to accumulate all constraints into a symmetric matrix $\mathbf A^\top \mathbf A$ of size $3^k \times 3^k$
\begin{equation}
    \mathbf A^\top \mathbf A = \sum_g \left(\hat{\mathbf D}(g)-\mathbf I\right)^{\top}\left(\hat{\mathbf D}(g)-\mathbf I\right) + \sum_{a} 2(\mathbf I-\hat P_{a,a+1}),
\end{equation}
where we made use of the property that permutation tensors swapping two elements are always symmetric and their own inverse ($\hat P_{a,a+1}^2 = \mathbf I$), which means that each quadratic contribution simplifies to
\begin{equation}
    (\hat P_{a,a+1}-\mathbf I)^{\top}(\hat P_{a,a+1}-\mathbf I) = 2(\mathbf I-\hat P_{a,a+1}) .
\end{equation}

Solving the linear constraints for $\mathbf A^\top \mathbf A$ rather than $\mathbf A$ limits the size of the array to $3^k \times 3^k$ and allows the null space to be determined using a fast symmetric eigensolver. Similar eigenvector and null-space approaches to determine symmetry-allowed tensor components are discussed in Ref.~\cite{Wu2022NeumannsPrincipleBased}.

The dimension of the resulting null space contains all independent degrees of freedom allowed by symmetry, while the eigenvectors $\mathbf B^{(\alpha)}$ define the relations between the tensor components that are strictly required by symmetry. Any physical tensor in this subspace is a linear superposition of the basis tensors:
\begin{equation}
    \mathbf T = \sum_{\alpha} c_\alpha \mathbf B^{(\alpha)} .
\end{equation}
Because the basis tensors are linearly independent, the coefficients $c_\alpha$ uniquely determine the tensor. These coefficients represent free physical amplitudes that are not fixed by symmetry alone and must be obtained from microscopic models, first-principles calculations, or experiments. 
When building the symmetry-adapted moment tensor, the frontend identifies the symmetry-enforced relations encoded in $\mathbf B^{(\alpha)}$ and displays them in the \emph{Moment Tensors} tab using matching symbols and colors. A step-by-step worked example for a rank-2 electric charge tensor is given in Supplementary Sec.~SII\,A, and its channel decomposition and spherical-harmonic slider weights are derived in Supplementary Sec.~SII\,B. A complementary rank-5 magnetic example for \ce{MnF2} illustrates how individual site-symmetry operations constrain tensor components (Supplementary Sec.~SII\,C and Supp.~Table~S1).

\subsection{No-SOC construction}
\label{subsec:nosoc_solver}

The full magnetic-group solve of Sec.~\ref{subsec:solver} couples space
and spin through a single joint constraint; without SOC this coupling is
absent, so the spatial and spin factors can be built independently.  For
a requested magnetic tensor of rank $k$, the backend first solves the
rank-$(k-1)$ electric problem using the symmetries of the
nuclear parent.  This parent group is either read directly from an uploaded structure,
together with its Wyckoff mapping, or, for a directly selected magnetic
space group, is set to the associated paramagnetic parent in the BNS setting.  Because the parent group's tabulated setting can differ from the magnetic structure's coordinate setting, the parent operations are first expressed in the magnetic structure's setting before the spatial null-space calculation of Sec.~\ref{subsec:solver} is applied.

After obtaining the tensor components of the nuclear parent, we obtain the allowed spin directions $\mathbf m$ in the crystal frame from the magnetic site symmetry, which we combine to obtain the magnetic tensor without SOC
\begin{equation}
\mathcal M^{(k)}_{\mathrm{noSOC}}
=\mathcal Q^{(k-1)}_{\mathrm{elec}}
\otimes\mathbf m^{(1)}_{\mathrm{spin}} .
\end{equation}

When the local tensors are propagated over the magnetic Wyckoff orbit, the spatial part is transformed like a k-1 electric tensor, and the spin is transformed as a time-odd axial vector in
Eq.~\eqref{eq:tensor_transformation}.  Linearly dependent products are removed.
For a selected Wyckoff position, the backend returns the representative-site
basis.  For ``System'', it sums the orbit tensors and keeps only
independent combinations.

The returned parent basis is shared by both interface representations.
The Channel and moment-visualizer outputs keep the spin index explicit and
decompose only the spatial part; the Irreducible output couples all displayed
indices.  After the Cartesian conversion below, a no-SOC magnetic toroidal
tensor of rank $k$ is obtained from the first dual of the no-SOC magnetic
parent of rank $k+1$, as in
Eq.~\eqref{eq:T_from_M}.

\subsection{Crystal, Cartesian, and local frames}
\label{subsec:metric}

The null-space solver yields tensor components in the conventional fractional
crystallographic basis of the space
group~\cite{Aroyo2006BilbaoCrystallographicServer,Aroyo2006BilbaoCrystallographicServerb}.
For spherical-harmonic projections, dualizations, and 3D visualization, every
tensor index is transformed to an orthonormal Cartesian frame, the standard transformation is given by
\begin{equation}
\mathbf{r}_{\mathrm{glob}} = \mathbf{M}\,\mathbf{r}_{\mathrm{cryst}},
\qquad
\mathcal{T}_{\mathrm{glob}} = \mathbf{M}^{\otimes k}\,\mathcal{T}_{\mathrm{cryst}},
\end{equation}
where $k$ is the tensor rank and
\begin{equation}
\label{eq:cryst_to_cart}
\mathbf{M} =
\begin{pmatrix}
a & b\cos\gamma & c\cos\beta \\[4pt]
0 & b\sin\gamma &
  c\,\dfrac{\cos\alpha-\cos\beta\cos\gamma}{\sin\gamma} \\[8pt]
0 & 0 & \dfrac{V}{ab\sin\gamma}
\end{pmatrix},
\end{equation}
with
\begin{equation*}
V = abc\Bigl[
1-\cos^{2}\alpha-\cos^{2}\beta-\cos^{2}\gamma
+2\cos\alpha\cos\beta\cos\gamma
\Bigr]^{1/2}.
\end{equation*}
Equation~\eqref{eq:cryst_to_cart} is evaluated using a symmetry-normalized
conventional cell. Imported structures are first brought to their conventional
setting, after which the lattice-vector lengths are normalized to one.
When a magnetic structure is given in a BNS setting different from that of
the parent nuclear structure, the normalization is performed in the
\emph{parent} cell .

Because both tensor components and the spherical-harmonic index $m$ depend
on the chosen Cartesian axes, the \textsf{Frame} selector provides additional choices (if applicable):
\begin{description}
\item[Crystal] Fixed by the lattice, with
$\hat{\mathbf{x}}\parallel\mathbf{a}$,
$\hat{\mathbf{z}}\parallel\mathbf{c}^{*}$, and
$\hat{\mathbf{y}}=\hat{\mathbf{z}}\times\hat{\mathbf{x}}$, corresponds to
Eq.~\eqref{eq:cryst_to_cart}. This frame is the default choice.

\item[Magnetic] Defined from the full magnetic point group.
The $\hat{\mathbf{z}}$ axis is chosen along the highest-order proper rotation
axis; $\hat{\mathbf{x}}$ is chosen along a perpendicular twofold axis or, if
none exists, normal to a mirror plane containing $\hat{\mathbf{z}}$; and
$\hat{\mathbf{y}}=\hat{\mathbf{z}}\times\hat{\mathbf{x}}$. This gives a single
symmetry-adapted frame for the full magnetic structure.

\item[Site] Defined by the same construction as the magnetic frame, but using the site symmetry of the selected Wyckoff position. 
\end{description}

The resulting Cartesian directions are displayed relative to the lattice parameters, for example
\textsf{Site axes ($x\parallel[1\bar{1}0]$, $z\parallel[001]$)}, so that the
orientation used to label the multipoles remains explicit. The selected
frame is applied consistently to the tensor components, spherical-harmonic
decomposition, 3D visualizer, and \emph{Ordering of Multipoles}. All sites
within a Wyckoff orbit are expressed in the same frame, allowing their
components to be compared directly.

\subsection{Channel--Irreducible coupling}
\label{subsec}

The interface provides both channel and irreducible decompositions of the multipoles in the \emph{Site Multipoles}, \emph{Ordering of Multipoles}, and 3D visualizer panels. The \emph{Site Multipoles} tab displays the full channel and irreducible composition for a representative atom of a Wyckoff orbit or for the system as a whole, while \emph{Ordering of Multipoles} (Sec.~\ref{subsec}) shows how these multipoles are distributed over the Wyckoff orbit.

For magnetic multipoles, this combined view provides a direct comparison between the non-relativistic and relativistic limits. Without SOC, spin and spatial transformations are independent, and the channel multipoles $Y_{\ell m}M_c$ form the symmetry-adapted basis, with their number giving the independent degrees of freedom. With SOC, different Cartesian channels become coupled, and the symmetry-adapted basis is instead given by the allowed irreducible ordinary, toroidal, and poloidal multipoles. Displaying both decompositions simultaneously therefore makes the onset of channel coupling under SOC directly visible and shows how independent channel components recombine into irreducible multipoles.

\subsection{Orbit propagation and ordering}
\label{subsec:orbit_alignment}

To implement the orbit propagation and classification derived in Sec.~\ref{subsec:local_system}, the program evaluates two complementary site-resolved properties across each Wyckoff orbit.

\paragraph*{Atom-card color groups}
In the \emph{Moment Tensors} tab, the moment tensor of a reference atom is decomposed into the allowed cartesian components and propagated along the orbit. Atom cards are grouped by comparing their transformed tensor bases: sites sharing identical or sign-reversed basis tensors receive matching color badges, automatically identifying collinear magnetic sublattices without requiring a separate color-group calculation.

\paragraph*{Alignment of multipoles}
In the \emph{Ordering of Multipoles} tab, the program evaluates the orbit sum of Eq.~\eqref{eq:orbit_sum_def} for each Channel and Irreducible component in the global Cartesian frame, reporting \emph{Anti-Ferro}, \emph{Ferro}, \emph{Mixed}, or \emph{NCL} badges. For macroscopic ``System'' calculations, the no-SOC tensor is formed by summing occupied magnetic orbits, whereas the SOC tensor is solved directly as the invariant null space of the full magnetic group.

\subsection{Visualization methodology}
\label{sec:visualization}
The interface provides two complementary 3D displays under \emph{3D Representation}: the \emph{Multipoles} view renders their angular dependence as lobes centered on atomic sites, while the \emph{Vector Field} view displays a representative vector field throughout the crystal cell.
In the \emph{Multipoles} view, slider coordinates adjust the spherical-harmonic expansion coefficients of each site tensor to render real spherical harmonics centered on atomic positions. The magnitude sets the radial extent of each lobe, and the sign determines its color, using a consistent color scale across the orbit to expose ferroic and antiferroic alignments directly. The amplitude of the sliders for each mode fixes the amplitude of the components in the underlying Cartesian tensor (Secs.~\ref{subsec:solver} and~\ref{subsec:nosoc_solver}). This enables users to directly observe the correspondence and symmetry between different channel multipoles and their relationship to the irreducible multipoles.
The \emph{Vector Field} view displays the representative vector field built from the irreducible multipole components. The angular contributions from the different multipoles on all sites are summed with an $r^{-(k+1)}$ radial decay ($r^{-2}$ for a rank-0 monopole). Lattice periodicity is included by summing over a balanced $3\times3\times3$ set of periodic images. Clicking an atom isolates its single-site field, while clicking outside restores the full periodic sum. Because the construction is linear, basis fields for each slider are precalculated once in a background worker, allowing for smooth updates of the displayed fields when the sliders are moved.

The \emph{Displayed field} control switches between the direct moment field $\boldsymbol\mu(\mathbf r)$ and its first and second duals, $\mathbf r\times\boldsymbol\mu(\mathbf r)$ and $\mathbf r\times[\mathbf r\times\boldsymbol\mu(\mathbf r)]$, respectively. The direct and first-dual views are illustrated for local toroidal order in the $\alpha$-\ce{Cr2O3} example (Fig.~\ref{fig:Cr2O3Toroidal}), where they connect circulating moment fields to the orientation and cancellation of the local toroidal dipoles. Camera presets align views along high-symmetry planes ($ab$, $ac$, $bc$), appearance controls adjust arrow density and thresholds, and scenes can be exported as PNG images.

\section{Examples}
\label{sec:examples}
We now apply the method to a sequence of materials of increasing multipolar complexity. We begin with electric dipole order (ferroic in \ce{BaTiO3} and antiferroic in \ce{PbZrO3}), then turn to magnetic multipoles: the ferroic rank-0 magnetoelectric monopole and rank-2 magnetic quadrupole of \ce{Cr2O3}, and the higher-rank octupole and triakontadipole order parameters of the altermagnets \ce{MnF2} ($d$-wave), \ce{MnTe} ($g$-wave), and the non-collinear \ce{Mn3IrSi}. For each we compare the symmetry-allowed multipoles with first-principles calculations; the detailed DFT results are provided in the Supplementary Material. Crystal and magnetic structure panels in this section were rendered with VESTA~\cite{Momma2011VESTA3Threedimensional}. All examples are bundled with the program and can be loaded directly from its
interface.

\subsection[Ferroic electric-dipole order in ferroelectric \ce{BaTiO3}]{Ferroic electric-dipole order in ferroelectric \ce{BaTiO3}}

\begin{figure}[tb]
\centering
\includegraphics[width=\linewidth]{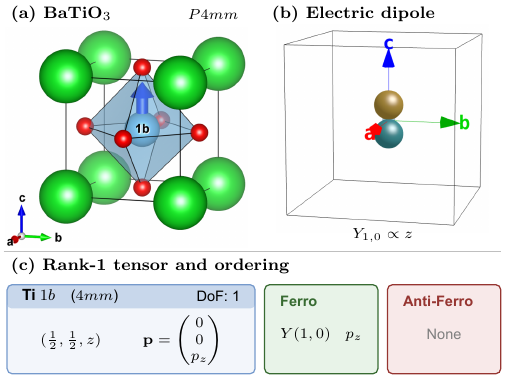}
\caption[Electric-dipole order in tetragonal BaTiO3]{\textbf{Ferroic electric-dipole order in tetragonal \ce{BaTiO3}.}
(a) $P4mm$ structure, with Ba, Ti, and O in green, blue, and red.
(b) Spherical-harmonic representation of the electric dipole ($Y_{1,0} \propto z$), oriented along the polar $c$ axis.
(c) Symmetry-allowed rank-1 tensor at the Ti $1b$ site and its ferroic ordering. The only allowed local component is $p_z$; the system polarization likewise has a single allowed component along $c$.}
\label{fig:BaTiO3}
\end{figure}

We begin with \ce{BaTiO3}, which illustrates ferroic electric-dipole order.
Figure~\ref{fig:BaTiO3} connects its tetragonal structure to the allowed
system polarization and the corresponding rank-1 output.
At rank 1, the electric moment tensor is simply the electric dipole vector.
The high-temperature cubic paraelectric phase ($Pm\bar{3}m$, No.~221) is
centrosymmetric and forbids a macroscopic polarization. On cooling through
the ferroelectric transition, a polar distortion breaks inversion symmetry,
producing the tetragonal ferroelectric phase ($P4mm$, No.~99) with
spontaneous polarization along $c$~\cite{1945CrystalStructureBarium,Kay1949XCVSymmetryChanges,
Devonshire1949XCVITheoryBarium,devonshireTheoryFerroelectrics1954,
VonHippel1950FerroelectricityDomainStructure}.

\textit{MagSymMultipoles} finds a single allowed local component,
$\mathbf p=(0,0,p_z)$, on every occupied orbit. Ba ($1a$), Ti ($1b$), and
apical O(1) ($1b$) have site symmetry $4mm$; equatorial O(2) ($2c$) has
site symmetry $2mm$. All $P4mm$ operations preserve the polar $c$ direction,
so symmetry-equivalent sites carry identical dipoles. Each orbit therefore
supports ferroic dipole order; in particular, the two O(2) dipoles add.

Summing the local contributions with their orbit multiplicities gives,
using Eq.~\eqref{eq:system_tensor},
\begin{equation}
\label{eq:BaTiO3_P}
\begin{aligned}
    \mathbf P_{\mathrm{sys}}
    &=\mathcal D^{(1)}_{\mathrm{sys}}=(0,0,P_z),\\
    P_z&=\frac{p_{z,\mathrm{Ba}}+p_{z,\mathrm{Ti}}
                  +p_{z,\mathrm{O1}}+2p_{z,\mathrm{O2}}}
                 {\Omega_{\mathrm{cell}}}.
\end{aligned}
\end{equation}
Here $p_{z,\alpha}$ is the dipole per site on orbit $\alpha$.
The result matches the single allowed component of the system point group
$4mm$ (Fig.~\ref{fig:BaTiO3}). Symmetry fixes the common axis and the
relations within each orbit, but leaves the amplitudes and relative signs
between distinct orbits undetermined.

The Ti-centered electronic charge-density projection contains a nonzero
$z$-like dipolar component (Supp.~Table~S2),
consistent with this local symmetry assignment. The macroscopic
polarization requires the full electronic and ionic contributions.

\subsection[Antipolar electric-dipole order in antiferroelectric \ce{PbZrO3}]{Antipolar electric-dipole order in antiferroelectric \ce{PbZrO3}}

\begin{figure}[htb!]
\centering
\includegraphics[width=.9\linewidth]{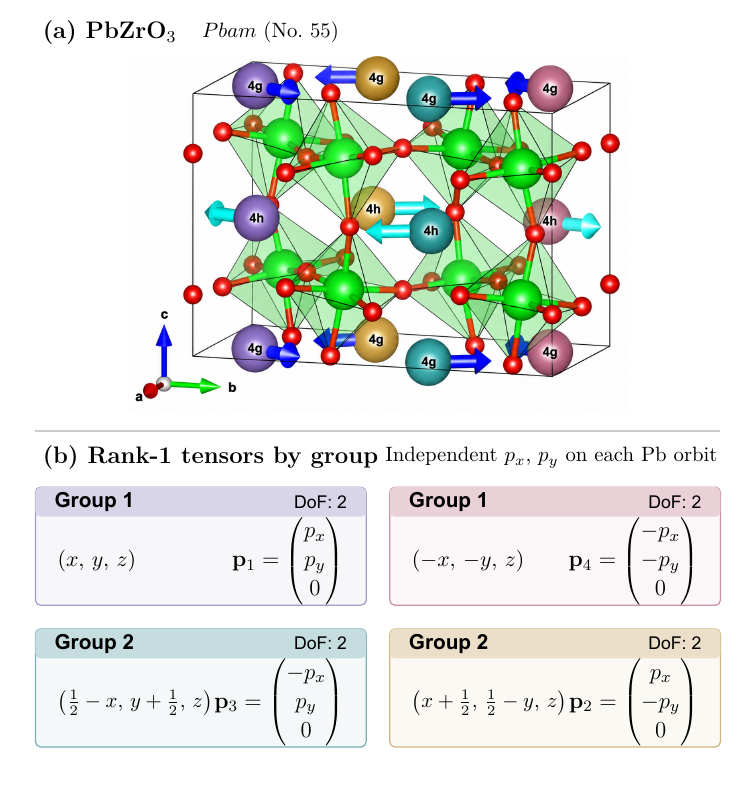}
\caption{\textbf{Local electric dipoles and symmetry-enforced cancellation in antiferroelectric \ce{PbZrO3}.}
(a) Orthorhombic $Pbam$ crystal structure showing Pb atoms in the $4g$ ($z=0$) and $4h$ ($z=\frac{1}{2}$) Wyckoff orbits with site symmetry $..m$ ($\mathrm{DoF}=2$ per orbit). Blue and cyan arrows show a possible in-plane direction of the local polarization on the $4g$ and $4h$ sites, respectively, with Pb atoms color-coded according to the four site groups in (b).
(b) \textit{MagSymMultipoles} rank-1 tensor cards for the four symmetry-related groups. Inversion symmetry pairs Groups 1--4 ($\mathbf p_4 = -\mathbf p_1$) and Groups 2--3 ($\mathbf p_3 = -\mathbf p_2$), ensuring exact pairwise cancellation on each orbit and yielding a vanishing net system dipole ($\mathbf P_{\mathrm{sys}} = 0$, $\mathrm{DoF}=0$).}
\label{fig:PbZrO3}
\end{figure}

\ce{PbZrO3} illustrates the complementary case to \ce{BaTiO3}: local
electric dipoles are allowed, but their orbit sums vanish. Its orthorhombic
antiferroelectric phase has space group $Pbam$ (No.~55), whose
centrosymmetric point group $mmm$ forbids a macroscopic polarization,
$\mathbf P_{\mathrm{sys}}=0$~\cite{kittelTheoryAntiferroelectricCrystals1951,
Tagantsev2013OriginAntiferroelectricityPbZrO3}. This constraint does not
require the local dipoles to vanish. Figure~\ref{fig:PbZrO3}(a) shows the
Pb orbits and the allowed local dipoles whose symmetry-enforced
cancellation produces this vanishing system polarization.

The Pb atoms occupy two distinct Wyckoff orbits, $4g$ ($z=0$) and $4h$
($z=\frac{1}{2}$), both with site symmetry $..m$. The mirror plane
perpendicular to $c$ restricts the local dipole to the basal plane,
$\mathbf p_{\mathrm{loc}}=(p_x,p_y,0)$ ($\mathrm{DoF}=2$ per orbit).
In \textit{MagSymMultipoles}, propagating this representative dipole
partitions each orbit into four symmetry-related site groups
[Fig.~\ref{fig:PbZrO3}(b)]. Inversion symmetry pairs Group~1 at $(x,y,z)$
with Group~4 at $(-x,-y,z)$ ($\mathbf p_4 = -\mathbf p_1$), and Group~2
with Group~3 ($\mathbf p_3 = -\mathbf p_2$). These pairwise relations
enforce exact dipole cancellation separately on each orbit:
\begin{equation}
    \sum_{a\in 4g}\mathbf p(a)=0,
    \qquad
    \sum_{a\in 4h}\mathbf p(a)=0.
\end{equation}
Because the in-plane amplitudes $p_x$ and $p_y$ are independent between
the $4g$ and $4h$ orbits, cancellation holds separately for each,
whatever those microscopic amplitudes are.

The contrast with \ce{BaTiO3} lies in orbit propagation: its allowed local
dipoles add within each orbit, whereas those on the Pb orbits cancel by
symmetry. The calculation therefore identifies the allowed local
antipolar pattern hidden by the vanishing system dipole, while leaving
its microscopic amplitudes undetermined.

\subsection{Ferroic magnetoelectric multipoles in $\alpha$-\ce{Cr2O3}}

\begin{figure*}[htb]
    \centering
    \includegraphics[width=.70\linewidth]{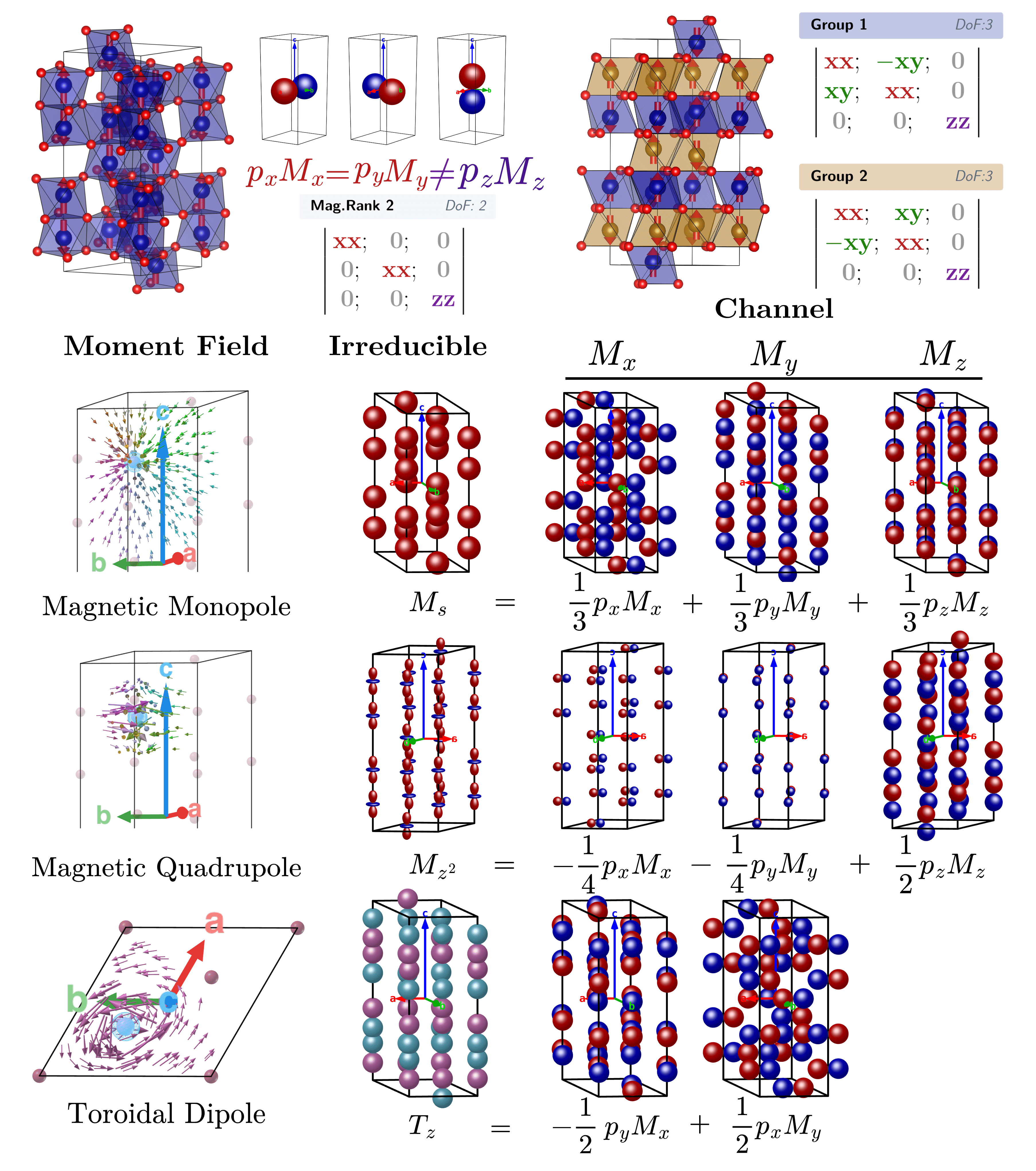}
    \caption{\textbf{Rank-2 magnetic moment tensor and irreducible multipoles
    in $\alpha$-\ce{Cr2O3} from \textit{MagSymMultipoles}.} (Color online)
    Upper part: magnetic structure with dipole moments (red arrows, left),
    system tensor with $p_xM_x=p_yM_y\neq p_zM_z$ (\texttt{DoF: 2}, center),
    and local site-symmetry partition into \texttt{Group 1} and
    \texttt{Group 2} (\texttt{DoF: 3}) with highlighted atomic positions (right).
    Lower part: rows show the magnetic monopole $M_s$, the $z^2$ magnetic
    quadrupole $M_{z^2}$, and the local toroidal dipole $T_z$.
    Columns show the moment field, irreducible multipole texture, and
    underlying microscopic channels $p_iM_j$ with their weights.}
    \label{fig:Cr2O3System}
\end{figure*}

$\alpha$-\ce{Cr2O3} is the prototypical linear magnetoelectric antiferromagnet~\cite{Rado1962MagnetoelectricEffectsAntiferromagneticsa,bousquetSignLinearMagnetoelectric2024}. Below $T_N=307$~K, it belongs to the magnetic space group $R\bar{3}'c'$ (\#167.106), preserving $\mathcal{PT}$ while breaking $\mathcal{P}$ and $\mathcal{T}$ individually. The four Cr spins along $c$ alternate as [$+--+$] (Fig.~\ref{fig:Cr2O3System}, upper left), compensating the magnetic dipole moment ($\mathcal{M}^{(1)}_{\mathrm{sys}}=0$). The orbit-summed rank-2 magnetic moment tensor is nevertheless nonzero and shares the symmetry of the linear magnetoelectric response $\alpha_{ij}$ (Jahn symbol $ae[V^1]V$).

As obtained from \textit{MagSymMultipoles} (Fig.~\ref{fig:Cr2O3System}, upper center), the system tensor has two independent diagonal degrees of freedom (\texttt{DoF: 2}) corresponding to the channel products $p_xM_x=p_yM_y\neq p_zM_z$,
\begin{equation}
\label{eq:Cr2O3_system_tensor}
\mathcal{M}^{(2)}_{\mathrm{sys}}
=
\begin{pmatrix}
p_xM_x & 0 & 0\\
0 & p_xM_x & 0\\
0 & 0 & p_zM_z
\end{pmatrix}.
\end{equation}
This diagonal tensor decomposes into the ferroic magnetoelectric monopole
\begin{equation}
\label{eq:Cr2O3_monopole}
M_s = \frac{1}{3}p_xM_x + \frac{1}{3}p_yM_y + \frac{1}{3}p_zM_z
\end{equation}
(scalar trace)~\cite{spaldinMonopolebasedFormalismDiagonal2013} and the ferroic $z^2$ magnetic quadrupole
\begin{equation}
\label{eq:Cr2O3_quadrupole}
M_{z^2} = -\frac{1}{4}p_xM_x - \frac{1}{4}p_yM_y + \frac{1}{2}p_zM_z
\end{equation}
(symmetric-traceless part). Both components order ferroically. Together, they describe the symmetry of the transverse ($\alpha_{xx}$) and longitudinal ($\alpha_{zz}$) magnetoelectric responses.

\begin{figure}
    \centering
    \includegraphics[width=1\linewidth]{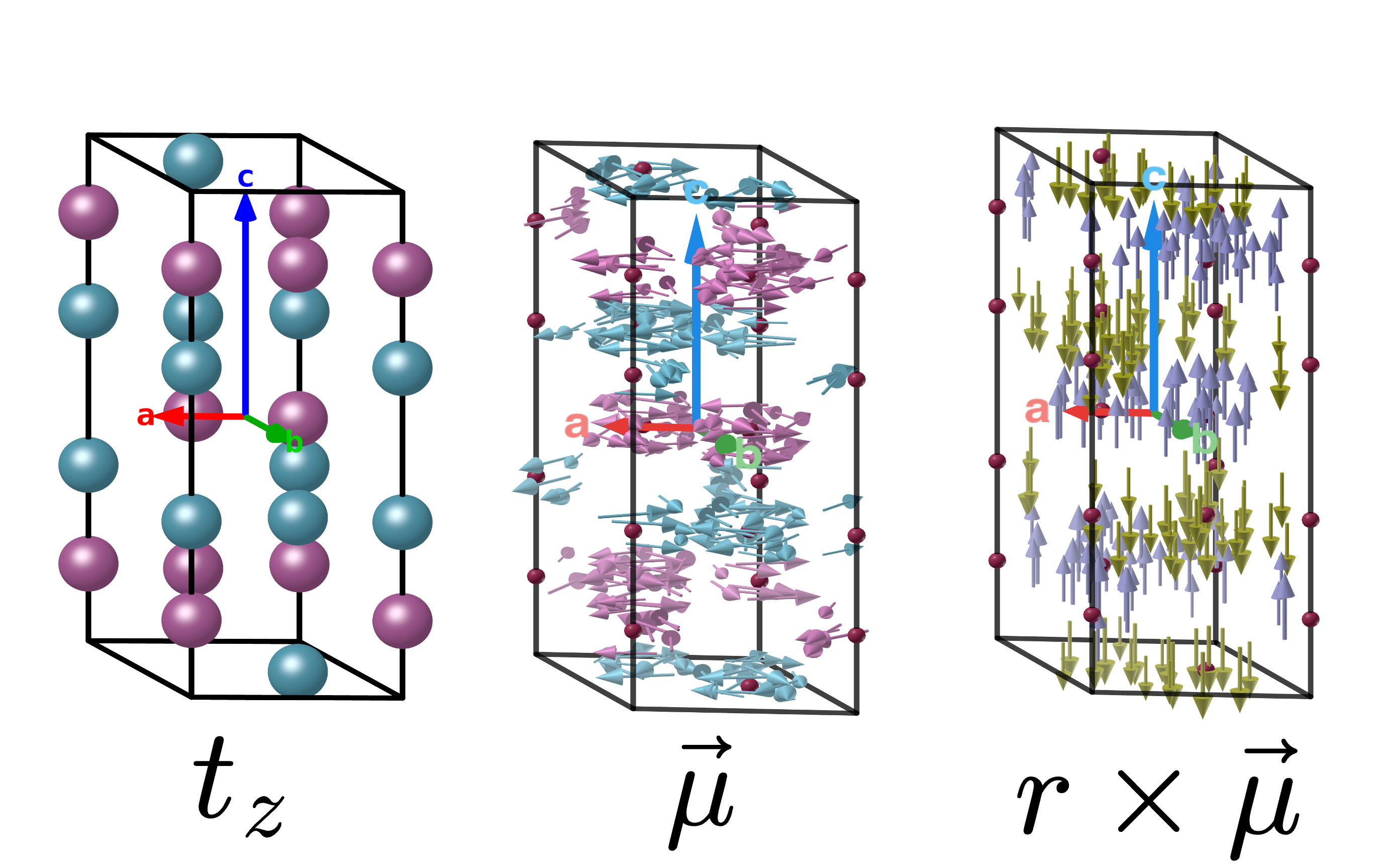}
    \caption{\textbf{Antiferroic local toroidal order of $\alpha$-\ce{Cr2O3} in the direct and first-dual vector fields.}
    (Color online) \textbf{Left:} The Cr $12c$ orbit partitioned into two symmetry-related subsets carrying opposite local magnetic toroidal dipoles, $\pm t_z$.
    \textbf{Center:} Magnetic vector field $\boldsymbol\mu(\mathbf r)$ of the antisymmetric part of the local rank-2 tensor, with magenta and cyan indicating the two sublattices, showing the toroidal magnetic moment distribution.
    \textbf{Right:} First-dual field $\mathbf r\times\boldsymbol\mu(\mathbf r)$, which exposes the opposite $c$-axis toroidal dipoles}
    \label{fig:Cr2O3Toroidal}
\end{figure}

The site symmetry in Fig.~\ref{fig:Cr2O3System} (upper right) reveals how the system symmetry emerges from the Cr $12c$ orbit. The sites partition into two symmetry-related sublattices (\texttt{Group 1} and \texttt{Group 2}, each with \texttt{DoF: 3}), with tensors~\cite{urruMagneticOctupoleTensor2022,Verbeek2023HiddenOrdersAntimagnetoelectric}
\begin{equation}
\label{eq:Cr2O3_local_tensor}
\mathcal{M}^{(2)}
=
\begin{pmatrix}
p_xM_x & \pm p_xM_y & 0\\
\pm p_yM_x & p_xM_x & 0\\
0 & 0 & p_zM_z
\end{pmatrix},
\end{equation}
where the off-diagonal entries satisfy $p_yM_x = -p_xM_y$ on each site. Together, these two transverse channels form the local magnetic toroidal dipole
\begin{equation}
\label{eq:Cr2O3_toroidal}
T_z = -\frac{1}{2}p_yM_x + \frac{1}{2}p_xM_y.
\end{equation}
As shown in the upper-right part of Fig.~\ref{fig:Cr2O3System}, the crystal structure displayed alongside each group card encodes the spatial positions of its constituent sites, revealing that \texttt{Group 1} and \texttt{Group 2} form alternating slabs stacked along the $c$-axis. The diagonal channels generating the monopole $M_s$ and quadrupole $M_{z^2}$ are identical across both slabs and survive the orbit sum. In contrast, the antisymmetric off-diagonal channels take opposite signs on the two sublattices ($+p_xM_y, -p_yM_x$ on \texttt{Group 1} versus $-p_xM_y, +p_yM_x$ on \texttt{Group 2}) and cancel between adjacent slabs in the sum, yielding antiferroic toroidal order (Fig.~\ref{fig:Cr2O3Toroidal}). The direct moment field in the center panel displays opposite circulations on the two subsets of Cr sites. The first-dual field $\mathbf r\times\boldsymbol\mu$ in the right panel exposes their opposite $c$-directed toroidal dipoles, making their cancellation in the orbit sum explicit.

The microscopic channel decomposition in the lower part of Fig.~\ref{fig:Cr2O3System} explains this hierarchy. In the non-relativistic collinear limit, the alternating local polar distortion $p_z$ and magnetic moment $M_z$ multiply to give a purely ferroic product $p_zM_z$. This single channel contributes simultaneously to $M_s$ and $M_{z^2}$, driving the longitudinal response $\alpha_{zz}$ already without spin--orbit coupling~\cite{Rado1961MechanismMagnetoelectricEffect,Date1961OriginMagnetoelectricEffect}. Relativistic SOC then enables the transverse diagonal products $p_xM_x=p_yM_y$ responsible for $\alpha_\perp$, as well as the antisymmetric combination $T_z=-\frac{1}{2}p_yM_x+\frac{1}{2}p_xM_y$ that generates the antiferroic local toroidal dipole. As visualized in the lower part of Fig.~\ref{fig:Cr2O3System}, each irreducible multipole isolates these distinct channel combinations, exhibiting isotropic ferroic textures for $M_s$, oriented $d_{z^2}$ lobes for $M_{z^2}$, and vortex-like antiferroic cancellation for $T_z$.

The component-resolved orbit sum is reported in Supp.~Table~S3, and the corresponding DFT magnetic densities are shown in Supp.~Fig.~S1.

\subsection{Magnetic octupolar order: $d$-wave altermagnetism in \ce{MnF2}}
\begin{figure}[tb]
    \centering
    \includegraphics[width=1\linewidth]{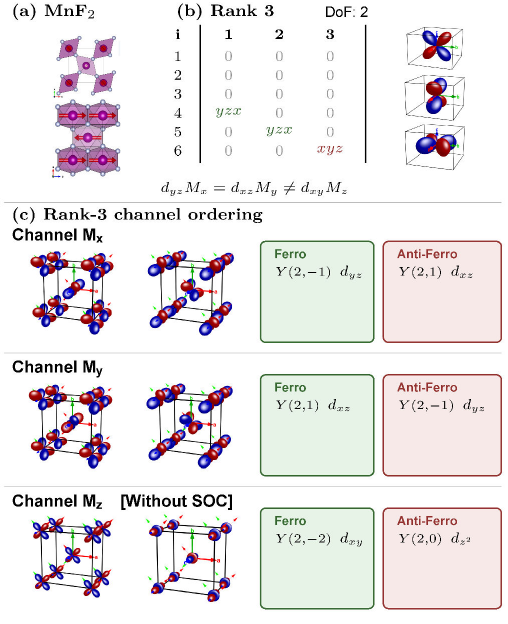}
    \caption{\textbf{Magnetic octupolar order in $d$-wave altermagnetic \ce{MnF2}.}
    (a) Magnetic structure. (b) System tensor $\mathcal M^{(3)}$ ($\mathrm{DoF}=2$), showing the tetragonal equality $d_{yz}M_x=d_{xz}M_y$ (green) and longitudinal component $d_{xy}M_z$ (red), alongside the corresponding spherical harmonics. (c) Channel-resolved octupoles and their ferroic or antiferroic classifications; the $M_z$ row shows the result without SOC.}
    \label{fig:MnF2System}
\end{figure}

Our next example is the prototypical $d$-wave altermagnet \ce{MnF2}, which exhibits non-relativistic momentum-dependent band splitting~\cite{smejkalEmergingResearchLandscape2022,Bai2024AltermagnetismExploringNew}. In its paramagnetic phase, rutile \ce{MnF2} belongs to the $P4_2/mnm$ space group. At $T_N \approx 67$~K it undergoes a phase transition into the collinear
antiferromagnetic space group $P4_2'/mnm'$ (BNS~136.499), with Mn moments aligned
antiparallel along the $c$ axis~\cite{Yamani2010NeutronScatteringStudy,Griffel1950MagneticAnisotropyManganous,Erickson1953NeutronDiffractionStudies,Stout1942MagnetismThirdLaw}.

Analyzing the system with \textit{MagSymMultipoles}, we find that while the magnetic dipoles compensate exactly ($\mathcal M^{(1)}_{\mathrm{sys}} = 0$), a macroscopic magnetic octupole is allowed. The corner and body-center Mn sites are coordinated by
\ce{F6} octahedra rotated by $90^\circ$ relative to one another about $c$
[Fig.~\ref{fig:MnF2System}(a)], inducing a local electric quadrupole of $d_{xy}$
symmetry. Under the fourfold screw rotation $\{C_{4z}|\frac{1}{2}\frac{1}{2}\frac{1}{2}\}$,
the $90^\circ$ rotation flips the quadrupole sign ($d_{xy} \to -d_{xy}$) while the
antiparallel moment reverses ($M_z \to -M_z$). Their product $(-d_{xy})(-M_z) = +d_{xy}M_z$
is invariant across both sites, adding constructively into a macroscopic ferroic
magnetic octupole~\cite{bhowalFerroicallyOrderedMagnetic2024} [Fig.~\ref{fig:MnF2System}(b)--(c)].
Using the program, we find that the magnetic rank-3 system tensor $\mathcal M^{(3)}_{\mathrm{sys}}$ has three nonzero components [Fig.~\ref{fig:MnF2System}(b)].
Without SOC, only the longitudinal channel $M_z$ is active,
yielding a single Channel multipole [degree of freedom ($\mathrm{DoF} = 1$)]:
$\mathcal M^{(3)}_{\mathrm{sys},xy,z} = \mathcal M^{(3)}_{\mathrm{sys},yx,z}$ ($d_{xy}M_z$).
Turning on SOC ($P4_2'/mnm'$), relativistic coupling activates the transverse channels
$M_x$ and $M_y$, while tetragonal symmetry enforces $\mathcal M^{(3)}_{\mathrm{sys},yz,x} = \mathcal M^{(3)}_{\mathrm{sys},xz,y}$
($d_{yz}M_x = d_{xz}M_y$)~\cite{Buiarelli2025NoncollinearMagneticMultipoles}. Thus, the four nonzero channel octupoles correspond to only two independent degrees of freedom ($\mathrm{DoF} = 2$). We further identify three antiferro-ordered multipoles, pointed out in a previous report~\cite{bhowalFerroicallyOrderedMagnetic2024}.
DFT calculations confirm the presence of these exact multipoles in the magnetization density as predicted by \textit{MagSymMultipoles} (Supp.~Table~S4 and Supp.~Fig.~S2). The symmetry-derived harmonic decompositions are shown in Supp.~Tables~S5 and~S6 for ranks 1 and 3, and rank 5, respectively; the corresponding DFT decompositions are shown in Supp.~Tables~S7 and~S8.

\begin{figure}[tb]
    \centering
    \includegraphics[width=.8\linewidth]{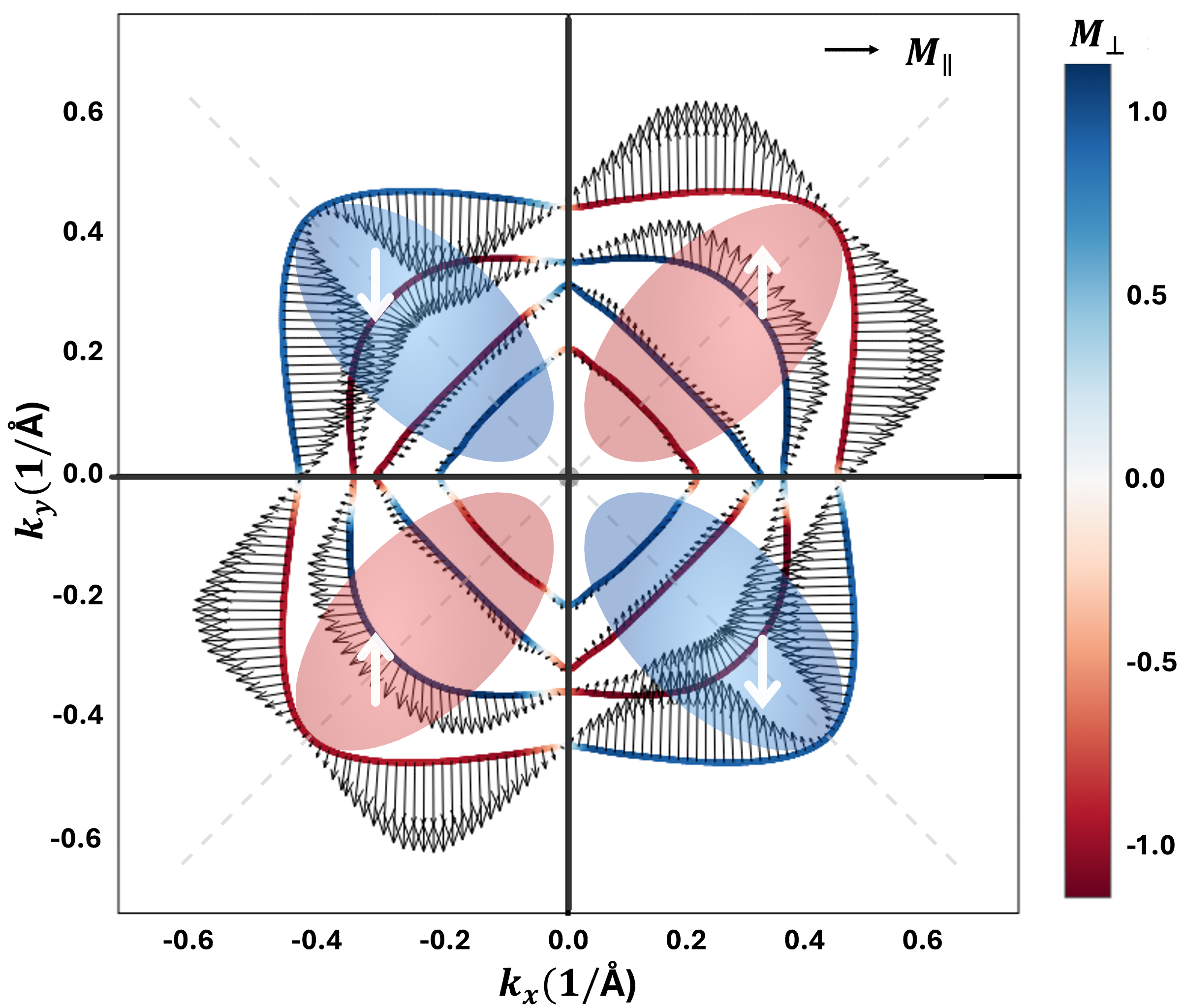}
    \caption{\textbf{Momentum-resolved spin texture of \ce{MnF2} in a $k_x$--$k_y$ slice at $-8.90$~eV relative to the Fermi level}~\cite{Ganose2021IFermiPythonLibrary}.
    Color map: out-of-plane spin component $M_\perp \equiv M_z$ governed by the non-relativistic octupole $d_{xy}M_z$ [$C_{\parallel} k_x k_y \sigma_z$ in Eq.~\eqref{eq:MnF2_H_spin}].
    Black arrows: in-plane texture $\mathbf{M}_\parallel=(M_x,M_y)$ induced by SOC through transverse octupoles $d_{yz}M_x$ and $d_{xz}M_y$ [$C_{\perp}(k_y k_z \sigma_x + k_x k_z \sigma_y)$].}
    \label{fig:spin_texture}
\end{figure}

The nonzero ferroic octupoles in the different magnetization channels can be directly related to the symmetry-allowed low-energy spin Hamiltonian describing the altermagnetism
around $\Gamma$~\cite{Yu2026IdentifyingOrientedSpin}:
\begin{align}
\label{eq:MnF2_H_spin}
 H_{\mathrm{spin}}^{\text{no SOC}}(\mathbf k)
    &= C_{\parallel}^{(0)}
       \underbrace{k_x k_y \sigma_z}_{d_{xy}M_z},
       \\
 H_{\mathrm{spin}}^{\text{SOC}}(\mathbf k)
    &= C_{\perp}\Bigl(
       \underbrace{k_y k_z \sigma_x}_{d_{yz}M_x}
       + \underbrace{k_x k_z \sigma_y}_{d_{xz}M_y}
       \Bigr)
       \notag\\
    &\quad
       + C_{\parallel}
       \underbrace{k_x k_y \sigma_z}_{d_{xy}M_z},
\end{align}
where we marked each term with the corresponding real-space multipole.
Without SOC, the $k_xk_y\sigma_z$ term gives the $d$-wave band splitting
with out-of-plane polarization $M_z$: its sign reverses when either $k_x$
or $k_y$ changes sign. Figure~\ref{fig:spin_texture} illustrates the
corresponding alternating out-of-plane spin polarization in a momentum-space
slice at $-8.90$~eV relative to the Fermi level. The black arrows resolve
the transverse spin components allowed by SOC. In the expansion around
$\Gamma$, these components arise from the $k_yk_z\sigma_x$ and
$k_xk_z\sigma_y$ terms and vanish at $k_z=0$ to the order retained.
The expansion identifies the leading symmetry-allowed angular dependence;
it does not by itself establish a quantitative fit to the displayed
constant-energy texture.
Because $\mathcal M^{(3)}_{\mathrm{sys}}$ matches the symmetry of the piezomagnetic tensor $\Lambda$~\cite{Radaelli2024TensorialApproachAltermagnetism}, the existence of this multipole also permits the stress-induced magnetization $M_z = \Lambda_{xy,z}\sigma_{xy}$.

Beyond the symmetric octupolar sector, Cartesian contractions isolate local multipoles
that cancel in the macroscopic sum. Successive dual contractions with the Levi-Civita
tensor (Sec.~\ref{subsec:reducibility}) map the
antisymmetric components to the toroidal intermediate $\mathfrak T^{(2)}_{xy} = \mathcal M^{(3)}_{xz,x} - \mathcal M^{(3)}_{xx,z}$
and then to a local $z$-directed poloidal dipole $\boldsymbol{\mathcal H}^{(1)} = (0, 0, 2\mathfrak T^{(2)}_{xy})$.
The field textures in Fig.~\ref{fig:MnF2Poloidal} illustrate this local
poloidal dipole in \ce{MnF2}.
Under the fourfold screw, $\mathfrak T^{(2)}_{xy}$ reverses
sign between sublattices, producing an antiferroic poloidal order that cancels across
the unit cell (Supp.~Tables~S4 and~S9, with the matrix conventions given in Supplementary Sec.~SIII\,C\,2), cleanly separating the macroscopic ferroic octupolar order parameter from the compensated internal poloidal circulation.

\subsection{Magnetic triakontadipolar order: $g$-wave altermagnetism in MnTe}
\label{subsec:MnTe}
\begin{figure}[t]
    \centering
    \includegraphics[width=1\linewidth]{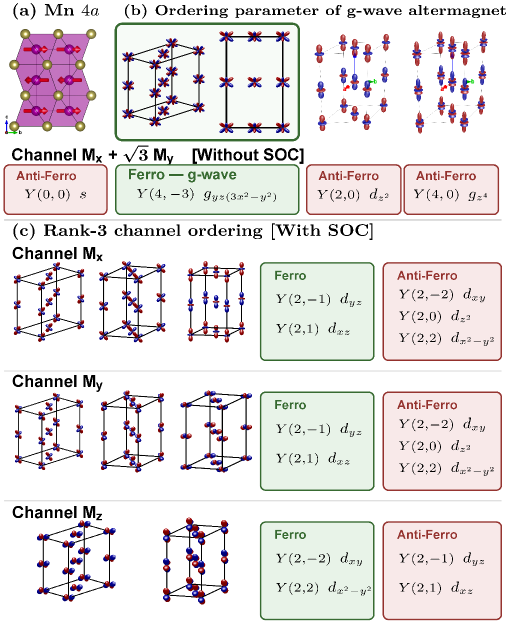}
    \caption{\textbf{Multipolar ordering in the $g$-wave altermagnet \ce{MnTe}.}
    (a) Compensated Mn $4a$ magnetic structure. (b) Without SOC, surviving ferroic rank-5 triakontadipole $g_{yz(3x^2-y^2)}$ (green outline), with in-plane projections linked by $M_y/M_x=\sqrt{3}$, alongside canceling antiferroic channels. (c) Channel-resolved rank-3 octupoles under SOC, showing allowed ferroic (green) and antiferroic (red) harmonics.}
    \label{fig:MnTeSystem}
\end{figure}

\ce{MnTe} is a prototypical $g$-wave altermagnet~\cite{
krempaskyAltermagneticLiftingKramers2024,
Amin2024MnTeAltermagnetism,
Sunko2025LinearMagnetobirefringenceProbe}.
Its paramagnetic parent has the NiAs-type structure with
space group $P6_3/mmc$. Below $T_N \approx 310$~K, the Mn
moments order collinearly in the basal plane. Because the
in-plane anisotropy is weak, the orientation of the moments
can depend on the sample. Here, we select $[110]$, one of
the orientations that retain maximal magnetic symmetry,
yielding the magnetic space group $Cmcm$ with point group
$mmm$ [Fig.~\ref{fig:MnTeSystem}(a)]. The Mn sites occupy
the $4a$ Wyckoff position in this orthorhombic setting.
An alternative maximal-symmetry orientation along
$[1\bar{1}0]$ yields $Cm'c'm$ (MAGNDATA~\cite{Gallego2016MAGNDATADatabaseMagnetic} entry 0.800)
and has been observed in epitaxial thin films~\cite{
kriegnerMultiplestableAnisotropicMagnetoresistance2016}.
For the $Cmcm$ configuration, the magnetic point group
forbids a net dipole moment,
$\mathcal M^{(1)}_{\mathrm{sys}}=0$.

In the non-relativistic limit, the symmetry relations between
the opposite-spin sublattices enforce cancellation of the
odd-rank magnetic multipoles below rank~5.
The lowest-rank surviving order parameter is therefore the
rank-5 magnetic triakontadipole
[Fig.~\ref{fig:MnTeSystem}(b)].
As identified by \textit{MagSymMultipoles}, this triakontadipole combines
an electric hexadecapole $g_{yz(3x^2-y^2)}$ that is antiferroically ordered across the Mn sites with the alternating magnetic moments.
The resulting composite rank-5 triakontadipole
$g_{yz(3x^2-y^2)}\otimes(M_x,M_y)$ adds constructively
in the orbit sum~\cite{
verbeekNonrelativisticFerromagnetotriakontadipolarOrder2024,
mcclartyLandauTheoryAltermagnetism2024}.
For moments along $[110]$, the geometric ratio
$M_y/M_x=\sqrt{3}$ in the Cartesian coordinates used here
links the two in-plane projections in
Fig.~\ref{fig:MnTeSystem}(b) as components of a single
ferroic order parameter. Additionally, we find two more antiferroically ordered magnetic multipoles allowed in the system: $d_{z^2}$ and $g_{z^4}$.

Enabling spin--orbit coupling, we find that several ferroic rank-3
(octupolar) multipoles appear [Fig.~\ref{fig:MnTeSystem}(c)].
In $Cmcm$, \textit{MagSymMultipoles} identifies three
independent degrees of freedom ($\mathrm{DoF}=3$),
distributed among the ferroic $d_{yz}$ and $d_{xz}$
harmonics in the $M_x$ and $M_y$ channels and the
$d_{xy}$ and $d_{x^2-y^2}$ harmonics in the $M_z$ channel.
In practice, SOC removes the requirement that all channel multipoles
follow the collinear ratio $M_y/M_x=\sqrt{3}$, while
retaining the symmetry relations imposed by the magnetic space group.

For the complete list of symmetry-adapted multipoles obtained by \textit{MagSymMultipoles}, we refer to Supp.~Table~S10. The comparison between the predicted multipoles and multipoles obtained from DFT is reported in Supp.~Table~S11. The SOC-induced rank-3 and rank-5 multipolar distributions are illustrated in Supp.~Table~S12.

The orbit-summed multipoles can be directly related to
the symmetry-allowed terms in the low-energy spin
Hamiltonian around $\Gamma$~\cite{Yu2026IdentifyingOrientedSpin}:
\begin{align}
 H_{\mathrm{spin}}^{\text{noSOC}}(\mathbf k)
    &=
    -\frac{A^{(0)}}{3}
    \underbrace{k_y k_z(3k_x^2-k_y^2)}_{g_{yz(3x^2-y^2)}}
    \Bigl(
      \frac{1}{\sqrt{3}}\underbrace{\sigma_x}_{M_x}
      +\underbrace{\sigma_y}_{M_y}
    \Bigr),
    \label{eq:MnTe_H_spin_noSOC}
    \\
 H_{\mathrm{spin}}^{\text{SOC}}(\mathbf k)
    &=\Bigl(
      C_1\underbrace{k_y k_z}_{d_{yz}}
      +\frac{\sqrt{3}}{2}C_2
        \underbrace{k_x k_z}_{d_{xz}}
    \Bigr)\underbrace{\sigma_x}_{M_x}
    \notag\\
    &\quad
    +\Bigl(
      (C_2-C_1)
        \underbrace{k_x k_z}_{d_{xz}}
      -\frac{\sqrt{3}}{2}C_2
        \underbrace{k_y k_z}_{d_{yz}}
    \Bigr)\underbrace{\sigma_y}_{M_y}
    \notag\\
    &\quad
    -C_3\Bigl(
      \frac{\sqrt{3}}{2}
        \underbrace{(k_x^2-k_y^2)}_{d_{x^2-y^2}}
      +\underbrace{k_x k_y}_{d_{xy}}
    \Bigr)\underbrace{\sigma_z}_{M_z}.
    \label{eq:MnTe_H_spin}
\end{align}
We thus see that, without SOC, the triakontadipoles are responsible for the $g$-wave splitting. With SOC, three independent coefficients $C_1$, $C_2$, and $C_3$ become allowed, corresponding to the ferroic ordering of magnetic octupoles. Consequently, the SOC-induced octupoles produce an additional relativistic $d$-wave band splitting in \ce{MnTe}~\cite{
autieriRelativisticSpinMomentum2026,hirakidaMultipoleAnalysisSpin2026}. These terms can also be related to the $d$-wave-like splitting induced in strained \ce{MnTe}~\cite{belashchenkoGiantStrainInducedSpin2025}.
Further, while pure non-relativistic $g$-wave order forbids linear piezomagnetism~\cite{
Khodas2026TuningAltermagnetismStrain},
the SOC-induced octupolar sector has the tensor character
$ea[V^2]V$ of the linear piezomagnetic response.
This provides a symmetry connection to the experimentally
observed linear piezomagnetism in \ce{MnTe}~\cite{
Aoyama2024PiezomagneticPropertiesAltermagnetic}.

\subsection[Magnetic octupolar order in non-collinear \ce{Mn3IrSi}]{Magnetic octupolar order in non-collinear \ce{Mn3IrSi}}
\label{sec:Mn3IrSi}

\begin{figure}
    \centering
    \includegraphics[width=1\linewidth]{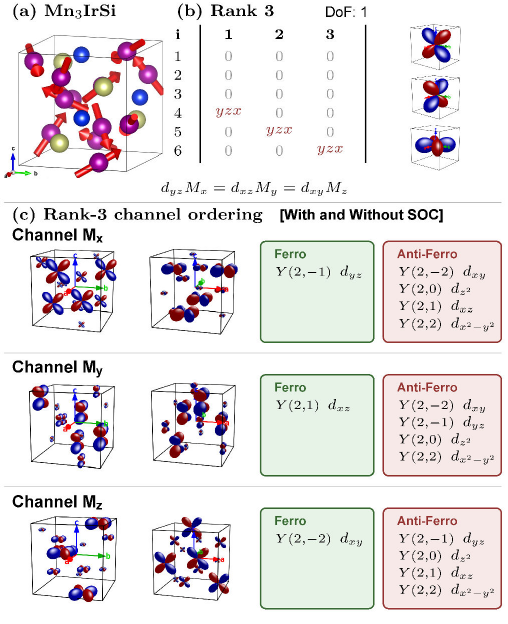}
    \caption[Magnetic octupolar order in non-collinear Mn3IrSi]{\textbf{Magnetic dipoles and octupoles in \ce{Mn3IrSi}.}
    The Mn moments sum to zero, while a ferroic $xyz$ magnetic octupole is
    symmetry-allowed. The local channels $d_{yz}M_x$, $d_{xz}M_y$, and
    $d_{xy}M_z$ have independent amplitudes at a representative site;
    crystal rotations relate them across the orbit and enforce equal
    system components. (a) Magnetic structure. (b) Corresponding
    piezomagnetic tensor and orbital shapes. (c) Channel-resolved orbital
    examples and ferroic/antiferroic harmonic lists with and without SOC.}
    \label{fig:Mn3IrSiSystem}
\end{figure}

\ce{Mn3IrSi} is a non-collinear antiferromagnet with a cubic, chiral crystal
structure. Below $T_N=210$~K, the Mn moments on the $12b$ orbit order in
the magnetic space group $P2_13$ (BNS~198.9), with point group
$23$~\cite{Eriksson2004CrystalMagneticStructure,Eriksson2004CrystalStructureMagnetic}.
The twelve moments form four symmetry-related groups of three
parallel spins, oriented close to four tetrahedral body-diagonal
directions [Fig.~\ref{fig:Mn3IrSiSystem}(a)]. Their sum vanishes,
$\mathcal M^{(1)}_{\mathrm{sys}}=0$, while the magnetic symmetry allows
ferroic octupolar order.

As obtained from \textit{MagSymMultipoles}
(Fig.~\ref{fig:Mn3IrSiSystem}), the orbit-summed rank-3 magnetic moment
tensor has one independent degree of freedom (\texttt{DoF: 1}),
\begin{equation}
\label{eq:Mn3IrSi_system_octupole}
    \mathcal M^{(3)}_{\mathrm{sys},yz,x}
    =\mathcal M^{(3)}_{\mathrm{sys},xz,y}
    =\mathcal M^{(3)}_{\mathrm{sys},xy,z}.
\end{equation}
All other entries vanish apart from spatial-index permutations.
The resulting tensor is fully symmetric and traceless and therefore
corresponds to a ferroic ordinary magnetic octupole of $xyz$ symmetry
(Supp.~Tables~S13
and~S14).

The local channels in Fig.~\ref{fig:Mn3IrSiSystem}(c) reveal how this
ferroic component emerges from the Mn $12b$ orbit. The site symmetry
$1$ allows independent amplitudes for the three local channels
$d_{yz}M_x$, $d_{xz}M_y$, and $d_{xy}M_z$. Their symmetrized combination
is the local ordinary $xyz$ magnetic octupole. Threefold rotations
permute these channels between symmetry-related sites, while twofold
rotations preserve each channel through simultaneous sign changes of
its spatial and spin factors. The local $xyz$ octupole is consequently
identical on all twelve Mn sites and orders ferroically, despite their
different dipole directions. The orbit sum retains this component and
enforces the equalities in Eq.~\eqref{eq:Mn3IrSi_system_octupole}.
The resulting system tensor is consistent with the color-symmetry
analysis of Radaelli and Gurung~\cite{Radaelli2025ColorSymmetryAltermagneticlike}.

The calculator obtains the same rank-3 system tensor with and without
SOC for the reported structure. The corresponding altermagnetic-like
spin texture is therefore symmetry-allowed without SOC, with a leading
even-in-$\mathbf k$, time-odd term
\begin{equation}
\label{eq:Mn3IrSi_H_spin}
    \Delta H(\mathbf k)=C\Bigl(
    \underbrace{k_y k_z \sigma_x}_{d_{yz}M_x}
    + \underbrace{k_x k_z \sigma_y}_{d_{xz}M_y}
    + \underbrace{k_x k_y \sigma_z}_{d_{xy}M_z}
    \Bigr).
\end{equation}
This $d$-wave term couples the three spatial harmonics to three spin
components, with an amplitude $C$ undetermined by symmetry. The same
tensor form governs the allowed piezomagnetic response shown in
Fig.~\ref{fig:Mn3IrSiSystem}(b). SOC can modify the amplitudes but introduces
no additional rank-3 system components while the $P2_13$ magnetic symmetry
is retained.

\section{Conclusion}
In this work, we revisited the theoretical foundations of electric
and magnetic multipoles and developed a unified Cartesian framework
that determines the symmetry-allowed components of moment tensors
of arbitrary rank and their decomposition into ordinary, toroidal,
and poloidal multipoles. We connect symmetry-allowed local multipoles,
their ferroic, antiferroic, or non-collinear arrangements across
crystallographic orbits, and macroscopic responses in both relativistic
and non-relativistic settings, distinguishing effects enabled by
spin--orbit coupling from those allowed without it.

Our central result is the web tool \textit{MagSymMultipoles}, which
implements this framework by automating the construction and
decomposition of symmetry-allowed moment tensors up to rank~5
for all 1{,}651 magnetic space groups. The program applies the
generalized Neumann principle to calculate symmetry-allowed tensor
relations in the conventional crystallographic setting and
distinguishes system-level tensors from local Wyckoff-site tensors
and their orbit ordering.

We presented several representative examples demonstrating how our workflow
enables the visualization of complex multipolar orders. Specifically, we identified the
ferroic magnetoelectric monopole and magnetic quadrupole of
\ce{Cr2O3}, together with its antiferroically ordered local toroidal
dipoles. We also characterized the rank-3 octupolar order parameter
of $d$-wave altermagnetism in \ce{MnF2}, and demonstrated how the multipoles in
each channel can be directly related to the non-relativistic spin
splitting. 

We hope \textit{MagSymMultipoles} will prove useful to experimentalists
and theorists alike as an intuitive tool for the identification and interpretation of multipolar orders. By connecting Cartesian tensor components,
spherical-harmonic decompositions, physical property tensors,
and interactive 3D visualizations, the framework aims to make high-rank
multipolar symmetry analysis tractable.
The symmetry-allowed multipoles obtained from our method can also be directly compared with
multipole components extracted from charge and spin densities computed with density-functional theory, connecting the symmetry analysis to the calculated electronic structure. Together with
the mapping to physical property tensors, this helps relate
measured responses to compatible multipolar orders and,
conversely, identify the responses allowed by a given order,
supporting both the interpretation of experiments and the
prediction of material properties.

\section*{Acknowledgments}
The authors thank P. Radaelli and A. Cano for helpful discussions, and B. Campbell and H. Stokes for allowing the use of FINDSYM for the symmetry analysis in our backend. M.B. thanks B. Guster, E. Bousquet, and A. Urru for fruitful discussions. This work is supported by the France 2030 government investment plan managed by the French National Research Agency under grant reference PEPR SPIN–MALT (ANR-24-EXSP-0006). Computational resources were provided by the GRICAD supercomputing center of Université Grenoble Alpes and GENCI Grant No. 2025-AD010916740.

\bibliographystyle{apsrev4-2}
\bibliography{bib_qm,bib,references,spin_space_groups}

\appendix

\section{Tensor shorthand used in the tables}
\label{app:tensor_shorthand}

The solver applies the intrinsic tensor symmetries before presentation.
Charge moment tensors $\mathcal Q^{(k)}$ and $\mathcal N^{(k)}$ are fully
symmetric in all indices. Vector-valued moment tensors, including
$\mathcal D^{(k)}$ and $\mathcal M^{(k)}$, are symmetric in their leading
spatial indices while retaining a distinguished final vector index.
Dual-derived tensors retain the corresponding symmetry among their spatial indices. To keep high-rank matrices readable, the tables use a compact
shorthand in which the final displayed symmetric pair is replaced by a single
index, while all earlier indices remain explicit. This is a display convention
only and does not modify the tensor or its component symmetries. Ranks~0, 1,
and 2 require no compression. For higher ranks, the shorthand index is
\begin{center}
\begin{tabular}{c|cccccc}
$\nu$ & $1$ & $2$ & $3$ & $4$ & $5$ & $6$ \\
\hline
$(ij)$ & $11$ & $22$ & $33$ & $23/32$ & $13/31$ & $12/21$
\end{tabular}%

\end{center}
For the compact display, the distinguished vector index is placed first.
Thus, the rank-three tensor $T_{ij,k}$ used in the main text is displayed as
$T_{k,\nu(ij)}$, with row $k$ and column $\nu(ij)$. This reordering is a
display convention and does not imply symmetry between the vector index and
a spatial index. At higher ranks, the compressed pair is likewise chosen
only from the symmetric spatial indices. Explicitly, $T_{ijk,l}$ is displayed
with row $l$ and column $i\nu(jk)$, and $T_{ijkl,m}$ with row $m$ and column
$ij\nu(kl)$.

\section{Explicit tensor operations}
\label{app:tensor_formulas}

This appendix collects the explicit Cartesian-tensor operations used in
Sec.~\ref{subsec:reducibility}:
index symmetrization, the symmetric trace-free (STF) decomposition, the
embedding of a lower-rank STF tensor into the traces of a higher-rank
tensor, and the STF representation of the real spherical harmonics. The
conventions follow Ref.~\cite{thorneMultipoleExpansionsGravitational1980}.

\paragraph*{Symmetrization}
Round parentheses around a group of indices denote symmetrization over
those indices, normalized by the number of permutations. For example,
\begin{equation}
\label{eq:symmetrization}
\begin{split}
    T_{ab(cde)}
    =
    \tfrac{1}{3!}\bigl(
    &T_{abcde} + T_{abced} + T_{abdce} \\
    + {}&T_{abdec} + T_{abecd} + T_{abedc}
    \bigr) ,
\end{split}
\end{equation}
where only the indices inside the parentheses are symmetrized. We write
$\mathcal T^{(k)}_{\mathrm{sym}}$ for the fully symmetric part of a
rank-$k$ tensor $\mathcal T^{(k)}$, obtained by symmetrizing over all $k$
indices,
\begin{equation}
\label{eq:Tsym_def}
    \mathcal T^{(k)}_{\mathrm{sym},\,i_1\cdots i_k}
    =
    \mathcal T^{(k)}_{(i_1\cdots i_k)}
    =
    \frac{1}{k!}\sum_{\pi\in S_k}
    \mathcal T^{(k)}_{i_{\pi(1)}\cdots i_{\pi(k)}} ,
\end{equation}
the sum running over all $k!$ permutations $\pi$ of the indices. When
several Kronecker deltas and a lower-rank tensor appear inside a single
pair of parentheses, the symmetrization runs over all free indices
distributed across the product.

\paragraph*{STF decomposition}
The STF projection of a symmetric rank-$k$ tensor
$\mathcal T^{(k)}_{\mathrm{sym}}$ onto its leading $\ell=k$ trace-free
component is
\begin{equation}
\label{eq:STF_projection}
\begin{split}
    \mathcal T^{(k)}_{\mathrm{STF},\,i_1\cdots i_k}
    =
    \sum_{n=0}^{\lfloor k/2 \rfloor}
    a^{n}_{k}\,
    &\delta_{(i_1 i_2}\cdots\delta_{i_{2n-1} i_{2n}} \\
    &\times\,
    \mathcal T^{(k)}_{\mathrm{sym},\,i_{2n+1}\cdots i_k) j_1 j_1 \cdots j_n j_n} ,
\end{split}
\end{equation}
with coefficients
\begin{equation}
\label{eq:a_nk}
    a^{n}_{k}
    =
    (-1)^{n}\,
    \frac{k!\,(2k-2n-1)!!}{(k-2n)!\,(2k-1)!!\,(2n)!!} ,
\end{equation}
where the repeated indices $j_1 j_1\cdots j_n j_n$ are summed and take
$n$ traces of $\mathcal T^{(k)}_{\mathrm{sym}}$. The lower-rank STF
components in Eq.~\eqref{eq:sym_decomposition} follow by applying the same
projection to the successive traces: contracting $p$ index pairs yields a
symmetric rank-$(k-2p)$ tensor, whose STF part is the multipole of angular
momentum $\ell=k-2p$.

\paragraph*{Embedding of a lower-rank STF tensor}
Conversely, a rank-$k$ symmetric tensor is reconstructed from its STF
components by re-embedding each one into the rank-$k$ index space, placing
$p$ Kronecker deltas at the index pairs over which the traces were taken. The
embedded $\ell=k-2p$ component, written $\mathcal T^{(k)}_{\mathrm{sym},(\ell)}$
in Sec.~\ref{subsec:reducibility}, is the rank-$\ell$ STF tensor
carried by $p$ symmetrized Kronecker deltas,
\begin{equation}
\label{eq:embedded_component}
\begin{split}
    \mathcal T^{(k)}_{\mathrm{STF},(\ell),\,i_1\cdots i_k}
    =
    c_{k,p}\,
    &\delta_{(i_1 i_2}\cdots\delta_{i_{2p-1} i_{2p}} \\
    &\times\,
    \mathcal T^{(\ell)}_{\mathrm{STF},\,i_{2p+1}\cdots i_k)} ,
    \quad
    p=\tfrac{k-\ell}{2},
\end{split}
\end{equation}
each $\delta_{i_a i_b}$ acting as a trace placeholder for the index pair
that would be contracted in the corresponding trace extraction. The
full symmetric tensor is the sum of these components over all ranks,
\begin{equation}
\label{eq:STF_embedding}
\begin{split}
    \mathcal T^{(k)}_{\mathrm{sym},\,i_1\cdots i_k}
    &=
    \sum_{\ell = k,\,k-2,\,\ldots}
    \mathcal T^{(k)}_{\mathrm{STF},(\ell),\,i_1\cdots i_k} \\
    &=
    \sum_{p = 0}^{\lfloor k/2 \rfloor}
    c_{k,p}\,
    \delta_{(i_1 i_2}\cdots\delta_{i_{2p-1} i_{2p}} \\
    &\hphantom{{}={}\sum_{p = 0}^{\lfloor k/2 \rfloor}}
    \times\,
    \mathcal T^{(k-2p)}_{\mathrm{STF},\,i_{2p+1}\cdots i_k)} ,
\end{split}
\end{equation}
with coefficients
\begin{equation}
\label{eq:c_kp}
    c_{k,p}
    =
    \frac{k!\,(2k-4p+1)!!}{(k-2p)!\,(2p)!!\,(2k-2p+1)!!} .
\end{equation}
The $c_{k,p}$ are fixed by requiring that successive traces of
Eq.~\eqref{eq:STF_embedding} reproduce the STF components of the
corresponding ranks; equivalently, they invert the projection
Eq.~\eqref{eq:STF_projection}, with $c_{k,0}=1$.

\section{Antisymmetric part of the poloidal double dual}
\label{app:poloidal_antisymmetry}

We take a general moment tensor $\mathcal T^{(k)}_{i_1\cdots i_{k-1},b}$ that is symmetric in the
first $k-1$ indices. Its first Levi--Civita dual,
\begin{equation}
\label{eq:app_first_dual}
    \tau^{(k-1)}_{i_1\cdots i_{k-2},c}
    =
    \epsilon_{cab}\,
    \mathcal T^{(k)}_{i_1\cdots i_{k-2}a,b},
\end{equation}
is symmetric in its first $k-2$ indices, and any trace between its final
index ($c$) and one of the position indices vanishes,
\begin{equation}
\label{eq:app_tau_mixed_trace}
    \tau^{(k-1)}_{i_1\cdots i_{k-3}p,p}
    = 0,
\end{equation}
because $\mathcal T^{(k)}$ is symmetric and $\epsilon_{pab}$ antisymmetric
under $p\leftrightarrow a$.  The poloidal moment is the second dual
\begin{equation}
\label{eq:app_second_dual}
    \pi^{(k-2)}_{i_1\cdots i_{k-3},d}
    =
    \epsilon_{dpc}\,
    \tau^{(k-1)}_{i_1\cdots i_{k-3}p,c}.
\end{equation}

For $k\ge4$, the poloidal moment has a generally nonvanishing antisymmetric
part, so we can apply a third dual transformation, and obtain
\begin{align}
    \epsilon_{qsd}\,
    \pi^{(k-2)}_{i_1\cdots i_{k-4}s,d}
    &=
    \left(
        \delta_{qp}\delta_{sc}
        -
        \delta_{qc}\delta_{sp}
    \right)
    \tau^{(k-1)}_{i_1\cdots i_{k-4}sp,c}\\
    &=
    -\tau^{(k-1)}_{i_1\cdots i_{k-4}pp,q},
    \label{eq:app_antisymmetric_proof}
\end{align}
where the mixed trace $\tau^{(k-1)}_{i_1\cdots i_{k-4}sq,s}$ vanishes by
Eq.~\eqref{eq:app_tau_mixed_trace}.  The right-hand side corresponds to the rank-$(k-3)$
toroidal multipole:
\begin{equation}
\label{eq:app_lower_toroidal_component}
    \tau^{(k-3)}_{i_1\cdots i_{k-4},q}
    \equiv
    \tau^{(k-1)}_{i_1\cdots i_{k-4}pp,q}.
\end{equation}
The antisymmetric part of the rank-$(k-2)$ poloidal moment is therefore
fixed completely by the $\tau^{(k-3)}$ toroidal multipole and is already present in the
decomposition of the rank-$k$ parent tensor.  The antisymmetric part of the second dual thus defines no further independent multipoles.

\end{document}